\documentclass[numberedappendix]{emulateapj}
\usepackage{apjfonts}
\usepackage[a4paper,pdftex]{geometry}
\usepackage{graphicx}

\newcommand{\msun}{M_{\odot}}
\def\Tiny{\fontsize{8pt}{8pt}\selectfont}

\slugcomment{Accepted for publication in the Astrophysical Journal}

\shorttitle{The SINS-zCOSMOS Project}
\shortauthors{C. Mancini et al. 2011}

\begin{document}

\title{The zCOSMOS-SINFONI Project I.: sample selection and natural-seeing observations\,\altaffilmark{*}}

\author{C. Mancini\altaffilmark{1},N. M. F\"orster Schreiber\altaffilmark{2},A. Renzini\altaffilmark{1}, G. Cresci\altaffilmark{3},E. K. S. Hicks\altaffilmark{2,4}, Y. Peng\altaffilmark{5}, D. Vergani\altaffilmark{6}, S. Lilly\altaffilmark{5},M. Carollo\altaffilmark{5},L. Pozzetti\altaffilmark{6},G. Zamorani\altaffilmark{6} , E. Daddi\altaffilmark{7}, R. Genzel\altaffilmark{2}, C. Maraston\altaffilmark{8}, H. J. McCracken\altaffilmark{9}, L. Tacconi\altaffilmark{2}, N. Bouch\'e\altaffilmark{10,11,12}, R. Davies\altaffilmark{2}, P. Oesch\altaffilmark{13}, K. Shapiro\altaffilmark{14}, V. Mainieri\altaffilmark{15},  D. Lutz\altaffilmark{2}, M. Mignoli\altaffilmark{6}, A. Sternberg\altaffilmark{16}}

\altaffiltext{*}{Based on observations obtained at the Very Large Telescope (VLT) of the European Southern Observatory, Paranal, Chile (ESO Programme IDs 079.A-0341, 081.A-0672, and 183.A-0781). Also based on observations with the NASA/ESA {\it Hubble Space Telescope} (HST), obtained at the Space Telescope Science
Institute, and with the {\it Spitzer} Space Telescope, which is operated by the Jet Propulsion Laboratory, California Institute of Technology.}

\altaffiltext{1}{INAF-OAPD, Osservatorio Astronomico di Padova Vicolo Osservatorio 5, I-35122, Padova, Italy}
\altaffiltext{2}{Max-Planck-Institut f\"ur Extraterrestrische Physik, Giessenbachstrasse, D-85748 Garching, Germany}
\altaffiltext{3}{Osservatorio Astrofisico di Arcetri (OAF), INAF-Firenze, Largo E. Fermi 5, I-50125 Firenze}
\altaffiltext{4}{Department of Astronomy, University of Washington, Seattle, WA 98195-1580, USA}
\altaffiltext{5}{Institute of Astronomy, Department of Physiscs, Eidgenossische Technische Hochschule, ETH Zurich, CH-8093, Switzerland}
\altaffiltext{6}{INAF-Bologna, Via Ranzani, I-40127 Bologna, Italy}
\altaffiltext{7}{CEA-Saclay,DSM/DAPNIA/Service d'Astrophysique, F-91191 Gif-Sur Yvette Cedex, France}
\altaffiltext{8}{Institute of Cosmology and Gravitation, University of Portsmouth, Dennis Sciama Building, Burnaby Road, PO1 3HE Portsmouth, UK}
\altaffiltext{9}{IAP, 98bis bd Arago, F-75014, Paris, France}
\altaffiltext{10}{Department of Physics, University of California, Santa Barbara, CA 93106, USA; Marie Curie Fellow}
\altaffiltext{11}{CNRS; Institut de Recherche en Astrophysique et Plan\'etologie [IRAP] de Toulouse, 14 Avenue E. Belin, F-31400 Toulouse, France}
\altaffiltext{12}{Universit\'e Paul Sabatier de Toulouse; UPS-OMP; IRAP; F-31400 Toulouse, France}
\altaffiltext{13}{UCO/Lick Observatory, University of California, Santa Cruz, CA 95064}
\altaffiltext{14}{Aerospace Research Laboratories, Northrop Grumman Aerospace Systems, Redondo Beach, CA 90278, USA}
\altaffiltext{15}{ESO, Karl-Schwarzschild-Strasse 2, 85748 Garching bei M\"unchen, Germany}
\altaffiltext{16}{Tel Aviv University, Sackler School of Physics \& Astronomy, Ramat Aviv 69978, Israel}
\clearpage

\begin{abstract}
The {\it zCOSMOS SINFONI project} is aimed at studying the physical and kinematical properties of a sample of massive $z\sim 1.4-2.5$ star-forming galaxies, through SINFONI near-IR integral field spectroscopy (IFS), combined with the multi-wavelength information from the zCOSMOS (COSMOS) survey. 
 The project is based on 1 hour of natural-seeing observations per target, and Adaptive Optics (AO) follow-up for a major part of the sample, which includes 30 galaxies selected from the zCOSMOS/VIMOS spectroscopic survey.\\
This first paper presents the sample selection, and the global physical characterization of the target galaxies from multicolor photometry, i.e., star formation rate (SFR), stellar mass, age, etc. The H$\alpha$ integrated properties such as, flux, velocity dispersion, and size, are derived from the natural-seeing observations, while the follow up AO observations will be presented in the next paper of this series.
Our sample appears to be well representative of star-forming galaxies at $z\sim 2$, covering a wide range in mass and SFR. 
The H$\alpha$ integrated properties of the 25 H$\alpha$ detected galaxies are similar to those of other IFS samples at the same redshifts. 
Good agreement is found among the SFRs derived from H$\alpha$ luminosity and other diagnostic methods, provided the extinction affecting the H$\alpha$ luminosity is about twice that affecting the continuum.
A preliminary kinematic analysis, based on the maximum observed velocity difference across the source, and on the integrated velocity dispersion, indicates that the sample splits nearly 50-50 into rotation-dominated and velocity dispersion-dominated galaxies, in good agreement with previous surveys. 
\end{abstract}


\maketitle

\section{Introduction}
For some time, the high star formation rates (SFR) and the ensuing stellar mass growth of massive galaxies at early cosmic epochs have been widely attributed to violent major mergers.  Motivated by a number of recent observational findings, our picture is now changing towards one in which smoother yet efficient modes of gas accretion fueling intense star formation play an important, if not dominant, role at $z \sim 2$.  In fact, spatially- and spectrally-resolved integral-field spectroscopy has revealed large rotating disks at $z \sim 2$ with SFRs as high as $\sim 100~M_{\odot}\,yr^{-1}$,  or more, but without any sign of ongoing major merging \citep{2006ApJ...645.1062F,2009ApJ...706.1364F,2009ApJ...697..115C,2006Natur.442..786G,2008ApJ...687...59G, 2011ApJ...733..101G, 2008ApJ...682..231S}. 
In parallel, multiwavelength galaxy surveys have shown that the SFR correlates tightly with stellar mass $M_{\star}$ and is nearly proportional to it, with a small scatter ($\rm \sim 0.3~dex$ in logarithmic units), and that the overall SFR$-M_{\star}$ relation steadily declines from $z \sim 2.5$ to $z \sim 0$ \citep{2007ApJ...670..156D,2007A&A...468...33E,2007ApJ...660L..43N,2009ApJ...698L.116P,2009MNRAS.394....3D, 2009A&A...504..751S,2009ApJ...690..937D}, although somewhat flatter SFR$-M_\star$ relations have also been derived \citep{2010A&A...518L..25R,2011ApJ...730...61K,2011arXiv1107.0317W}.  
Moreover, recent high spatial resolution observations of CO emission from ``normal'' (i.e., non-major merging) massive star-forming galaxies at $z\sim 1 - 2$ have uncovered large molecular gas mass fractions, three to ten times higher than in local massive disks, and indicate that the star formation efficiency (i.e., the surface star formation rate density to surface gas density ratio) does not evolve strongly with cosmic time \citep{2010ApJ...713..686D,2010Natur.463..781T, 2010MNRAS.407.2091G}.

 Thus, the majority of star-forming galaxies appears to be continuously fed by gas, promoting and maintaining star formation, rather than occasionally bursting as a result of a (major) merger.  
The regularity and simplicity of the SFR$(M_{\star} ,t)$ relation all the way to $z\sim 0$ \citep[e.g., ][]{2010ApJ...721..193P,2010ApJ...718.1001B,2010MNRAS.405.1690D}, the mass-metallicity relation \citep{2006ApJ...644..813E}, and the tight fundamental metallicity relation (FMR), between $M_{\star}$, metallicity, and SFR \citep{2010MNRAS.408.2115M}, actually suggest that this gas accretion process is the dominant mechanism driving the mass growth of galaxies over a wide range of redshifts.
This new empirical evidence matches remarkably well with recent hydrodynamical simulations in which massive galaxies acquire a large fraction of their baryonic mass via quasi steady, narrow cold flows or streams that penetrate effectively through the shock-heated media of massive dark matter halos \citep[e.g.,][]{2004MNRAS.347.1093B,2005MNRAS.363....2K,2006MNRAS.368....2D, 2008MNRAS.383..119D,  2009Natur.457..451D}. 
These cold streams consist of smoothly flowing material (whose clumps may be seen as minor merger events) that can sustain elevated SFRs over much longer timescales than major mergers.  Under these conditions, the angular momentum is largely preserved as matter is accreted, early (thick) disks can survive and be replenished, and internal dynamical processes can drive secular evolution of disks and formation of bulges/spheroids at high $z$. Indeed, fragmentation in such turbulent, gas-rich disks tend to form massive self-gravitating clumps that could migrate inwards and coalesce to form a young bulge in $\rm \sim 1 - 2~Gyr$, as argued both on observational \citep[e.g.][]{2008ApJ...687...59G,2011ApJ...733..101G,2009ApJ...692...12E} and on theoretical grounds \citep[e.g.][]{1999ApJ...514...77N, 2002A&A...392...83B, 2004A&A...413..547I,2004ApJ...611...20I, 2007ApJ...658..960C, 2007ApJ...670..237B, 2008ApJ...688...67E, 2009Natur.457..451D,2010MNRAS.404.2151C}. In contrast, other studies suggest that giant clumps may not survive long enough to migrate to the galaxy center since they could be rapidly disrupted by vigorous stellar feedback \citep{2010ApJ...709..191M, 2010arXiv1011.0433G}. 
 
Spatially-resolved gas kinematics of high redshift objects has provided key insights into the mechanisms driving early galaxy evolution in the above framework.
The ``SINS'' project, carried out with the SINFONI integral field spectrograph at ESO VLT, was the first and largest survey with full 2D mapping of the spatial distribution and kinematics of the H$\alpha$ line emission
for 62 massive $z \sim 1.3 - 2.6$ star-forming galaxies,
on typical resolved scales of $\sim4 - 5$ kpc and down to $\sim$1.5~kpc for a subset observed with adaptive optics (AO) \citep[][hereafter FS09, and references therein]{2009ApJ...706.1364F}.
About 1/3 of the SINS galaxies exhibit clear kinematic signatures of ordered rotation in a disk, despite generally irregular H$\alpha$ morphologies. Key properties of these early disks are large intrinsic velocity dispersions and high gas mass fractions of $\sim 30\%$ or more.  Concurring evidence has been found from other dynamical and/or morphological studies \citep[e.g., ][]{2007ApJ...658...78W,2009ApJ...699..421W, 2007ApJ...658..763E, 2008A&A...486..741B, 2008A&A...488...99V,2007ApJ...669..929L,2009ApJ...697.2057L,2008Natur.455..775S,2009A&A...504..789E,2010MNRAS.404.1247J,2011A&A...528A..88G}. 

While the results from SINS and other recent studies have provided new and significant insights into galaxy formation/evolution, the picture remains very incomplete.  Indeed, another key outcome of SINS and other IFS studies is the diversity of dynamical properties among the sample galaxies. Along with large massive disks, these studies have revealed a population of compact gas rich systems, frequent at lower masses, that appear dominated by random motions at the current resolution of 8m class telescopes \citep[][FS09]{2007ApJ...669..929L,2009ApJ...697.2057L}, as well as a number of merging/interacting systems.   
The nature of the dispersion dominated galaxies, and the relationship between all three dynamical classes of high-$z$ galaxies are still unclear. The present new series of SINFONI observations have been conceived to address these, and other related open issues, such as: why these objects do follow the same SFR$(M_{\star} ,t)$ relation, in spite of their morphological and dynamical diversities? and how they relate to the cold-stream  paradigm? 
In an even broader perspective, with these SINFONI observations we would like to start investigating issues such as the formation of galactic bulges, star formation quenching in massive galaxies, and ultimately how the diversity and complexity of the physical processes at work may result in the remarkable {\it simplicity}  in the evolution of galaxy populations as extensively illustrated by \citet{2010ApJ...721..193P}. 

To make progress, it is now important to expand the currently still limited
sample of galaxies at $z\sim 2$ with high S/N information on the kinematics
and distribution of star formation resolved on $\sim 1$~kpc scales.
High sensitivity and high spatial resolution are necessary to address
the ultimate science goals outlined above.  More specifically, these goals
require to study separately structural components within galaxies (disk/bulge
regions, massive star-forming clumps), resolve merging units and more compact
lower-mass systems, and detect kpc-scale perturbations in the velocity fields
and dispersion maps as well as faint emission line profile components signaling
dynamical interactions or processes such as radial inflows/outflows.

This can be achieved with AO-assisted near-IR integral field spectroscopy
with spectral resolution of
$R \equiv \Delta\lambda / \lambda \sim 3000 - 5000$
but it remains prohibitive in terms of telescope time to collect such data
for samples of galaxies of size comparable to that of current spectroscopic
surveys.

Our aim is therefore to observe with SINFONI and AO a controlled set of galaxies at $z\sim 2$, i.e., around the peak of the cosmic SFR density \citep[e.g.][]{2006ApJ...651..142H,2009A&A...504..727L} and AGN activity \citep{2001AJ....121...54F,2004MNRAS.349.1397C}, with well-characterized photometric, spectral, and morphological properties. This will allow a detailed study of their kinematics in conjunction with their stellar mass, star formation, dust extinction, and structural properties. This set of {\it benchmark $z\sim 2$ star-forming galaxies will help to understand the processes at play at the heyday of massive galaxy formation in the much broader sample of galaxies in current spectroscopic and photometric surveys}. In this perspective, we have collected a set of 30 $1.4 \lesssim z \lesssim 2.5$ massive star-forming galaxies from the zCOSMOS-Deep database \citep[][hereafter L07]{2007ApJS..172...70L} spanning the mass range $3\times 10^9 \lesssim (M_{\star}/M_{\odot}) \lesssim 2\times 10^{11}$ and probing the ``main-sequence'' of star-forming galaxies over this mass range.  zCOSMOS-Deep is the most appropriate survey in terms of the combination of field size and spectroscopic completeness at $z \sim 2$. It thus provides the best opportunity to cull our benchmark sample, which needs to satisfy the stringent criteria for feasible AO-assisted SINFONI observations: accurate redshifts at $z \sim1.4 - 2.5$, and the proximity of a bright star for AO correction.  In addition, the extensive multiwavelength observations available for the COSMOS field from the X-ray to the radio regimes provide rich complementary data sets on each target, as well as for other $\sim$ 10,000 star forming galaxies in the same redshift interval.

Our SINFONI programs includes two parts: (i) the natural-seeing {\it ``pre-imaging'' observations}, carried out without adaptive optics correction (no-AO mode) for a total sample of 30 zCOSMOS star-forming galaxies at $1.4\lesssim z \lesssim 2.5$, typically with one hour of total integration time per object, and (ii) the AO follow-up observations for a subset of about 20 sources, selected based on the strength of the H$\alpha$ line emission in the pre-imaging data, and with integration times in the range $\sim$ 4-10 hours per target.  Combined with AO data of another eight galaxies drawn from the SINS survey, the total benchmark AO sample will reach nearly 30 objects for an increase by factors of at least two over currently published AO samples at $z \sim 2$.
 In this paper we present the general properties of the full sample, along with the first results  of our SINFONI+VLT {\it pre-imaging} observations consisting of one hour integrations under natural seeing conditions.  The follow-up  AO-assisted SINFONI observations are now well under way, and  will be presented in a series of forthcoming papers.
The present paper is primarily meant to fully characterize the target galaxies, and compare them to the general galaxy population at the same redshift. This new sample of 30 objects with natural-seeing SINFONI data expands
significantly on similar existing samples at $z \sim 2$ and constitutes the second largest one after the SINS H$\alpha$ survey of 62 objects (FS09).
The results from the seeing-limited SINFONI data discussed in this paper are therefore valuable in their own right, notably in view of future
seeing-limited surveys of larger samples with the near-IR multi-integral
field spectrograph KMOS at the VLT.

The paper is organized as follows. Section~\ref{sec:sample}  describes  the sample selection criteria and Section~\ref{sec:properties} illustrates  the properties of each galaxy in the sample as derived by their photometric spectral energy distribution (SED), and compare them to the properties of other similar galaxy samples from the literature. Section~\ref{sec:sinfoni} focuses on the SINFONI observations, data reduction and the H$\alpha$ line integrated properties. In Section~\ref{sec:ha_prop} we compare the SFR derived from the H$\alpha$ luminosity with those obtained through other SFR indicators. Finally, in Section~\ref{kinem} we perform a classification into rotation-, and dispersion-dominated sources, based on the seeing-limited {\it pre-imaging observations}. Section~\ref{sec:summary} is the summary and conclusions.     
Throughout this paper we adopt a cosmology with H$_0$=71~km\,s$^{-1}$\,Mpc$^{-1}$, $\Omega_{M}$=0.27, and $\Omega_{\Lambda}$=0.73.

\section{Target selection: the zC-SINF sample}\label{sec:sample}
The zCOSMOS-SINFONI sample (hereafter zC-SINF) was extracted from the spectroscopic zCOSMOS-Deep survey collected with VIMOS at the VLT (L07; \citealt{2009ApJS..184..218L}; and in prep.), covering the central~1 deg$^2$ of the whole 2 deg$^2$ COSMOS Survey \citep[]{2007ApJS..172...38S}. With about 10,000 spectra having been obtained, and reliable redshifts derived for over 50\% of them, zCOSMOS-Deep is the largest spectroscopic redshift survey of $1.4<z<3$ galaxies. It is part of the larger zCOSMOS program, including also the `zCOSMOS-Bright' component, i.e. $\sim$20,000 galaxy spectra covering the whole COSMOS field area (1.7 deg$^2$ effective area), and the redshift range $0.1< z\lesssim 1.2$ \citep[L07][and in prep.]{2009ApJS..184..218L}.

\begin{figure}[htbp]
\centering
\includegraphics[width=\columnwidth]{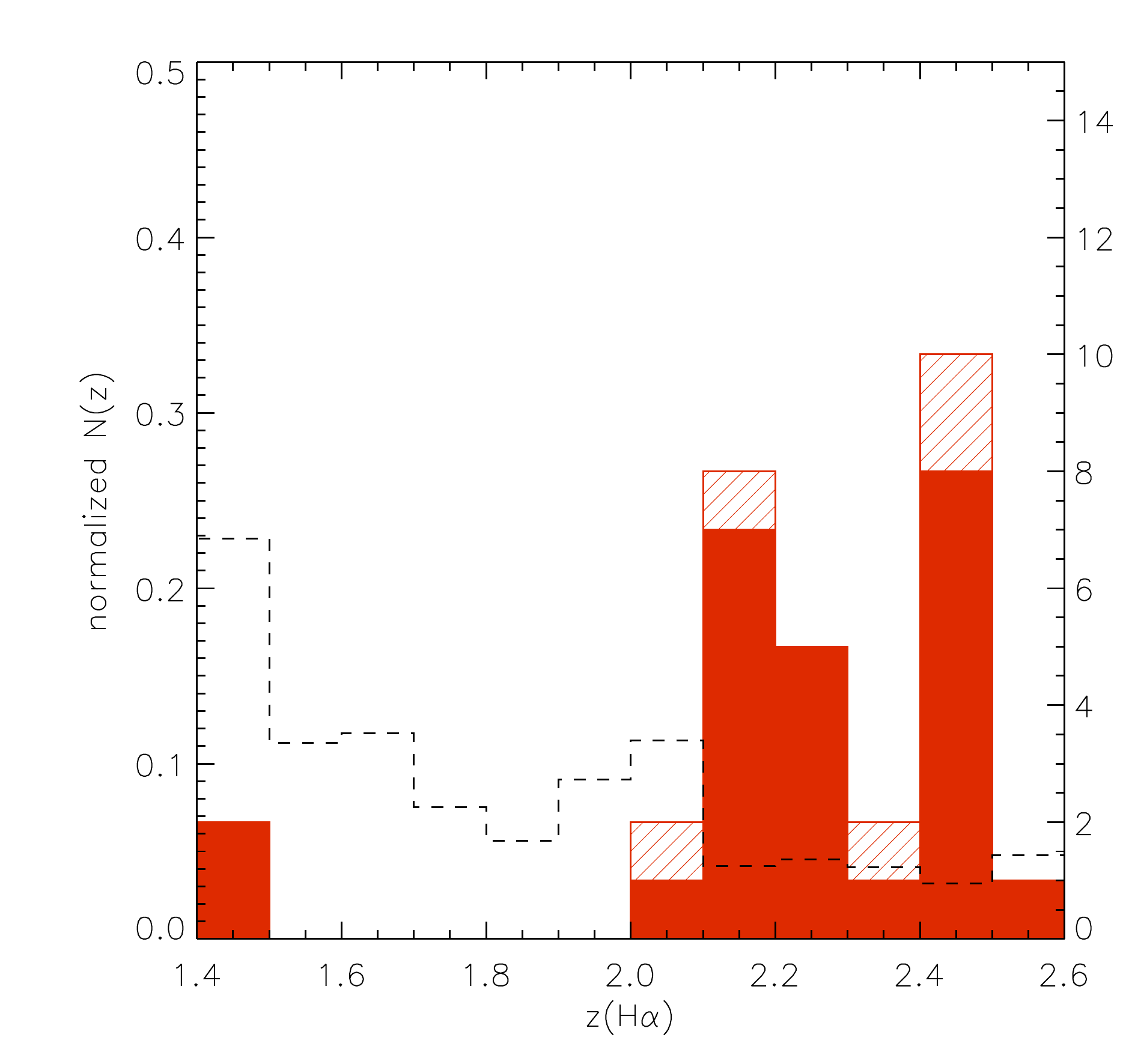}
\caption{\small {The redshift distribution of the  sample of 30 zC-SINF targets (red  dashed histogram, right scale), among which 25 have H$\alpha$ detection with SINFONI (red filled area), is compared with the photometric redshift distribution of all the $\sim 30,000$  BzK-selected  galaxies in COSMOS (normalized black dashed line histogram, left scale). The latter histogram was built by matching the high-$z$ BzK \citet{2010ApJ...708..202M} catalog with the \citet{2009ApJ...690.1236I} photometric redshifts. The observed peaks around $z\sim 1.5$ and $z\sim 2-2.5$ in the red histogram are produced by the constraint of having the  H$\alpha$ line within the SINFONI $H$- or $K$-band, respectively, combined with the other selection criteria, as discussed at the end of Section \ref{sec:sample}.}}\label{fig:histo}
\end{figure}

For the early zCOSMOS-Deep observations, undertaken in 2005, targets were selected using a relatively shallow $K$-band imaging and catalog \citep[$K_{AB}<21.85,$][]{2007ApJS..172...99C}. However, the bulk of the zCOSMOS-Deep targets were observed in 2007 (see L07), by taking advantage of the deeper COSMOS Wircam+CFHT $K$-band imaging, and of the relative $K$-band selected catalog \citep[][]{2010ApJ...708..202M}. This catalog is $\gtrsim 70\%$ complete down to $K_{\rm AB}=23$ for extended sources ($\gtrsim 50\%$ down to $K_{\rm AB}=23.5$). 
The pre-selection of zCOSMOS-Deep spectroscopic targets was based on two-color criteria, both very efficient in selecting $1.4<z<2.5$ galaxies: the $BzK$ \citep{2004ApJ...617..746D}, and the $UGR$ BM/BX optical color criteria \citep{2004ApJ...604..534S}. The combination of both criteria was initially necessary to ensure a sufficient surface density of suitable targets, to take full advantage of the VIMOS instrument multiplex.
The $BzK$ criterion, coupled with a deep $K$-band selection (i.e., $K<23.5$, for zCOSMOS-Deep), enables one to select both passive galaxies ($pBzK$) and star-forming galaxies ($sBzK$, across a range of dust obscuration) at $1.4 \lesssim z \lesssim 2.5$. The most massive ($K$-brightest) galaxies are largely missed by optical (UV rest-frame) color selection criteria \citep[see][for more details]{2004ApJ...617..746D}. 
The $sBzK$-selected objects, the relevant star-forming population in the context of this work, have on average higher mass, reddening, and SFR compared to the $UGR$-selected galaxies at the same redshift ($z\sim2$) \citep[cf.][]{2004ApJ...617..746D, 2005ApJ...633..748R}. Whereas the $BzK$ criterion with deep $K$-band data picks also the vast majority of the BM/BX objects, it could miss some younger and less massive galaxies in the redshift range of interest, and therefore the $UGR$ BX and BM selection criteria  were also used to recover this minor part of the population. In zCOSMOS-Deep $B$-band magnitude cuts were also applied to ensure an adequate S/N ratio in the optical continuum of VIMOS spectra, i.e., $B_{AB}<24.75$, and $B_{AB}<25.25$ for sources with photometric redshift $z_{phot}<2$, and $z_{phot}>2$, respectively. This different depth was dictated by the need of measuring redshifts from relatively low S/N spectra which is harder in the lower redshift interval when Lyman-$\alpha$ is not included in the VIMOS spectral range.
Moreover, the zC-SINF targets were selected to match the following requirements: 

\begin{enumerate}

\item Proximity of a star suitable for natural guide star (NGS) adaptive optics
correction (with $g_{AB}<17$ mag, and within 30$''$ of the galaxy).
A total of 622 of the photometric candidate objects satisfied this criterion,
forming the basis for culling the SINFONI targets according to the next
three selection steps.

\item Spectroscopic redshift reliability: only spectra with the best L07
Confidence Classes, e.g., 4, 3, and 2 were used.  The three classes indicate,
respectively, ``completely secure redshifts'' (i.e., based on unambiguous
multiple spectral features of the expected relative strength that leave no
room for any doubt about the redshift), ``very secure redshifts'' (for which
the classifiers recognize a remote possibility for error, e.g., because
supporting features are in a noisy part of the spectrum), and ``less secure
redshifts'' (i.e., for which the  claimed redshift is the most likely, but
a not negligible possibility remains that the redshift is not correct).
A subset of 249 galaxies out of the 622 candidates close to a suitable
NGS reference star met this requirement.  

\item H$\alpha$ line ``observability:'' we then considered only the sources
with redshifted H$\alpha$ line falling either in the SINFONI $H$- or $K$-band,
within spectral regions of high atmospheric transmission, and at least
$400~km\,s^{-1}$ away from OH airglow lines with typical
intensity $\ga 10\%$ of that of the brightest one in either
$H$ or $K$ band as appropriate.

This criterion constrains the target redshifts either in the range
$z\sim 1.3-1.7$, or $z\sim 2-2.5$,
respectively, as shown in Figure~\ref{fig:histo}, where the spectroscopic
redshift distribution of the zC-SINF sample (red histogram) is compared
with the photometric redshift distribution of the total sample of COSMOS
$BzK$ galaxies at $1.4<z<2.5$
\citep[photometric redshifts from][]{2009ApJ...690.1236I}. 
Together with the OH avoidance requirement for H$\alpha$, this
led to a sample of 65 candidates.
 
\item Based on modeling of the optical to mid-IR spectral energy distribution
(using \citealt{2003MNRAS.344.1000B} with constant SFR, see next Section)
we selected targets so as to probe the widest possible range in stellar mass
($\sim 3\times 10^9-2\times10^{11}\msun$) and
SFR ($\sim 10-300\msun$/yr).
The lower SFR limit was meant to ensure H$\alpha$ detection in 
$\sim 1$~hr integration time with SINFONI in seeing-limited mode
\citep[see, e.g.,][hereafter FS09]{2009ApJ...706.1364F}, and
the resulting sample of candidates included 62 galaxies.\addtocounter{footnote}{-17}\footnote{
The impact of the (a priori unknown) H$\alpha$ size or surface
brightness distribution on source detectability at a given integrated
line flux or SFR is discussed in \S~\ref{sec:syst_unc}.
}

\end{enumerate} 

The final zC-SINF sample we observed with SINFONI is composed
of 30 zCOSMOS-Deep sources.  Among the 62 candidates that satisfy
all criteria listed above, they represent the 30 best ones in terms of
combined optical redshift reliability, H$\alpha$ line observability,
and coverage in $M_{\star}$ and SFR parameter space.
Four targets were observed in earlier campaigns as part of MPE SINFONI
guaranteed time observations (GTO) in the spring of 2007 (P79), and used
as a `pilot sample' for the project (see Table~\ref{tbl-observ-1}). These four
objects are also part of the `SINS H$\alpha$ sample' presented by FS09
and \citet{2009ApJ...697..115C}. 
Although we did not apply any selection criterion based on the galaxy size, relatively extended, low surface brightness sources are likely to suffer a negative bias, when selected as zCOSMOS targets on the base of magnitudes, as for them getting redshifts from VIMOS data was likely harder than for more compact sources.

As shown in Figure~\ref{fig:histo}, 25/30 objects are detected in H$\alpha$
(red histogram). The remaining objects (dashed area in Figure~\ref{fig:histo})
are undetected possibly due to higher noise levels associated with the proximity
in wavelength of night sky lines\footnote{Despite our OH avoidance criterion, it is possible that for some cases, weaker OH lines affected the noise levels sufficiently to hamper detection of a faint H$\alpha$ emission line.  It is also possible that the optical redshift differs from the H$\alpha$ redshift by a few 100 km/s (as for some of the detected objects, see Table~5) so that the H$\alpha$ line falls closer than expected to a bright OH line, preventing detection.},
H$\alpha$ fluxes lower than expected from the spectral energy distributions modeling, surface brightness limitations for more extended sources, or a combination of these effects.
The redshift distribution of zC-SINF targets, shown in Figure~\ref{fig:histo}, is much different from that of the general COSMOS population (dashed-line histogram), due to a combination of biases coming from the selection criteria described above.
First, the requirements to detect  H$\alpha$ in the SINFONI $H$- or $K$-band, coupled with the requirement to avoid atmospheric contamination to H$\alpha$ excludes galaxies in the $1.6<z<1.9$ redshift range. Moreover, the peak observed at $z\sim2.1$ reflects the peak of the spectroscopic redshift distribution of zCOSMOS-Deep galaxies (Lilly et al in prep.), resulting from the convolution of the actual redshift distribution of zCOSMOS-Deep targets with the success rate in redshift determination. Indeed, the most visible consequence of this issue, is the deficit of galaxies at $z=1.4-1.6$ in the final zC-SINF sample (only including two of them). Nevertheless, considering that in the SINS sample there is a fair number of galaxies in the redshift range $1.4<z<1.6$, these two galaxies should not be regarded "in isolation", but as complementing and expanding the sample of galaxies with resolved H$\alpha$ kinematics at these redshifts.

\section{Properties of the targets compared to the broader population of star-forming galaxies at $z\sim 2$}\label{sec:properties} 
In this section we present the multi-band photometric data, and the general properties of the zC-SINF sample, as derived from the Spectral Energy Distribution (SED) analysis. We test the reliability of masses and SFRs derived through SED fitting against the assumptions on stellar population models, star formation histories, and photometric catalogs. Then, we compare the zC-SINF galaxy properties with those of the whole sample of $z\sim 2$ COSMOS galaxies to establish whether our sample is representative of the star-forming galaxy population in the same redshift range. 

\subsection{Multi-band photometry}\label{photometry}
We used the multi-band photometry from the deep WirCam+CFHT (Canada-France-Hawaii Telescope) $K$-band selected catalog  
of \citet{2010ApJ...708..202M}, integrated with the IR/Mid-IR photometry of \citet{2009ApJ...690.1236I} and \citet{2009ApJ...703..222L}. While the published \citet{2010ApJ...708..202M} catalog provides magnitudes for few bands (i.e., $B_J$, $i^+$, $z^+$, $J$, and $K$, see that paper for more details), in this work we used an updated version of the catalog (E. Daddi, and H.J. McCracken private communication), complemented with the WirCam+CFHT $H$-band, and four more optical bands, i.e., the Mega-Prime+CFHT $u^*$-band, and the Suprime-Cam+Subaru $g^+$-, $V_J$-, and $r^+$-band data \citep[][]{2007ApJS..172...99C}.
The u$^*$, g$^+$, V$_J$, r$^+$, and $H$ photometry was derived using the same procedure described in \citet{2010ApJ...708..202M} for the other bands. Magnitudes were measured in 2$''$ diameter aperture, and corrected to total by applying an offset derived for point-like sources within a 6$''$ diameter aperture. For the four new optical bands, the following corrections were applied:

\begin{eqnarray*}
u_{tot}^*= u^* - 0.212 \\
g_{tot}^+= g^+ - 0.245 \nonumber \\
V_{J, tot}= V_J -0.16 \nonumber \\ 
r_{tot}^+=r^+ -0.125\nonumber  
\end{eqnarray*}

\noindent The $H$-band aperture correction depends sensitively on the position in the COSMOS field, varying from $-0.145$ to $-0.215$ magnitudes across the field (E. Daddi private communication).

For the other optical and near-IR bands we refer to the corrections reported in \citet{2010ApJ...708..202M}. However, the above corrections provide the total magnitude for point-like sources, while extended objects need a further correction obtained by comparing the $K$-band aperture total magnitude with the MAG\_AUTO provided by SExtractor \citep{1996A&AS..117..393B}, i.e., the magnitude measured in a Kron-like elliptical aperture (including more than 90\% of the flux), and generally called `auto-offset'. The `auto-offsets' derived in $K$-band for each galaxy, were equally applied to all the optical and near-IR bands, and, for our sample, they typically range from -0.05 to -0.35 magnitudes. 
As mentioned above, the catalog was cross-matched with the four {\it Spitzer}+IRAC channels of \citet{2009ApJ...690.1236I}, covering from 3.6 ~$\mu$m to 8.0~$\mu$m, and with the 24~$\mu$m  {\it Spitzer}+MIPS deep catalog \citep{2009ApJ...703..222L}. The final multi-band photometry from optical to mid-IR for all the zC-SINF sample is reported in Table~\ref{tbl-phot-2} and~\ref{tbl-phot-3}.  

\subsection{SED fitting analysis}\label{sedfitting}

\subsubsection{Assumptions and procedures}\label{sec:tredueuno}
The optical-to-IRAC photometric data reported in Tables~\ref{tbl-phot-2} and~\ref{tbl-phot-3} were used to perform SED fitting analysis to estimate galaxy stellar mass, SFR, dust extinction (A$_V$), and age (i.e., the time elapsed since the beginning of star formation).  In a few cases the sources were blended in some of the IRAC bands (due to the poorer angular resolution compared to the other bands), and such IRAC data were excluded from the fit.
We used the {\it HyperZmass} software, i.e., a modified version of the public {\it HyperZ} code \citep{2000A&A...363..476B}, which fits photometric data points with synthetic stellar population models, and picks the best fit parameters by minimizing the $\chi^2$ between observed and model fluxes. The stellar mass is obtained by integrating the star formation rate over the galaxy age, and correcting for the mass loss in the course of stellar evolution ($\sim 40\%$ for a galaxy of $\sim$ 1~Gyr and a Chabrier IMF) \citep[see][]{2007A&A...474..443P}.
The best fit results strongly depend on the assumptions concerning metallicity, initial mass function (IMF), stellar population models, extinction law and star formation history (SFH). In order to minimize the number of free parameters in this work we used only models built with solar metallicity, and adopted the \citet{2003PASP..115..763C} IMF. 
We fixed the redshift to the spectroscopic values, used the \citet{2000ApJ...533..682C} law to account for dust extinction, and the prescription of \citet{1995ApJ...441...18M} to treat the Lyman-$\alpha$ forest flux decrement. The dust extinction parameter \citep[i.e., $\rm A_V=E(B-V) \times 4.05$, cf.][]{2000ApJ...533..682C} was allowed to range from $A_V=0$ to $A_V=3$ in steps of 0.1. 

As recently discussed in \citet[][hereafter MA10]{2010MNRAS.407..830M}, the assumption on the SFH is crucial to derive astrophysically plausible  best fit parameters, and should be chosen based on the expected properties of the galaxy sample. In particular, MA10 argued that models with constant SFR (CSFR), and the so-called ``$\tau$-models'' characterized by a SFR exponentially decreasing with time as $e^{-t/\tau}$, do not give a realistic representation of $z\sim 2$ star-forming galaxies. When age is left as a free parameter both models lead to  very low ages ($\lesssim 0.1-0.3$~Gyr), which would imply a formation redshift almost identical to the redshift at which the galaxies are observed. As extensively discussed in MA10, such unrealistically young ages are the consequence of the outshining effect by the most recently formed stellar populations, and should not be considered real. In an attempt to circumvent this difficulty, a lower limit (of $\sim$0.1~Gyr) is often imposed to the galaxy age, with the result that many (most) galaxies are found to cluster at such imposed minimum age.  This is because in such fits the current SFR and $A_{\rm V}$ are  dictated by the rest-frame UV part of the spectrum, but within the  assumed SFH  such a SFR would largely overproduce mass (then  violating  the near-IR flux constraints) if not for very short ages. Such unrealistically short ages suffice to demonstrate that the average past SFR must have been much lower than the current values, hence SFR (on average) must has been secularly increasing with time. Thus, the best-fit ages derived using either CSFR or $\tau$-models should rather be interpreted as those of stars providing the bulk of the light in those bands for which the photometry is affected by the smallest errors (as it is built-in in the $\chi^2$ procedure), i.e. in the optical, which corresponds to the rest-frame UV.

Moreover,  ``$\tau$-models'' automatically assume that all galaxies are caught at the minimum of their SFR, while it appears more plausible that $z\sim 2$ star-forming galaxies are close to their SFR peak, given
that this is the cosmic epoch of maximum star formation activity and given the functional form of the SFR$(M_{\star} ,t)$ relation \citep[][MA10]{2009MNRAS.398L..58R, 2010ApJ...721..193P,2010ApJ...718.1001B}. 
Our zC-SINF sample includes by selection only star-forming galaxies at $1.4<z<2.5$, with very young stars boosting the H$\alpha$ emission line. We then adopted two different types of SFH: (i) CSFR with an age lower limit of 0.1~Gyr, using stellar population models from both \citet{2005MNRAS.362..799M}, and \citet[][hereafter MA05, and BC03, respectively]{2003MNRAS.344.1000B},  and (ii) SFR exponentially increasing with time (SFR $\propto e^{+t/\tau}$), i.e., the so-called ``inverted $\tau$-models'' (inv-$\tau$, cf. MA10), with fixed formation redshift ($z_f \simeq 5$, i.e. fixed ages, see Section~\ref{sec:const_vs_invt}), then using MA05 stellar population models. 
As MA05 templates (both CSFR  and inv-$\tau$) are built with the \citet{2001MNRAS.322..231K}, or \citet{1955ApJ...121..161S} IMF,
we used the following relations to correct galaxy stellar masses, according to the adopted \citet{2003PASP..115..763C} IMF: log$(M_{\star,\rm Chabrier})= {\rm log}(M_{\star,\rm Kroupa})-0.04$, and ${\rm log}(M_{\star,\rm Chabrier})= {\rm log}(M_{\star,\rm Salpeter})-0.23$. The same corrections apply to {\rm log}(SFR). 

In Appendix~A, Figure~\ref{fig:sed} (entirely published as online-only material) shows the best-fit SEDs that we have obtained with the three mentioned combinations of models and SFHs. The corresponding best-fit parameters are listed in Table \ref{tbl-sedfit-4}.
It is worth noticing that the best-fit SEDs obtained with different templates are very similar; those obtained with BC03+CSFR and MA05+CSFR models almost fully overlap each other. In the next sub-sections the SED  best-fit parameters (in particular $M*$ and SFR) obtained in the three cases are compared to each other, thus  showing what kind of systematics may affect the quantities derived from SED fitting procedures applied to our $z\sim 2$ galaxies.

\subsubsection{Constant SFR models with different population synthesis models}\label{sec:ma05vsbc03}

The main difference between MA05+CSFR and BC03+CSFR models is that the BC03 population synthesis models  do not include the contribution of the stars in the thermally pulsing asymptotic giant branch (TP-AGB) phase, which instead in MA05 models contribute a significant fraction of the luminosity (especially in the near-IR) for stellar population ages in the  range between  $\sim 0.2$ and $\sim 2$ Gyr. As a consequence, to compensate for the lack of this contribution, BC03+CSFR models tend to overestimate galaxy stellar masses and ages  with respect to MA05+CSFR models for galaxies in which stellar population in this age range dominate \citep[MA05,][etc.]{2006ApJ...652...85M, 2007ApJ...655...51W}.  Moreover, for ages in excess of $\sim 0.1$ Gyr MA05 models tend to have redder $B-V$ colors with respect to the BC03 models, which is due to differences in the adopted stellar evolutionary tracks that affect also other evolutionary phases (e.g., different treatment of overshooting in the stellar interiors; see the top panel of MA05 Figure 27, and also the top-left panel of  the \citealt{2007astro.ph..3052B} Figure 3). The redder colors of MA05 model spectra at given stellar age require younger best-fit ages in order to reproduce the observed colors, and this drives the results towards lower stellar masses and higher SFRs.  

\begin{figure*}[htbp]
\includegraphics[width=\textwidth]{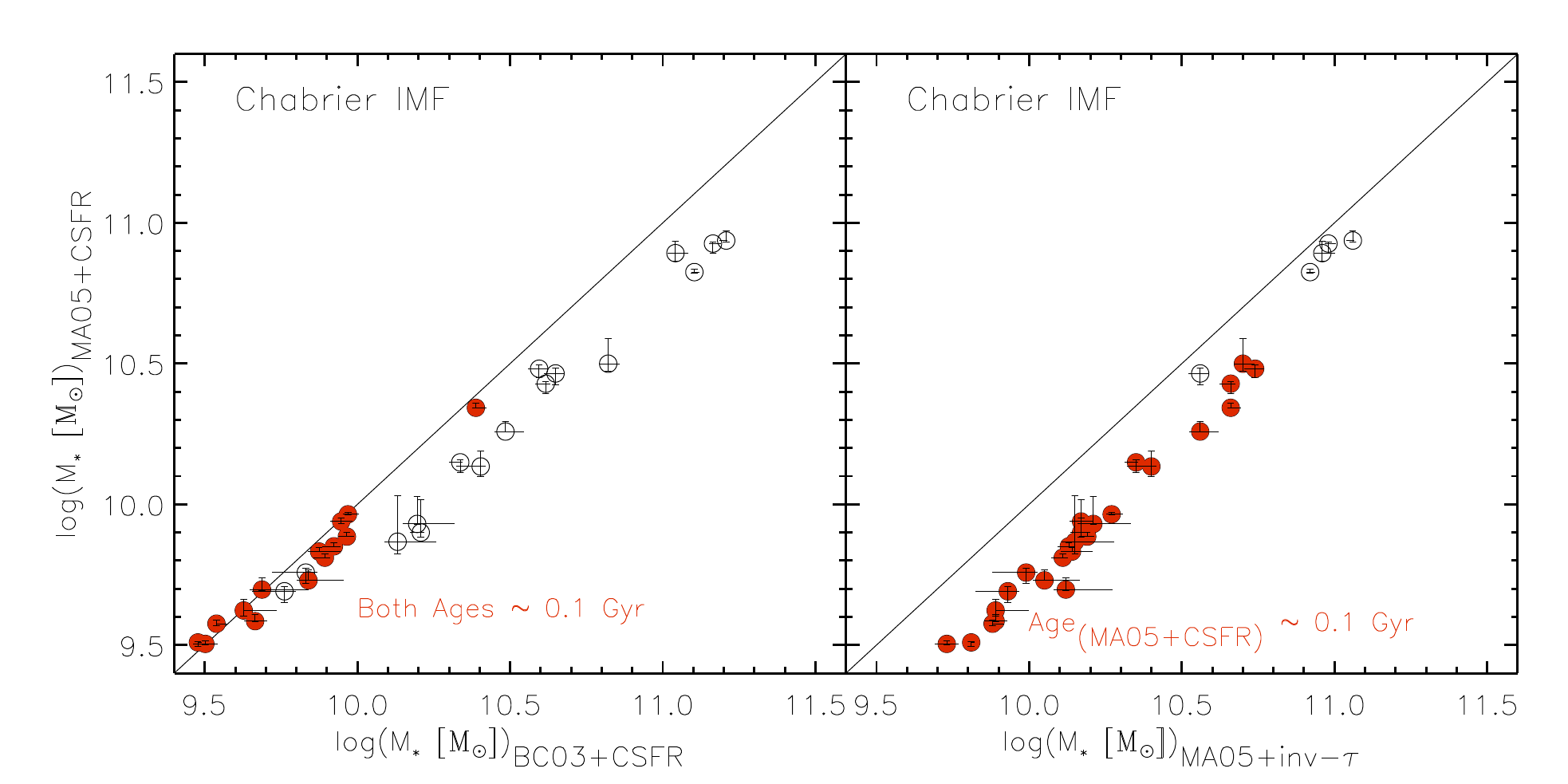}
\caption{\small{Comparison among zC-SINF galaxy stellar masses derived by using different stellar population models, and different SFHs. 
{\bf Left panel:} MA05+CSFR $vs$ BC03+CSFR. Red filled circles are galaxies with both Age$_{(\rm MA05)}$ and Age$_{(\rm BC03)}\simeq$ 0.1~Gyr, i.e. the imposed minimum age, and open circles galaxies with (at least) Age$_{(\rm BC03)}\gtrsim 0.1-0.2$~Gyr (see the discussion in the text).
{\bf Right panel:} MA05+CSFR $vs$ MA05+inv-$\tau$. Here the red filled circles indicate galaxies with Age$_{(\rm MA05+CSFR)}\simeq 0.1$ Gyr (i.e., artificially set to the imposed minimum value), and the open circles galaxies with Age$_{(\rm MA05)}\gtrsim 0.1-0.2$~Gyr. It is evident that the best-fit stellar masses of the latter better agree with masses estimated using inverted-$\tau$ models.}}\label{fig:comp-mass}
\end{figure*}

\begin{figure*}[htbp]
\includegraphics[width=\textwidth]{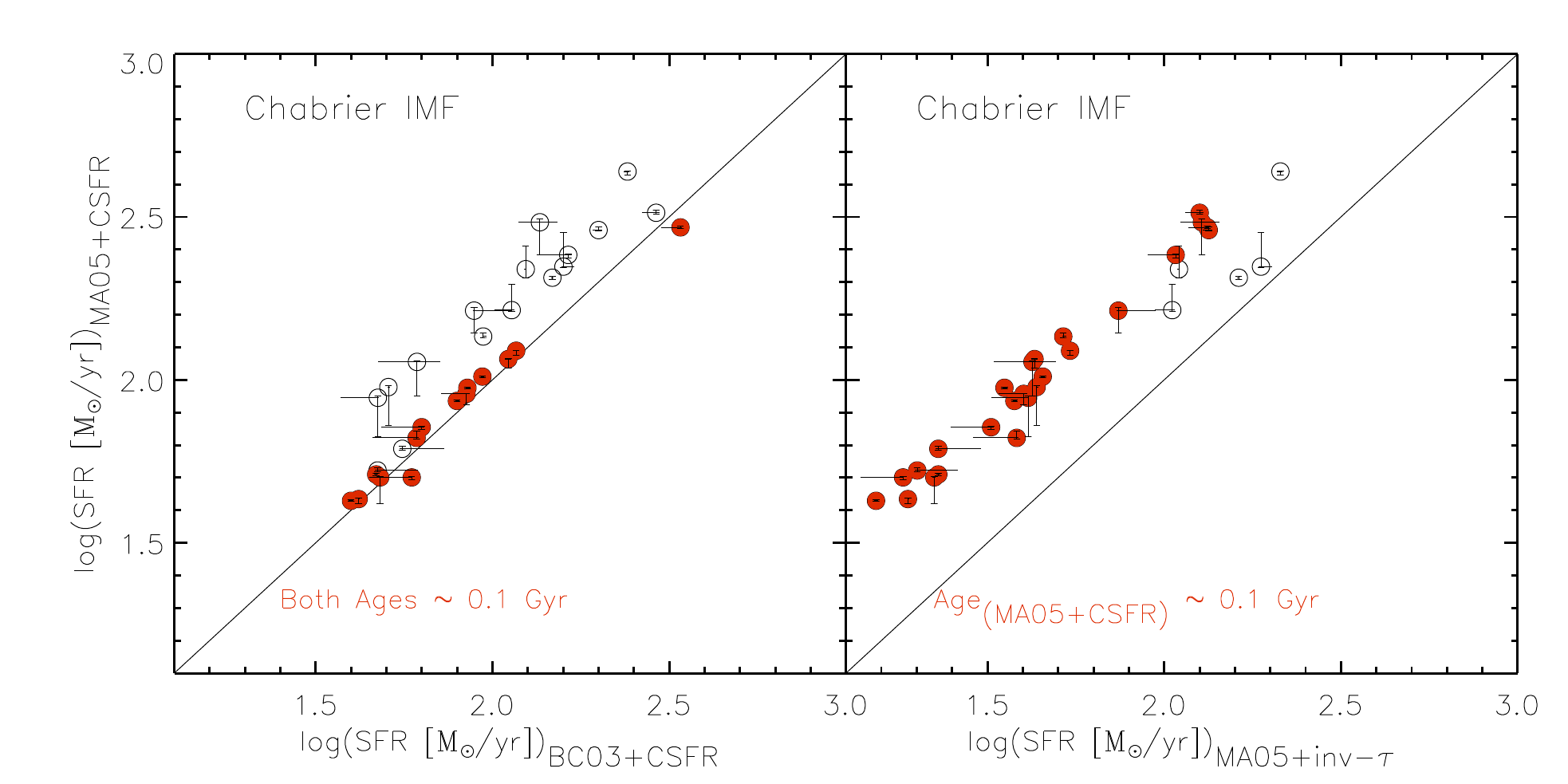}
\caption{\small{Same as Figure~\ref{fig:comp-mass}, but with the SFR, instead of mass. A comparison with Figure~\ref{fig:comp-mass} indicates that inv-$\tau$ models lead to a factor $\sim 3$ reduction in the {\it specific} SFR, compared to CSFR models.}}\label{fig:comp-sfr}
\end{figure*}

This is illustrated in the left panels of Figure~\ref{fig:comp-mass} and Figure~\ref{fig:comp-sfr}  comparing masses and SFRs obtained with the BC03+CSFR and MA05+CSFR models for the zC-SINF sample. 
Note that the ages derived with both MA05+CSFR and BC03+CSFR models are indeed very low for the majority of the zC-SINF galaxies, and tend to cluster to the artificially imposed minimum age (i.e., 0.1 Gyr, see the previous subsection and Table~\ref{tbl-sedfit-4}).  These galaxies are shown as red filled circles in the left panels of 
Figure~\ref{fig:comp-mass} and Figure~\ref{fig:comp-sfr}, and for them MA05+CSFR and BC03+CSFR models 
give consistent results. Instead, results differ for galaxies for which at least BC03+CSFR ages exceed 0.2 Gyr:
MA05+CSFR masses are smaller, and SFRs are higher, both 
by a factor of $\sim 1.5$, compared to those obtained with BC03+CSFR models. 
 These discrepancies can be ascribed to a combination of the two effects mentioned above, MA05 optical colors being redder than BC03 ones for ages older than $\sim 0.1$ Gyr, and near-IR color being redder for ages older than $\sim 0.2$ Gyr.

\subsubsection{Constant SFR models  vs. models with exponentially increasing SFR}\label{sec:const_vs_invt}
 
An option for avoiding excessively young ages from the best fit procedures is to fix the galaxy formation redshift ($z_{\rm f}$, as the beginning of star formation) to some high value. The age then follows from the cosmic time difference between  the formation redshift $z_{\rm f}$ and the redshift $z_{\rm spec}$ at which the galaxy is observed. Galaxies indeed abound at $z>2$ (and even at $z\gg2$), and are the likely precursors of those  we observe at $z\sim 2$.  This latter procedure was adopted in MA10 in conjunction with ``inverted-$\tau$'' models, by fixing $z_{\rm f}\simeq 5$, and $\tau>0.3$ Gyr, and the same recipe is used here, then comparing the results to those of CSFR models. The outcome is fairly insensitive to the precise value of $z_{\rm f}$, because most stars are formed in the last $\tau$ time interval in these inverted-$\tau$ models (MA10).  In practice, since most of the galaxies in our sample are clustered at $z\sim 2-2.5$,  the assumption that star formation begins at $z \simeq 5$ implies galaxy ages in the range $\sim 1.1-2.1$ Gyr. Clearly, the two objects at $z\sim 1.4$ turn out to be older, with ages $\sim 3$~Gyr. The values of $\tau$ selected by the best-fit procedure range from $\sim 0.345$ to 0.89 Gyr.

Stellar masses and SFRs derived assuming a constant SFR and those obtained from inverted-$\tau$ models (in both cases using MA05 stellar population models) are compared in the right panels of Figure~\ref{fig:comp-mass} and Figure~\ref{fig:comp-sfr}. Stellar masses from inverted-$\tau$  models are on average $\sim 1.5$ times higher than those obtained assuming a constant SFR, and the SFRs are $\sim 2$ times lower.  Therefore, specific SFRs (i.e., SFR/$M_\star$) are on average a factor $\sim 3$ lower when adopting inverted-$\tau$ models. As discussed in the previous section, and reported in Table 4, best fit ages are typically very short with CSFR models. Since for such models $M_\star=$Age$\times$SFR (apart from the mass return fraction)
the best-fit procedure picks a higher SFR (and a higher $A_{\rm V}$) in order to match the $M_\star -$sensitive 
near-IR flux. In doing so, the ``best" compromise is one in which $M_\star$ is underestimated and SFR
and $A_{\rm V}$ are overestimated, compared to the inverted-$\tau$ models, which instead are allowed to accumulate mass for typically a $\sim 10$ times longer time.

Incidentally, it is worth noting that such a large difference in specific SFRs would have an impact on the comparison of the integral of the cosmic SFR density with the growth of cosmic stellar mass density \citep[cf.][]{2008MNRAS.385..687W}. Note that these systematic differences  are much smaller for the few objects for which CSFR+MA05 models give ages older than 0.2 Gyr (open circles  in Figures~\ref{fig:comp-mass} and~\ref{fig:comp-sfr}) compared to 
those for which  CSFR+MA05 models give ages below 0.2 Gyr (red filled circles).

For the reason mentioned above and in Sections~\ref{sec:tredueuno}, we favor inverted-$\tau$ models for the description of our $z\sim 2$ actively star-forming galaxies. However, in the following we keep reporting all three sets of SED-fitting results (CSFR with BC03 and MA05, and inv-$\tau$ with MA05) in order to better gauge the robustness (or the systematic uncertainties) of the results. We note in fact that the $\chi^2$ procedure often gives formal error in $M_{\star}$ or SFR which are just a few per cent. Such exceedingly small formal errors clearly contrast with the much larger systematic differences noted above. Given the {\it benchmark} nature of these SINFONI galaxies, quoting the results for various assumed SFHs will also favor intercomparisons with future datasets.

\subsection{Comparison with other samples and selection-bias}\label{sec:sel_bias}
\begin{figure}[htbp]
\centering
\includegraphics[width=\columnwidth]{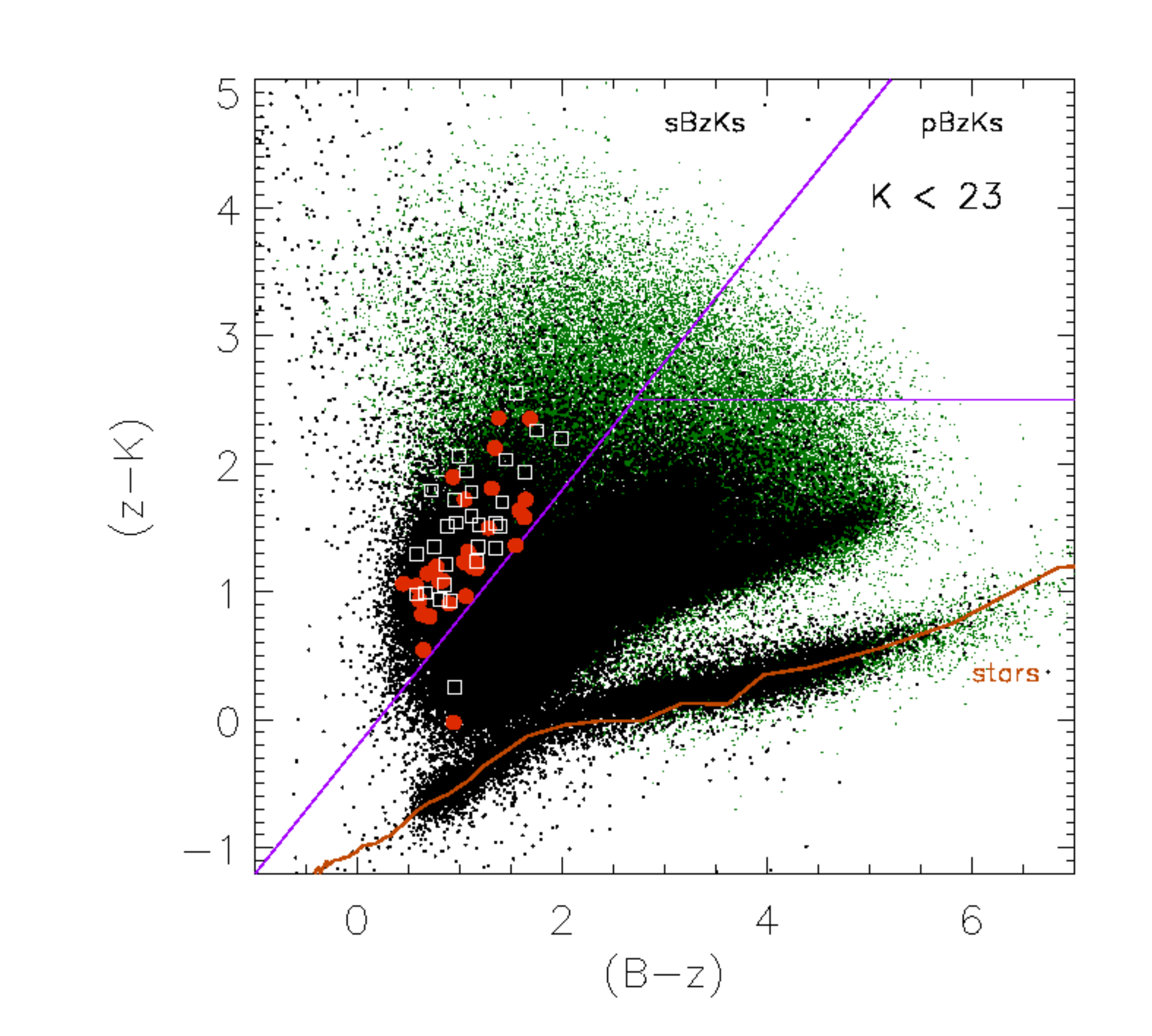}
\caption{\small{BzK diagram for the total 2 deg$^2$ COSMOS sample of $K$-selected sources ($K<23$, \citealt{2010ApJ...708..202M}). 
Following \citet{2004ApJ...617..746D}, the magenta lines mark off the three different regions of $sBzK$, $pBzK$, and low redshift galaxies.
The red curve in the bottom part of the diagram shows the expected colors for the stars \citep[from][]{1997A&AS..125..229L}, overlapping with the locus of real star
sequences. The zC-SINF targets are shown as red filled circles, and are all included in the $sBzK$ left part of the diagram, but one, i.e., 
an $UGR$/BX-selected star-forming galaxy at $z=2.361$, located in the low-redshift area (see discussion in the text). 
Black and green dots are $K<23$ COSMOS $BzKs$ \citep[from][]{2010ApJ...708..202M} with $B_{AB}<25$, and $B_{AB}>25$, 
respectively. The SINS sample of F09 (open white square) is also shown for comparison.}}\label{fig:BzK}
\end{figure}

In this section we compare properties such as colors, masses, and SFRs of the zC-SINF galaxies with those of the full sample of $K<23$ galaxies at $z\sim 2$ in COSMOS (E. Daddi and H.J. McCracken private communication), and with those of the SINS H$\alpha$ sample of FS09 (i.e., a similar sample at the same redshifts). 
These comparisons are merely meant to establish which are the main selection biases affecting the zC-SINF sample, and to which extent it is representative of the general population of galaxies at $z\sim 2$.
Besides the bias relative to the redshift distribution, discussed at the end of Section~\ref{sec:sample}, one of the most significant biases affecting the zC-SINF sample is due to the zCOSMOS-Deep $B_{AB}<24.75-25.25$ cutoff, applied to reach an adequate S/N ratio with VIMOS spectroscopy (see Section~\ref{sec:sample}).  Besides excluding virtually all passively evolving (quenched) galaxies at $z\ga 1.4$, this selection criterion also excludes the reddest $K<23$ star-forming galaxies at $1.4\lesssim z\lesssim 2.5$, i.e., objects highly obscured by dust, which are the majority of actively star forming and massive galaxies in this redshift range \citep{2009Msngr.137...41R}.
This is shown in Figure~\ref{fig:BzK}, where the positions of the zC-SINF galaxies in the BzK diagram (red filled circles) are compared to those of the  general population of galaxies in the COSMOS field from \citet{2010ApJ...708..202M}, down to the (70\% completeness) limit $ K_{\rm AB}<23$. Above the diagonal line, parallel to the reddening vector, lie the $sBzKs$, the upper-right wedge-shaped region is occupied by the $pBzKs$, while the lower right part of the plot is mostly populated by galaxies at $z<1.4$ and by the stellar sequence identified by the red line \citep{2004ApJ...617..746D}.
All the zC-SINF sources are in the $sBzK$ region, but one, i.e., object ZC403103 which was selected for VIMOS spectroscopy following the $UGR$/BX criterion ($u^*-g^+\sim 0.74$ and $g^+-r^+\sim 0.36$).
Although this object lies in the $z<1.4$ galaxy region, it has a solid spectroscopic redshift of $z=2.3613$ (Lilly et al. 2011, in prep.), and its location in the $BzK$ plot is as expected for very young, almost unreddened populations at $z>1.4$ \citep{2004ApJ...617..746D}.
To better illustrate the effect of the $B\lesssim 25$ cutoff used in zCOSMOS-Deep, the COSMOS sample is split in two sub-samples,  with $B_{AB}<25$ (black dots), and $B_{AB}>25$ (green dots), respectively.  Clearly the zC-SINF galaxies are among the bluest $sBzKs$, with $(z-K)$ and $(B-z)$ colors $<2-2.5$, whereas redder $sBzK$ (green dots in the top part of the diagram) are missed by the zCOSMOS-Deep selection.
These are the most dust-obscured galaxies, including the most massive/star-forming objects, which are excluded by the $B\lesssim 25$ constraint \citep[see][]{2009Msngr.137...41R}.
We also compare the colors of the zC-SINF sample with those of the FS09 SINS objects with measured $BzK$ colors (43 out of 62), 
represented by open squares in the figure. The SINS galaxies have $BzK$ colors very similar to the ones of our sample, because comparable optical magnitude cutoffs were also applied for the parent spectroscopic samples from which the SINS targets were drawn (cf. FS09). FS09 have argued that such optical selection excludes more than 50\% of the $z\sim 2 $ galaxies in the same mass range \citep[see also][]{2006ApJ...638L..59V, 2006ApJ...650..624R, 2007A&A...465..393G}, a loss fraction that may exceed $\sim 80\%$ for galaxies more massive than $\sim 10^{10}\,\msun$ in the COSMOS field \citep{2009Msngr.137...41R}.

\begin{figure*}[htbp]
\centering
\includegraphics[width=\textwidth]{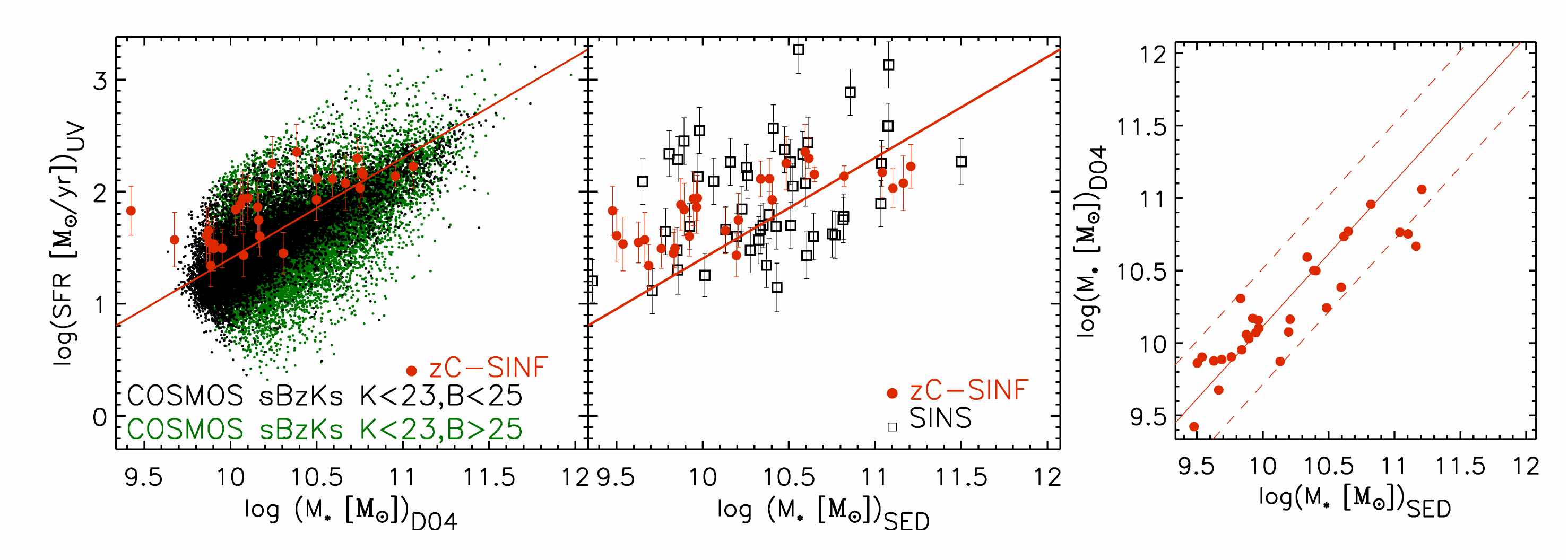}
\caption{\small{
{\bf Left and middle panels}: The SFR$_{UV}$-$M_{\star}$ relation found for the zC-SINF targets (red filled circles) is compared 
with that of the COSMOS $sBzKs$ (dots in the left panel),  and of the FS09 SINS sample (black open squares in the middle panel).
In the left panel the black dots are the COSMOS $sBzKs$ with $B_{AB}<25$, and the green ones are $sBzKs$ with $B_{AB}>25$ (as explained in the text). 
The red solid line show the best-fit of the SFR$_{UV}$-$M_{\star}$ relation for COSMOS $sBzKs$, also reported in the middle panel for comparison.
The zC-SINF targets have quite high SFRs compared to the total sample of $sBzK$, especially at low masses, and cover the full 
mass range of $ K<23$ selected $sBzK$. The resulting SFR-$M_{\star}$ for the zC-SINF sample is therefore appreciably flatter 
than that relative to the full COSMOS sample. Note also the zC-SINF galaxy with the lowest mass (the faintest $K$-band magnitude, i.e., ZC403103), which was selected according to the $UGR$ criterion. {\bf Right panel:} Comparison between zC-SINF stellar masses estimated with two different methods: (i) log$(M_{\star})_{D04}$, used in the left panel, derived from the empirical relation based on the $K$ magnitude 
\citep{2004ApJ...617..746D}; (ii) log$(M_{\star})_{\rm SED}$, used in the middle panel, derived through SED fitting (BC03+CSFR models). 
The solid and dashed lines show the best-fit of the relation between these quantities, and the 1~$\sigma$ scatter, respectively.}}\label{fig:m_vs_sfr}
\end{figure*}

Another bias is illustrated in the left and middle panels of Figure~\ref{fig:m_vs_sfr}, where the SFR-$M_{\star}$ relation of the zC-SINF sample is compared with that of the whole sample of COSMOS $sBzK$ galaxies (left panel), and with the SINS sample (open squares in the middle panel). While our zC-SINF sample covers, by selection, the same mass range of COSMOS $sBzKs$, at the low-mass end of the SFR-$M_{\star}$ relation, it includes preferentially galaxies with the highest SFR due the lower limit imposed to the SFR ($>10~M_{\odot}$/yr, i.e., see item 4 in Section~\ref{sec:sample}). On the other hand, at the high-mass end of the relation we are missing the most star-forming (dusty) objects with $B\gtrsim 25$ (i.e., the green dots in the top-right part of Fig.~\ref{fig:m_vs_sfr}, left panel). 
The combined effect of these selection criteria flattens the SFR-$M_{\star}$ relation of zC-SINF galaxies compared to that of the whole population at $1.4\lesssim z\lesssim 2.5$. 
We note that, in general this effect tends to appear whenever a selection criterion sensitive to the current SFR is applied on a sample of star-forming galaxies previously selected by mass. In fact, especially at low $M_{\star}$ values, galaxies with SFR above the average SFR-$M_{\star}$ relation will be preferentially selected, a typical Malmquist bias.
The middle panel of the same figure shows that the same effect is seen also in the SINS H$\alpha$ sample (open squares; see also FS09), that was selected with similar criteria, and then suffers the same bias of our zC-SINF sample.
All the SFRs reported in Figure~\ref{fig:m_vs_sfr} are derived from the UV rest-frame luminosity (SFR$_{UV}$), according to the relation of \citet{2004ApJ...617..746D}, and adjusted to a Chabrier IMF (as explained in Section~\ref{sfr}). The SFR$_{UV}$ derived for the zC-SINF sample are also listed in Table~\ref{tbl-lum_sfr-6}.
The stellar masses shown on the left panel were determined using the empirical relation based on the $K$ magnitude, also derived by \citet{2004ApJ...617..746D}, so to ensure a homogeneous comparison between zC-SINF galaxies and COSMOS $sBzKs$, for which masses from SED fitting were not available for the full sample.
As the \citet{2004ApJ...617..746D} relation was calibrated on stellar masses derived from SED fitting with BC03 models and Salpeter IMF, we applied the appropriate corrections to Chabrier IMF. 
 On the contrary, all the stellar masses reported in the middle panel are from SED fitting with BC03+CSFR models, to avoid systematics derived from different assumptions in the comparison with the SINS sample (open squares). 
The right panel of the same figure shows the comparison between masses computed with the two methods, which are on average in good agreement, with a scatter of $\sim 0.2$ dex. 

As a result, the zC-SINF sample offers an adequate set of benchmark galaxies for $2 \lesssim z \lesssim 2.5$, with a few selected objects probing later epochs ($z\sim1.5$).

\subsection{Comparing results from different catalogs}
\begin{figure}[htbp]
\includegraphics[width=\columnwidth]{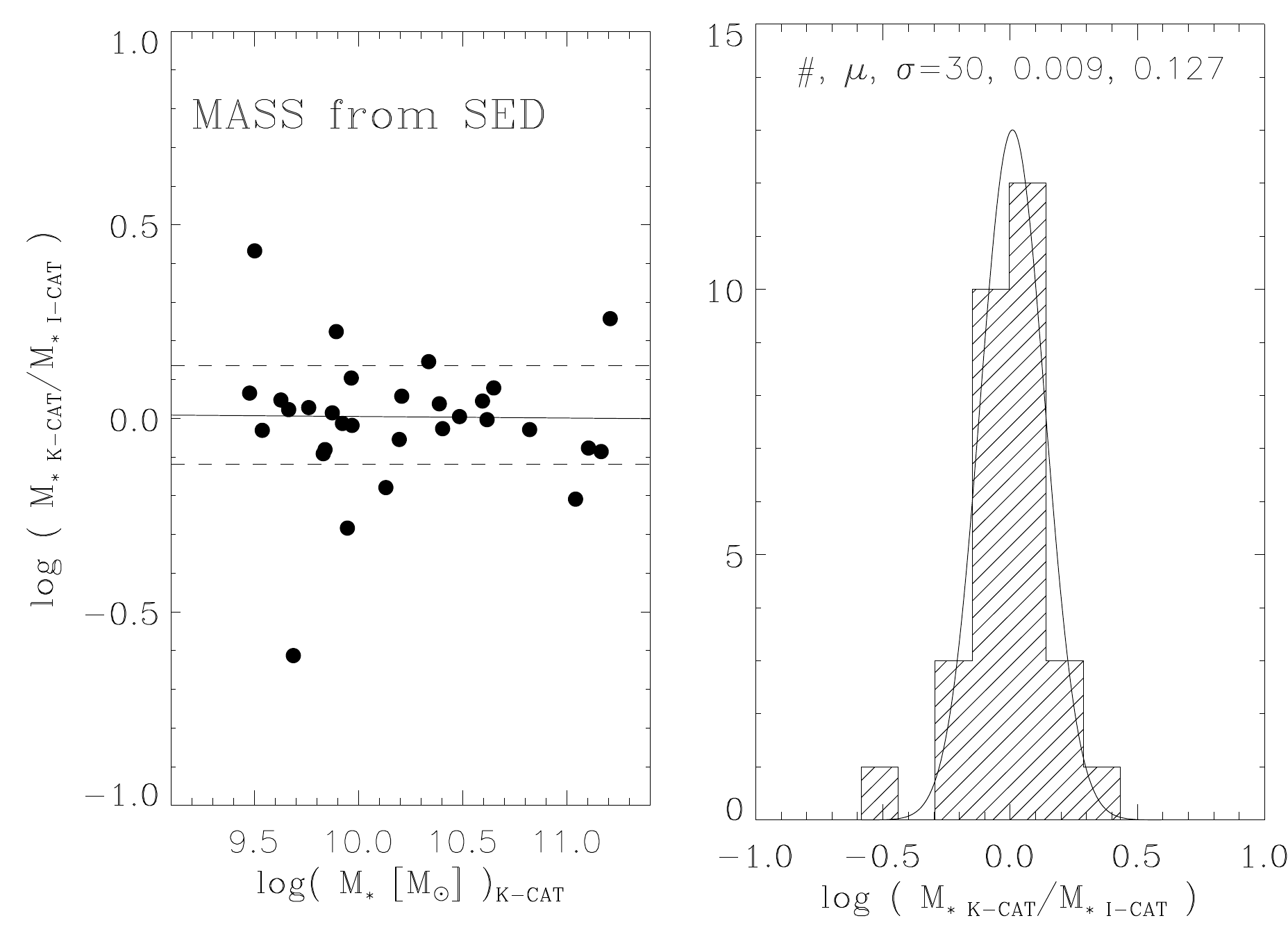}
\caption{\small{{\bf Left panel:} Comparison between stellar masses derived for the zC-SINF sample by fitting SEDs (BC03+CSFR models) built with \citet{2010ApJ...708..202M}, 
and \citet{2009ApJ...690.1236I} photometry (${\rm log}(M)_{K-cat}$, and ${\rm log}(M)_{I-cat}$, respectively).
{\bf Right panel:} the histogram shows the mass-difference distribution of the sample, together with the corresponding Gaussian fit. 
On the top part of the diagram, the number of galaxies used in the fit (\#), the mean ($\mu$), and the standard deviation ($\sigma$) of 
the distribution are labeled.}}\label{fig:mass-Il_McC}  
\end{figure}

\begin{figure}[htbp]
\includegraphics[width=\columnwidth]{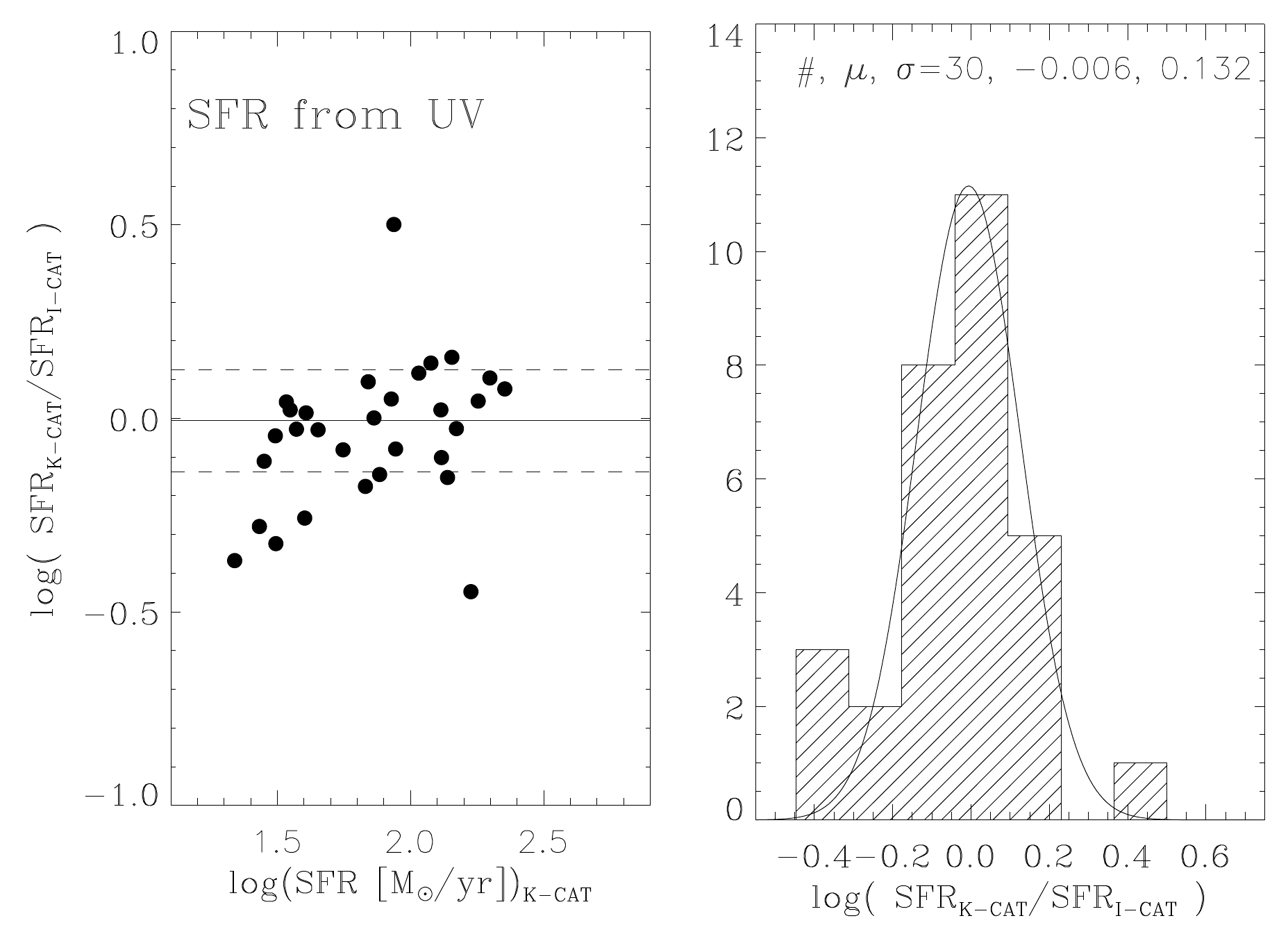}
\caption{\small{Same as Fig.~\ref{fig:mass-Il_McC}, but with SFR$_{UV}$ instead of mass. The SFR$_{UV}$ from \citet[][i.e., ${\rm log}(SFR)_{K-cat}$]{2010ApJ...708..202M} was derived from the $B$-band magnitudes, and that from \citet[][i.e., ${\rm log}(SFR)_{I-cat}$]{2009ApJ...690.1236I} was derived from $g$-band magnitudes as explained in the text.}}\label{fig:sfrUV-Il_McC}
\end{figure}

\begin{figure}[htbp]
\includegraphics[width=\columnwidth]{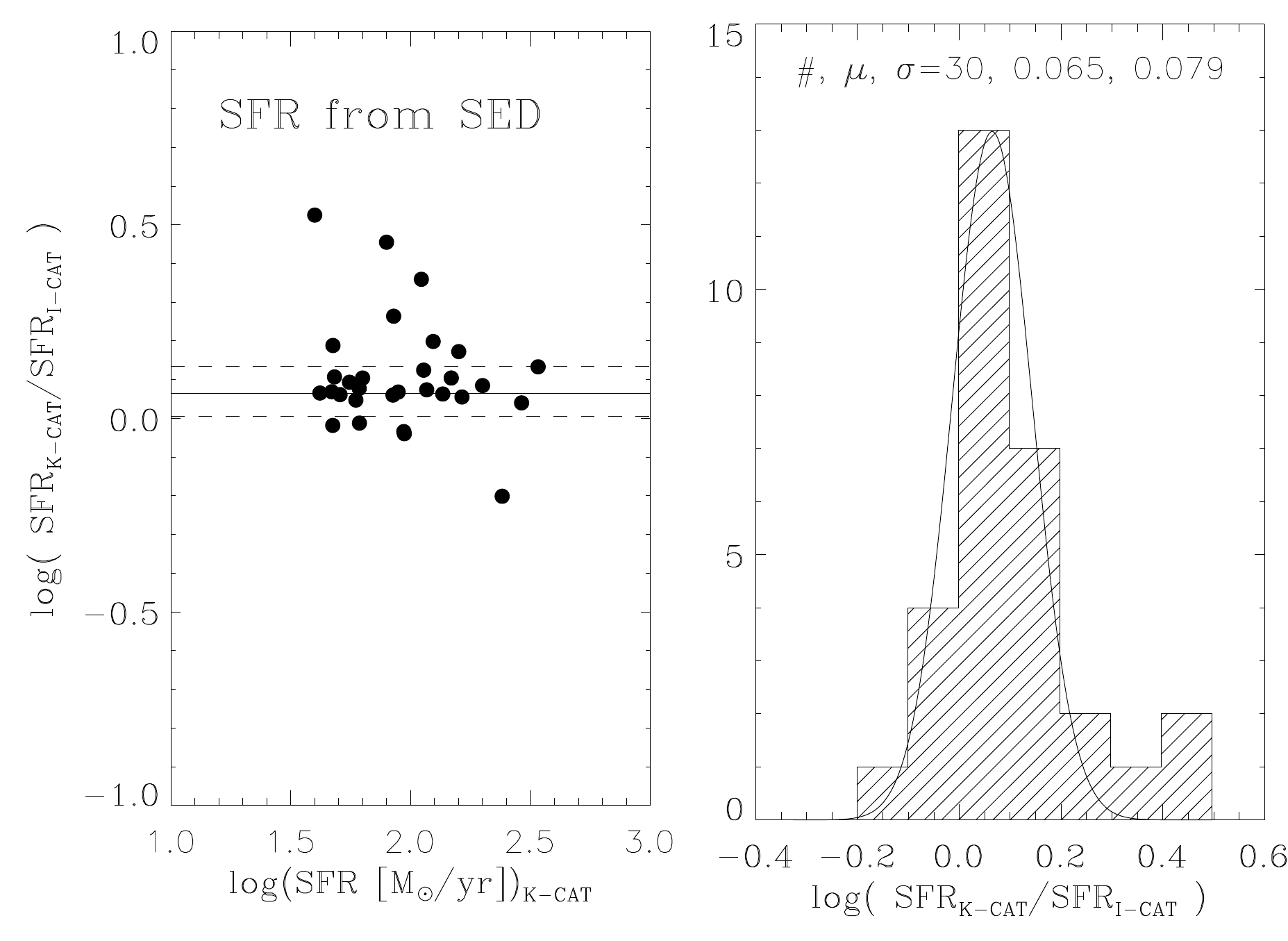}
\caption{\small{Same as Fig.~\ref{fig:mass-Il_McC} and Fig.~\ref{fig:sfrUV-Il_McC}, but with SFR derived from SED fitting (BC03+CSFR models).
}}\label{fig:sfr-Il_McC}
\end{figure}

A final check is relative to the possible impact on derived quantities (i.e., $M_{\star}$, and SFR) of the use of different photometric catalogs for galaxies in the COSMOS field.
We took advantage of the public COSMOS $I$-band selected catalog (hereafter, $I-CAT$) from \citet{2009ApJ...690.1236I} to
compare the results obtained using the $K$-band selected catalog ($K-CAT$) of  \citet{2010ApJ...708..202M}, and therefore quantify the differences in physical quantities when using these two catalogs.
The $I-CAT$ differs from the $K-CAT$ adopted in this work mostly in : (i) the optical selection ($I_{AB}<26$), and source extraction at the $I$-band positions; (ii) the PSF-matching: all the images (from $u^*$ to $K$-band) were degraded to the same PSF of $1\farcs5$ (i.e., the $K$-band PSF) as detailed in \citet{2007ApJS..172...99C}, while no PSF degradation was used in the $K-CAT$; (iii) the magnitudes measurement in 3'' diameter apertures (instead of 2''); (iv) the $I-CAT$ does not include $H$-band photometry. 
We used the same procedure described in Section~\ref{sedfitting} to fit the galaxy SEDs with the $I-CAT$ photometry, using the {\it HyperZmass} software, with BC03+CSFR models and Chabrier IMF.  We also computed the SFR from the UV rest-frame luminosity (i.e., in particular, from the observed $g$-band magnitudes), applying the same relation used for the $K-CAT$ (described in Section~\ref{sfr}). Then, we compare the resulting masses and SFRs with those obtained using the $K-CAT$, and show the results in Figures~\ref{fig:mass-Il_McC}, \ref{fig:sfrUV-Il_McC}, and \ref{fig:sfr-Il_McC}.

Stellar masses from SED fitting and SFR from UV derived using such different catalogs are generally in good agreement with each other, with an average offset of $\mu \left[ log \frac{M*_{K-CAT}}{M*_{I-CAT}}\right]=0.009$, and $\mu\left[log \frac{SFR_{UV, K-CAT}}{SFR_{UV, I-CAT}}\right]=-0.006$, and 1$\sigma$ dispersion $\sigma\left[log \frac{M*_{K-CAT}}{M*_{I-CAT}}\right]=0.127$, and $\sigma\left[log \frac{SFR_{UV, K-CAT}}{SFR_{UV, I-CAT}}\right]= 0.132$, respectively. The corresponding individual differences and their distributions are shown in Figures ~\ref{fig:mass-Il_McC}, and \ref{fig:sfrUV-Il_McC}. In addition to the systematic offset, one notes an apparent trend in the ratio of SFR$_{UV}$ estimates as a function of SFR$_{UV, K-cat}$. We investigated the possible cause of this trend, and found that it is due to a combination of two effects. First, systematic fainter $B$-band magnitudes from the $K-CAT$ with respect to the $g$ magnitudes from $I-CAT$ (with a mean offset of $<B_{K-CAT}-g_{I-CAT}> \sim 0.37$ mag) lead to a smaller $K-CAT$ UV rest-frame luminosity ($L_{1500}$). Second, dust extinction parameters, $A_{1500}$, derived from $I-CAT$ are smaller than those from $K-CAT$ for the most obscured sources (which are also the most star-forming galaxies), while the values are comparable for the less obscured sources. 

On the other hand, when using SFRs from SED fits, a slight systematic shift is present between the SFRs derived from the two different catalogs, i.e., $\mu\left[{\rm log} \frac{SFR_{SED, K-CAT}}{SFR_{SED, I-CAT}}\right]=0.065$, with a dispersion $\sigma\left[{\rm log} \frac{SFR_{K-CAT}}{SFR_{I-CAT}}\right]= 0.079$. This is due to systematic smaller values of the $E(B-V)$ parameter derived fitting SEDs with the $I-CAT$ compared to those obtained from the $K-CAT$. The corresponding individual differences and their distribution are shown in Figure~\ref{fig:sfr-Il_McC}. 

In summary, these comparisons offer an estimate of the reliability of the derived masses and SFRs of zC-SINF galaxies here reported in Table~\ref{tbl-sedfit-4}.
By taking into account the results of these sections, we can state that zC-SINF galaxy masses and SFRs derived from SED fitting can be considered reliable within $\sim 0.2-0.3$ dex (i.e., within a factor of 1.5-2.).

\section{SINFONI Data}\label{sec:sinfoni}
\subsection{Observations}\label{sec:obs}
   
SINFONI \citep{2003SPIE.4841.1548E, 2004Msngr.117...17B} is a near-infrared integral field spectrometer, mounted at the Cassegrain focus of UT4 at the Very
Large Telescope (VLT).  The instrument is a combination of the integral
field spectrograph SPIFFI \citep{2003Msngr.113...17E} and the MACAO
adaptive optics module \citep{2003SPIE.4839..329B}.  SPIFFI slices the
square field of view (FoV) into 32 slitlets and rearranges them into a
one-dimensional pseudo long-slit, which is then fed to the grating
spectrometer part of the instrument.  Each of the 32 slitlets is projected
onto 64 detector pixels, thus the spatial pixels (called {\it spaxels}) on
the sky are rectangular.  The angular size of the slitlets can be selected
between 250, 100, and 25 mas (corresponding to pixel scales of 125, 50,
and 12.5 mas, and FoV of $8''\times 8''$, $3''\times 3''$, and
$0\farcs8\times 0\farcs8$, respectively).  Spatial dithering of on-source
exposures by an odd or fractional number of pixels during the observations
ensures adequate sampling of the spatial axis perpendicular to the slitlets.
In this way, SINFONI records simultaneously the spectrum of every spaxel
across the FoV.  In the reduction procedure, the three-dimensional datacube
is then reconstructed, with two spatial dimensions and one spectral dimension.
SINFONI can be used with or without the AO module, and in the latter case the
spatial resolution is dictated by the natural seeing. 

The SINFONI data presented in this paper were mostly taken as part of an
ESO VLT Large Program (PI: A. Renzini), which
started in February 2009.  As already mentioned, our sample also includes
objects observed in earlier SINFONI programs (see Table~\ref{tbl-observ-1}):
four ``pilot'' zCOSMOS targets observed as part of the MPE SINFONI
Guaranteed Time Observations (GTO; data taken in spring/summer 2007, PI: N.M. F\"orster Schreiber), as well as the five zCOSMOS targets from a Normal Program (PI: A. Renzini) observed in February 2008.
The targets for all these three programs were selected and observed in the
same fashion, and with identical scientific goals.  They are therefore
presented and discussed together in this and future papers of this series. 
In this paper, we focus on the results for the first part of the program
and present the analysis of the no-AO data for the full zC-SINF sample.
While for the AO observations the intermediate SINFONI pixel scale, i.e. 50 mas, will be used to achieve the full gain in resolution afforded by AO, for the no-AO data the largest SINFONI pixel scale of 125 mas has been chosen. The reason is primarily for the higher surface brightness sensitivity, and for a better optimization of the AO follow-up observing strategy. 
Such optimization strategy includes pointing, field rotation, strategy for the sampling of the background emission (appropriate with the smaller FoV used in AO mode), and proper estimate of the total integration time required to reach sufficient S/N per resolution element to achieve our AO science goals. The AO integration time is based on the no-AO H$\alpha$ surface brightness distribution of each individual galaxy, and ranges from about 4 hours to 10 hours. We refer to the forthcoming AO papers for the exact details for each target. 
We used the $H$ or $K$ band gratings in order to map the rest-frame
6562.8~\AA\ H$\alpha$ emission line for galaxies in the redshift ranges
$1.4\lesssim z\lesssim 1.7$ and $2\lesssim z\lesssim 2.5$, respectively (also taking into account the OH avoidance).  With these choices of grating
and pixel scale, the nominal spectral resolution is $R\simeq 2900$ and 4500
in the $H$ and $K$ bands, respectively.

The natural-seeing observations were carried out in {\it Service Mode}
between February 2008 and April 2010 (see Table~\ref{tbl-observ-1}).  For each
object a typical science ``observing block'' (OB) included six on-source
dithered exposures, each one with exposure time of 600 s, for a total OB
integration time of 1\,hr.  For each pair of successive exposures, the
target was positioned in opposite quadrants or halves of the FoV; additional
small dither offsets of $\approx 1/10$ of the full FoV were applied to avoid
redundant source positions on the detector frame.  With this strategy, the
area with full integration time (hereafter the ``effective FoV'') ranges
from $\approx 4^{\prime\prime} \times 4^{\prime\prime}$ to
$\approx 4^{\prime\prime} \times 8^{\prime\prime}$, depending on the source.
Exposures of the acquisition stars used for blind offsetting to the galaxies
were taken just before the science target data to monitor the seeing and the
positional accuracy.  Standard stars for absolute flux calibration and
correction for atmospheric transmission (late-B, early-A, and G1V to G3V
stars, with near-IR magnitudes of 7-10 mag) were observed with the same
instrument setup, at similar airmass, and close in time to the science
targets.

\subsection{Data reduction}\label{sec:red}

We reduced the data with the {\it SPRED} software, specifically developed for
SINFONI observations \citep{2004ASPC..314..380S,2006NewAR..50..398A}, along
with additional custom routines.  We followed the procedure optimized for
data of faint, high-redshift targets described by FS09, where full details
can be found.  Notably, this procedure features the method developed by
\citet{2007MNRAS.375.1099D} for accurate wavelength calibration and
background subtraction, which substantially reduces residuals from night
sky emission lines.
Each individual exposure is reduced separately and reconstructed into a
wavelength-calibrated and distortion-corrected three-dimensional data cube.
The final reduced cube of each science target is obtained by combining the
spatially-aligned individually reduced data cubes.  We combined the cubes
by averaging them with a sigma-clipping algorithm (typically clipping at
the 2.5~$\sigma$ level).  This step also produces a ``sigma-cube,'' whose
pixel values correspond to the standard deviation of pixels at the same
spatial and spectral coordinate in the cubes for individual exposures
(see Section 4.2 and Appendix C of FS09).

The data of the acquisition and telluric standard stars were reduced
following the same procedure as for the science data.  For each target,
the spatial point-spread function (PSF) was measured on the two-dimensional
image of the acquisition star, obtained by averaging (with $\sigma$-clipping)
all the wavelength channels of the star's reduced data cube.  The stellar
images were fit with Gaussian profiles, which represent adequately the
SINFONI spatial PSF in no-AO mode (see Appendix B of FS09).  The effective
angular resolution of our data sets from the PSF full width at half-maximum (FWHM) ranges from $0\farcs 29$
to $0\farcs84$, with a mean and median of $0\farcs 60$ (corresponding to
$\rm \approx 2.5-7~kpc$ and $\rm \approx 5~kpc$, respectively, at $z\sim 2$),
and is listed in Table~\ref{tbl-observ-1}, together with information about the observing period, the used grating, and the total integration time (T$_{int}$[s]). The conversion from instrumental counts to flux units were derived by
comparing the integrated star counts from the synthesized broad-band image
of the star to its 2MASS catalog magnitude.

The effective spectral resolution of the reduced data was measured from
unblended, bright night sky lines based on ``sky'' data cubes created in
a similar way as the reduced science data cubes but without background
subtraction.  The night sky line spectra were extracted by integrating
over a $\sim$30 pixel-wide square aperture.
In reduced, seeing-limited data taken with the 125~mas pixel scale, the
night sky line profiles are well approximated by a Gaussian function and
show little dependence on spatial position.  The FWHM in wavelength units
is constant across each band and, for our data, corresponds to $\approx 120$ and
$\rm \approx 85 km\,s^{-1}$ in $H$- and $K$-band, respectively.

\subsection{Measurements of the H$\alpha$ emission line properties}\label{Ha}

\subsubsection{Integrated H$\alpha$ properties}

We detected the H$\alpha$ line emission for 25 of the 30 galaxies
observed.  For each detected target, we measured the global H$\alpha$
flux, systemic redshift, and velocity dispersion from spatially-integrated
spectra in circular apertures centered on the centroid of the line emission
determined from the line maps (extracted as described in Section~\ref{PV}).
The aperture size was chosen to enclose the total flux estimated from a
curve-of-growth analysis within the deepest part of the effective FoV.
Typical aperture radii ($r_{\rm ap}$) for our objects range from
$0\farcs 75$ to $1\farcs 5$, and are reported in Table~\ref{tbl-Ha-5}.  

The H$\alpha$ properties were derived using the code LINEFIT 
(developed by the SINS team and described by FS09 and \citealt{2011arXiv1108.0285D}) assuming that a single Gaussian profile represents the intrinsic emission line profile of the source.  In the fitting procedure, LINEFIT takes into account the instrumental spectral resolution by convolving a Gaussian profile with the spectral response function represented
by empirical night sky line profiles (see Section~\ref{sec:red}).
The resulting profile is fit to the observed spectrum by means of a
minimization procedure that accounts for the noise spectrum of the data
through Gaussian weighting, and the formal uncertainties of the best-fit
parameters are derived from 100 Monte Carlo simulations.  The noise
spectrum is computed from the ``sigma cube'' associated with each data
set (Section~\ref{sec:red}) and further accounts for deviations from
Gaussian scaling with aperture size (due to, e.g., correlated noise)
from an analysis of the effective noise properties of each data cube
(see Section 5.2 and Appendix C of FS09 for details).  A linear continuum
component underneath H$\alpha$ was simultaneously fit with the emission
line profile; in most cases, however, the continuum is essentially undetected in our $\rm \sim 1\,hr$ integration SINFONI data.

The total H$\alpha$ fluxes $F({\rm H\alpha})$ are given in Table~\ref{tbl-Ha-5},
along with the vacuum redshifts $z({\rm H\alpha})$ and integrated velocity
dispersions $\sigma({\rm H\alpha})_{\rm integrated}$ corresponding to the central
wavelength and width of the best-fit Gaussian profile.  For undetected
sources, the $3\,\sigma$ upper limits on the total H$\alpha$ fluxes are
reported.  These upper limits were derived based on the noise spectrum
for a fixed circular aperture of $1^{\prime\prime}$ radius centered at the
expected position of the source, assuming that the H$\alpha$ redshift is
equal to the redshift from the zCOSMOS optical spectroscopy and that the
intrinsic line width is $\sigma({\rm H\alpha})_{\rm integrated}=130~km\,s^{-1}$
(the average value for the detected sources).  
We emphasize that the velocity dispersions
listed in Table~\ref{tbl-Ha-5} and quoted throughout this paper are corrected
for the instrumental velocity broadening.  We also note that, given the
measurement procedure, the global $\sigma$(H$\alpha$)$_{\rm integrated}$ values in Table~\ref{tbl-Ha-5} include a contribution from internal velocity gradients
(e.g., due to rotation), as they are obtained from the integrated spectra
without any shift of individual pixels to the systemic velocity.
The uncertainties from the absolute flux calibration are estimated to
be $\sim 10\%$ and those from the wavelength calibration, $\la 5\%$.
Together with uncertainties from the continuum placement and the choice
of aperture for the measurements (see Section~\ref{sec:syst_unc}), we
estimate that systematic uncertainties amount to $30\%$ typically, and up to $50\%$ for the objects with the lowest S/N ratio.

Table~\ref{tbl-Ha-5} also lists the fractional contribution
$\rm frac_{BB}(H\alpha)$ from the integrated H$\alpha$ line
flux to the total broadband flux density from the observed magnitude
in $K$ or $H$ band (Table~\ref{tbl-phot-3}) for the targets at
$z > 2$ or $z < 2$, respectively.
The associated uncertainties reported in the Table correspond to the
formal measurement uncertainties of the H$\alpha$ flux and broadband
magnitudes.  The combined systematic uncertainties (including those of
the absolute flux calibration) are largely dominated by those of the
SINFONI H$\alpha$ data and are thus typically $30\%$, and up
to $50\%$ for the faintest objects.

\subsubsection{H$\alpha$ emission and kinematic maps, position-velocity diagrams, and sizes}\label{PV} 

We extracted maps of the velocity-integrated H$\alpha$ line flux,
velocity (relative to the systemic velocity), and velocity dispersions
for the zC-SINF sample in the same manner as described above for the
integrated properties, but now applying the procedure to the spectrum
of individual spaxels.  Prior to the fitting, the data cubes were lightly
smoothed with 3 pixel-wide median filtering in both spatial and in the
spectral dimensions, in order to improve the S/N ratio.  Because of the
short $\rm \approx 1\,hr$ integration time for our pre-imaging SINFONI data,
the kinematic maps for a large fraction of the sample are significantly
affected by low S/N even after median-filtering, and hamper reliable
extraction of the full two-dimensional information.
Higher S/N as well as higher spatial resolution will be achieved for
the subset of the sample followed-up with SINFONI$+$AO observations.

For the pre-imaging data sets, useful kinematic information can nevertheless
be obtained from position-velocity ($p-v$) diagrams, as well as profiles
of the velocities and velocity dispersions extracted along the major axis of
the objects.  For the $p-v$ diagrams, we used a synthetic slit with width
of 6 pixels ($0\farcs 75$) in all cases, integrating the spectra along
the direction perpendicular to the slit orientation.  
The slit was always positioned so as to pass through
the centroid of the H$\alpha$ emission based on the extracted line map.
Whenever possible,
the slit was aligned with the kinematic major axis identified from the
velocity fields, and defined as the direction of steepest observed velocity
gradient across the source.
In those cases, the H$\alpha$ kinematic major axis is generally consistent with the morphological major axis identified from the H$\alpha$ line maps as well as the {\it HST} ACS $I$-band imaging.
In the other cases (i.e., for data with too low S/N), the slit was aligned with the morphological major axis, and we verified with $p-v$ diagrams extracted along several different P.A.s that larger velocity differences were not measured for different directions than along the morphological P.A.
The maximum observed velocity difference, $v_{\rm obs} = v_{\rm max} - v_{\rm min}$, was derived from the
major axis velocity profile obtained from fits to spectra integrated in
contiguous, 6 pixel-wide apertures equally spaced along the major axis
(with centers separated by 3 pixels). 
Significant velocity differences could be measured in 17 of the 25 detected sources from the pre-imaging no-AO data sets presented in this paper.

To estimate the H$\alpha$ sizes of the detected zC-SINF objects,
we determined the half-light radius based on the curve-of-growth analysis
in circular apertures described in Section~\ref{Ha}.  The intrinsic sizes
$r_{1/2}(\rm H \alpha)$ are corrected for the effective spatial resolution
by subtracting in quadrature the PSF half width at half-maximum (HWHM) appropriate for each data set.
For one source (ZC415087), the observed half-light radius is smaller than
the PSF HWHM; since the PSF was not observed simultaneously with the data,
this could be attributed to variations in time of the seeing between the
observations of the acquisition star used as PSF calibrator and the
science target.  For this galaxy, we adopted the observed size as
upper limit to the intrinsic size. 
The formal uncertainties of the $r_{1/2}(\rm H \alpha)$ were estimated
as in FS09 (and turn out to be very similar as well), accounting for
variations of $\approx 20\%$ in the PSF FWHM between successive OBs and
for deviations from axisymmetry of the PSF profiles (the median ellipticity
of the PSFs associated with our data sets is $\approx 0.1$); the resulting
uncertainties on the $r_{1/2}(\rm H \alpha)$ measurements are typically
$\approx 35\%$ (median).  These uncertainties do not account for other
potentially important factors affecting size determinations, such as
the dependence on the sensitivity of the data, the actual intrinsic
surface distribution of the H$\alpha$ emission, and the measurement
method.  We address these issues further in the next subsection.

The velocity-integrated H$\alpha$ line maps and the position-velocity
diagrams of all detected objects in the zC-SINF sample are shown in 
Appendix~B, Figure~\ref{fig:sinfoni} (entirely published as online-only material), along with the ACS $I$-band images and the source-integrated spectra around H$\alpha$.  
The maximum observed velocity difference measurements are given in Table~\ref{tbl-Ha-5}.

\subsubsection{Impact of the surface brightness sensitivity on the measurements of the H$\alpha$ properties}\label{sec:syst_unc} 

In view of the relatively short integration times of about 1\,hr for the
zC-SINF pre-imaging SINFONI data sets, we examined the possible impact of
surface brightness sensitivity on the derived total H$\alpha$ fluxes and
half-light radii.  For the measurements presented in this work, we chose
a curve-of-growth analysis in circular apertures for its simplicity, and
because it does not rely on assumptions about the intrinsic spatial
distribution of the line emission.
This approach also allowed us to derive in a uniform and consistent manner
size estimates for all detected objects, which span a significant range
in apparent sizes and were observed under different seeing conditions.

We point out, however, that other methods would lead to more
robust size estimates for the better resolved sources with higher S/N
data, as discussed by Bouch\'e et al. (2011, in preparation; see also
FS09).  For instance, the use of elliptical apertures to construct the
curves-of-growth would be more appropriate for sources with elliptical
isophotes, such as inclined disk-like systems.  Our use of circular
apertures gives $r_{1/2}({\rm H\alpha})$ values that are analogous to
circularized half-light radii; the latter are by definition smaller than
the actual half-light radius by a factor of $\sqrt{b/a}$, where $b/a$ is
the ratio of the projected size along the minor and major axes of the
galaxy.  In addition, and more importantly, the curves-of-growth are
affected by the surface brightness sensitivity of the data.
Estimating the impact on the total flux measurements is complicated by
the {\it a priori} unknown details and extent of the intrinsic spatial
distribution of the H$\alpha$ line emission.

To gauge the impact of surface brightness sensitivity on the total
line fluxes, we used data of the SINS H$\alpha$ sample with a range of
integration times from typically a few hours and up to 10\,hr per object.
The curve-of-growth analysis for the 40 SINS objects with integrations longer
than 1\,hr (with mean and median of 4\,hr) was repeated after adjusting the
surface brightness limits of each data set to the equivalent limits for 1\,hr
integration time.
More specifically, we determined the average S/N per pixel at the
radius $r_{\rm ap}$ at which the curve-of-growth converges, based on the
data with full integration time $T_{\rm int}$, and used to compute the
total H$\alpha$ flux.  To mimic the shallower surface brightness limits
for 1\,hour integrations, we scaled the above average S/N per pixel by
$1 / \sqrt{T_{\rm int}}$ appropriate in the background-limited
regime and identified the new $r_{\rm ap}$ for 1\,hour as the radius
at which the S/N per pixel corresponds to the scaled value.  From
the curve-of-growth and total flux measured within this new radius,
we re-calculated the $r_{1/2}({\rm H\alpha})$.  This is clearly a
simplistic approach, but we verified for a few objects that consistent
results would be obtained if we had used cubes combining only a subset
of the exposures amounting to 1\,hour integration time to estimate the
$r_{1/2}({\rm H\alpha})$ in shallower data.

The results indicate that the total flux derived from the
shallower data is on average $\sim 10\%$ lower than for a typical integration
time of 4\,hr, albeit with significant scatter.  Unsurprisingly, the median
differences tend to increase for data sets with higher integrations, up to
$\sim 30\%$ for $\rm 7 - 10\,hr$.  The scatter also increases, probably as a
result of the diverse source morphologies.
This exercise further indicates that the half-light radii could be
underestimated by comparable amounts of $\sim 10\% - 30\%$.
This simple analysis thus suggests that while undoubtedly present, surface
brightness limitations in our zC-SINF pre-imaging data do not appear to lead
to substantial (i.e., by factors of two or more) underestimates of the total
H$\alpha$ fluxes or sizes of the targets, at the very least in a comparative
sense with the deeper SINS H$\alpha$ sample.
The longer integrations of the AO-assisted SINFONI observations planned for
the second part of our program is expected to provide better constraints,
in particular on the H$\alpha$ sizes.

\begin{figure*}[htbp] 
\includegraphics[width=\textwidth]{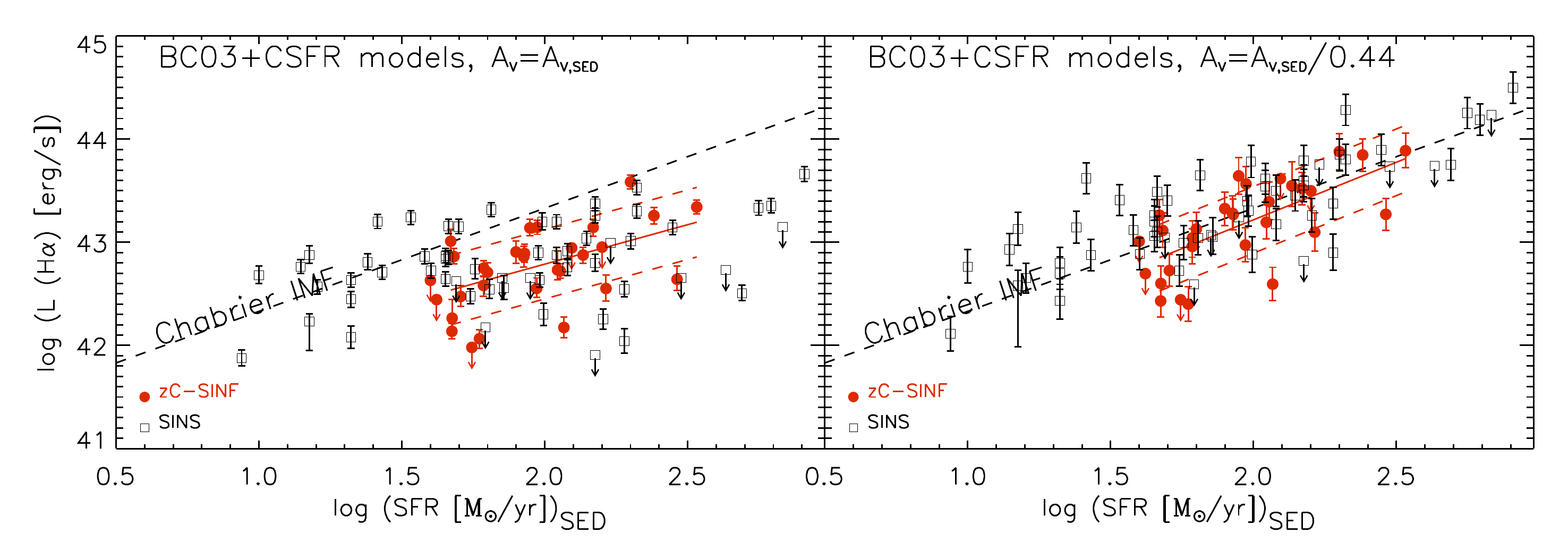}
\caption{\small{The H$\alpha$ luminosities (in erg s$^{-1}$) L(H$\alpha$) $vs$ the SFR obtained from SED fitting with BC03+CSFR models) 
for our zC-SINF sample (red filled circles) and SINS sample (black open squares). H$\alpha$ luminosities of sources undetected in SINFONI are plotted at their 3 $\sigma$ limits (cf. Table~\ref{tbl-lum_sfr-6}). 
{\bf Left panel}: H$\alpha$ luminosity corrected for dust extinction $A_{\rm V}$ identical to that derived from the SED fitting. {\bf Right-panel}: the same relation with enhanced extinction correction as indicated. 
In both the panels the black dashed line shows the SFR-H$\alpha$ luminosity relation from Brinchmann et al. (2004) for a Chabrier IMF. The 
best-fit and $\pm 1~\sigma$ scatter of our data-points are shown by the red solid line, and the red dashed lines, respectively. 
It is evident that in the right panel the best-fit slope better agrees with the one-to-one relation derived from Equation~\ref{eq:1}.}}\label{fig:sfr_bc03}
\end{figure*}

\begin{figure*}[htbp]  
\includegraphics[width=\textwidth]{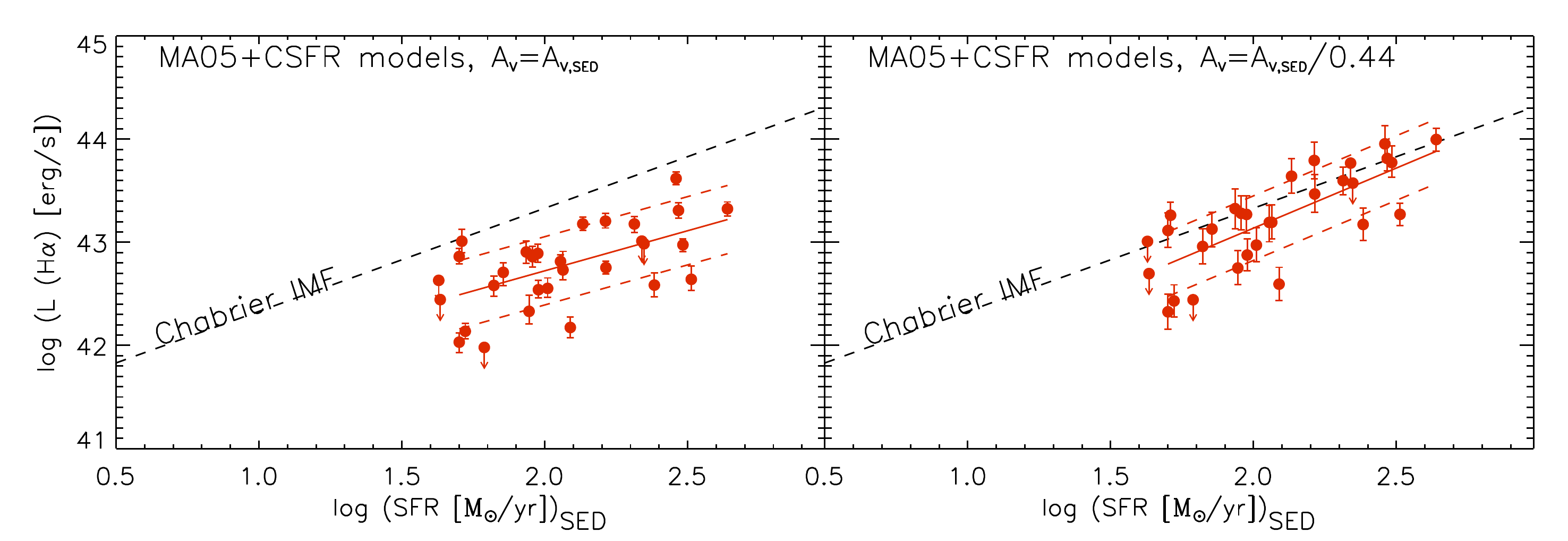}
\caption{\small{The same as in Figure~\ref{fig:sfr_bc03}, but using MA05+CSFR stellar population models. Here we show the results just for the zC-SINF sample, as the SFRs from SED fit are only given for BC03 models for the SINS FS09 H$\alpha$ sample.}}\label{fig:sfr_ma05}
\end{figure*}

\begin{figure*}[htbp]   
\includegraphics[width=\textwidth]{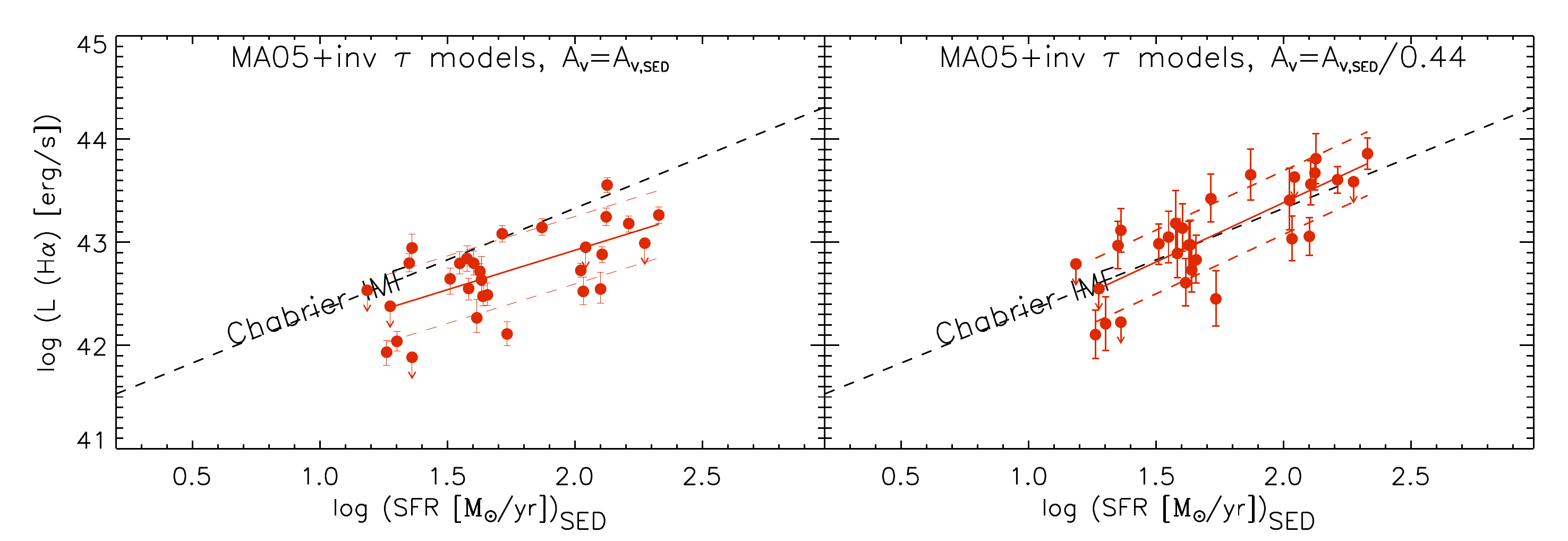}
\caption{\small{The same as in Figure~\ref{fig:sfr_ma05}, but using MA05 inverted-$\tau$ models.}}\label{fig:sfr_ma10}
\end{figure*}

\section{Constraints from H$\alpha$ Properties}\label{sec:ha_prop} 

\subsection{H$\alpha$ Luminosity, Dust Extinction, and Star Formation Rate}\label{Ha_luminosity} 

The observed H$\alpha$ luminosity $L_{obs}(\rm H\alpha$) of each galaxy was then derived from the H$\alpha$ fluxes extracted from the spatially integrated spectra, neglecting any Balmer absorption from the underlying stellar population \citep[][FS09]{2004MNRAS.351.1151B}.  
H\,{\small II} regions suffer dust extinction, hence to find the intrinsic luminosities we correct $L_{obs}(\rm H\alpha)$ for the radiation absorbed by dust. 
In general, the most direct estimate of the amount of dust extinction in the nebular regions is based on the recombination lines ratio H$\alpha$/H$\beta$. 
In this work we did not observe the H$\beta$ emission line, and therefore we used as alternative estimate the extinction based on the color excess (i.e., $E(B-V)$) derived from the galaxy SED modeling, as well as from the observed $B-z$ color, as described in what follows.
By studying local galaxies, several authors \citep{1988ApJ...334..665F,1999A&A...349..765M, 2004ApJ...600..188M,2005MNRAS.358..363C,1994ApJ...429..582C, 2000ApJ...533..682C} found that the Balmer emission  lines of H\,{\small II} regions are more extincted than the UV stellar emission (estimated through SED fitting) by an amount empirically given by the relation proposed by \citet{1994ApJ...429..582C,2000ApJ...533..682C}:
\begin{equation}
A_{\rm V,H{\small II}}=A_{\rm V,SED}/0.44.\label{eq:calz2000}
\end{equation}

In agreement with this prescription, FS09 found  that nebular regions are about two times more extincted than the stellar continuum in their SINS sample of star-forming galaxies at $1.3<z<2.6$.

The first direct measurements of the Balmer decrement (H$\alpha$/H$\beta$ flux ratio) in small $z \sim 1.5 - 2.5$ samples provide further support for extra attenuation towards the H{\small II} regions relative to the bulk of the stellar population that dominates the broad-band continuum light \citep[][Buschkamp et al. 2011, in prep.]{2010ApJ...718..112Y,2010ApJ...725..742M}. 
On the other hand, \citet{2010ApJ...712.1070R} derived similar extinction for H$\alpha$ fluxes and UV continuum for UV-selected objects at the same redshifts (see also Section~\ref{sfr}).  
Thus, in this work we consider both possibilities to correct for dust extinction, and, following the notation of FS09, we denote as $L^{0}(H\alpha)$ and $L^{00}(H\alpha)$ the luminosity corrected with $A_{\rm V}=A_{\rm V,SED}$, and $A_{\rm V}=A_{\rm V,H{\small \rm II}}$ (from Eq.~\ref{eq:calz2000}), respectively (i.e., $L(H\alpha)=L_{\rm obs}(\rm H\alpha)\times 10^{0.33\,A_{\rm V}}$, see Table~\ref{tbl-lum_sfr-6}).

The SFR [$M_{\odot}\ {\rm yr}^{-1}$] is directly related to the H$\alpha$ luminosity [erg\ s$^{-1}$] through the \citet{1998ARA&A..36..189K} relation, that is: 

\begin{equation}
{\rm log(SFR(H}\alpha))={\rm log}(L({\rm H}\alpha))-41.10-0.23, \label{eq:1} 
\end{equation}
 once the correction for the different IMF used in this paper (i.e., -0.23 dex, from Salpeter to Chabrier IMF; see Section~\ref{sedfitting}) has been applied. The resulting H$\alpha$ luminosities and SFRs are then reported in Table~\ref{tbl-lum_sfr-6}.

\subsubsection{Comparison among different Star Formation Rate indicators}\label{sfr}
One of the main advantages of our zC-SINF sample is the availability of both H$\alpha$ integral field spectroscopy, and multi-wavelength photometric information. This allows us to compare the derived H$\alpha$ properties with those estimated from the broadband photometry. 
In particular, in the following sections we compare the SFRs from the H$\alpha$ intrinsic luminosities with those obtained from the SED fits and from the UV rest-frame luminosity.

\subsubsection*{SFR from H$\alpha$ {\it vs} SFR from SED fitting} 
In Figures~\ref{fig:sfr_bc03}, \ref{fig:sfr_ma05}, and~\ref{fig:sfr_ma10} we plot $L^{0}({\rm H}\alpha)$, and $L^{00}({\rm H}\alpha)$, derived in the previous section, as a function of the SFR estimated by fitting the galaxy SEDs with BC03+CSFR, MA05+CSFR, and MA05+inv-$\tau$ templates, respectively (SFR$_{\rm SED}$). From these figures it is evident that for the zC-SINF sample (red filled circles) the slope of the best fit SFR$_{\rm SED}$-$L({\rm H}\alpha)$ relation better agrees with the one-to-one relation derived from Equation~\ref{eq:1} (black dashed line) when H$\alpha$ luminosities are corrected for dust extinction with $A_{\rm V, H{\small \rm II}}$ (Equation~\ref{eq:calz2000}) instead of $A_{\rm V,SED}$. This statement remains true also taking into account the relatively small fraction of H$\alpha$ flux missed by these 1 hr integration time SINFONI data (i.e., $\sim 10-30\%$), see Section~\ref{Ha_luminosity}. Our results show that this is the case regardless of the stellar population models adopted or the SFH assumed. This result is in agreement with the finding by FS09 for the SINS targets (overplotted as open squares in Figure~\ref{fig:sfr_bc03} for comparison), and in contrast to the conclusions by \citet{2006ApJ...647..128E} and \citet{2010ApJ...712.1070R}. 
As already mentioned in Section~\ref{Ha_luminosity}, these other studies of $z\sim2$ galaxies claimed that in their sample H$\alpha$ regions are as extincted as the stellar UV continuum, and no further correction is needed for L(H$\alpha$). 

\citet{2010ApJ...712.1070R} suggested that this discrepancy could be due to the lower stellar masses and SFRs of their sample with respect to the bulk of the SINS sample, and that it is possible that the nebular reddening may on average be larger for galaxies that are forming stars at a higher rate. This lower reddening for lower mass galaxies may be related to these galaxies naturally having lower metallicities. Alternatively, it could also be due to the presence of a more important older and, on average, less attenuated stellar population \citep[e.g., ][]{2009A&A...504..789E}. Our results are consistent with this explanation, as our zC-SINF sample includes  only objects comparable to those with the highest masses and H$\alpha$ luminosities (hence SFRs) of the SINS sample. 
From Figure~\ref{fig:sfr_bc03}, it appears that the SINS objects (open squares) with lower SFRs and masses do not necessarily need a larger dust extinction correction to match the L(H$\alpha$)-SFR$_{\rm SED}$ relation (dashed line), as instead needed for the most massive and active SINS galaxies.

\begin{figure*}[htbp] 								
\includegraphics[width=\textwidth]{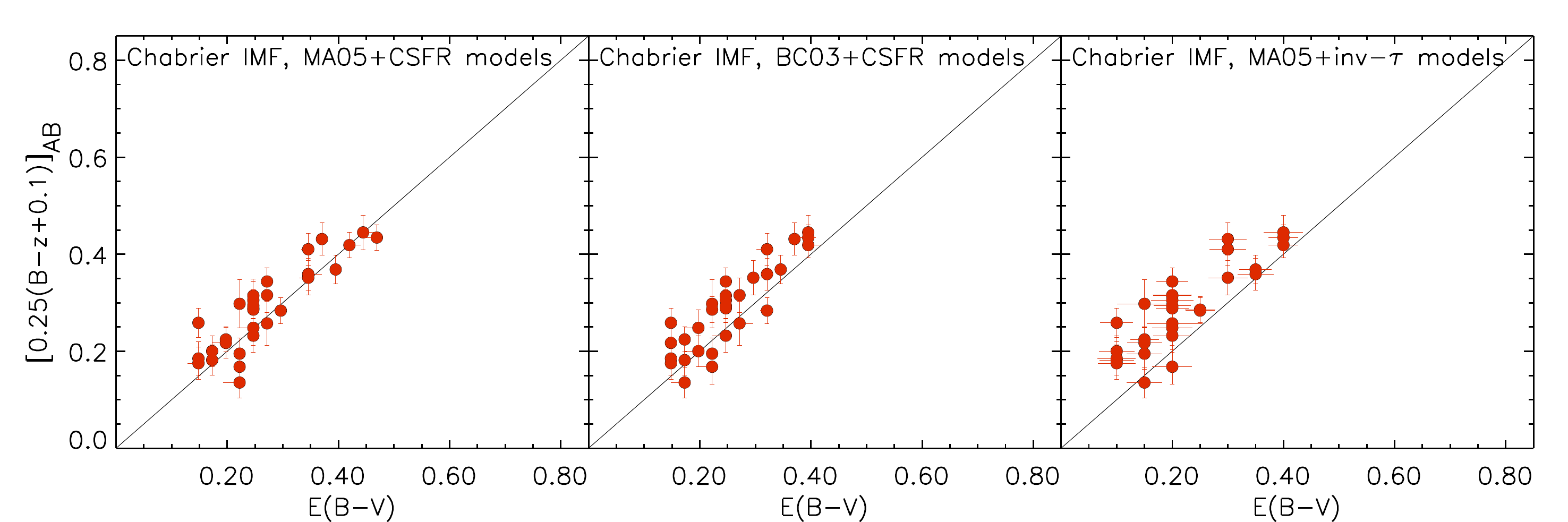}
\caption{\small{Comparison between the $E(B-V)$ reddening parameter computed from the $B-z$ colors with the \citet{2004ApJ...617..746D} relation, and the same parameter derived by fitting the galaxy SEDs with MA05+CSFR, BC03+CSFR, and MA05+inv-$\tau$ models, left, middle and right panel, respectively.}}\label{fig:comp-ebv}
\end{figure*}

\begin{figure*}[htbp]
\includegraphics[width=\textwidth]{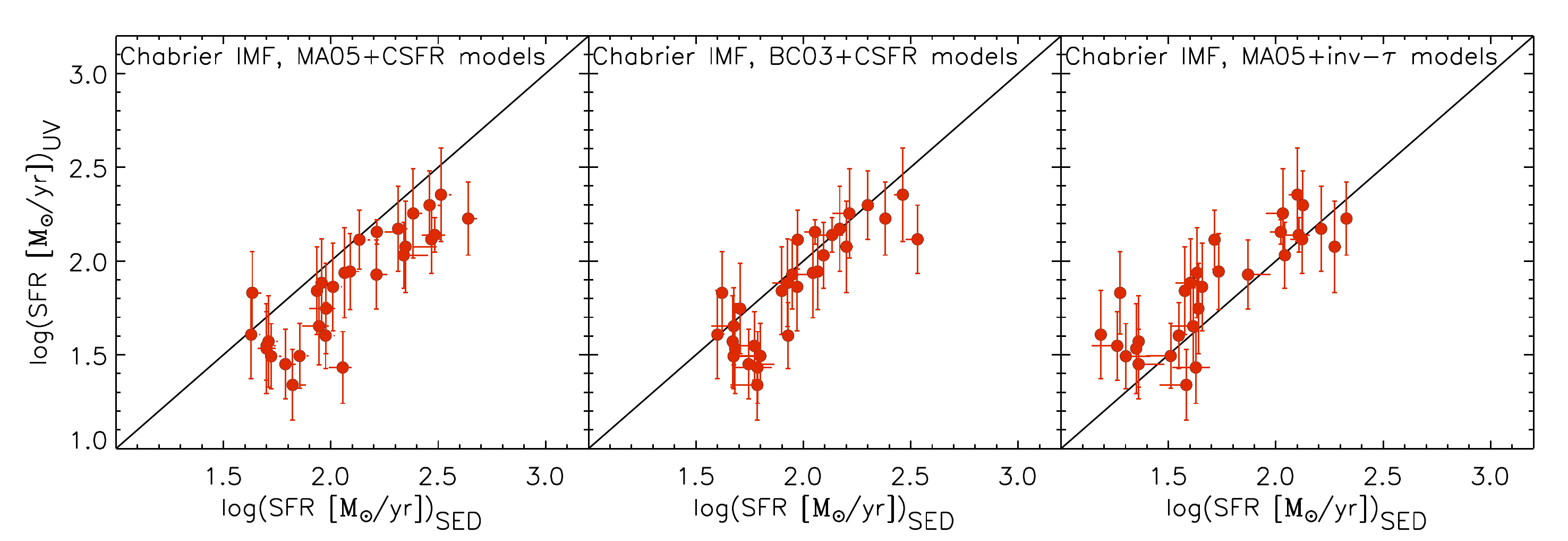}
\caption{\small{Comparison of the SFR derived from the rest-frame UV only (including extinction correction) and those derived from SED fitting with MA05+CSFR, BC03+CSFR, and MA05+inv-$\tau$ models, left, middle, and right panel,  respectively.}}\label{fig:comp-sfr_uv}
\end{figure*}

\begin{figure*}[htbp] 
\includegraphics[width=\textwidth]{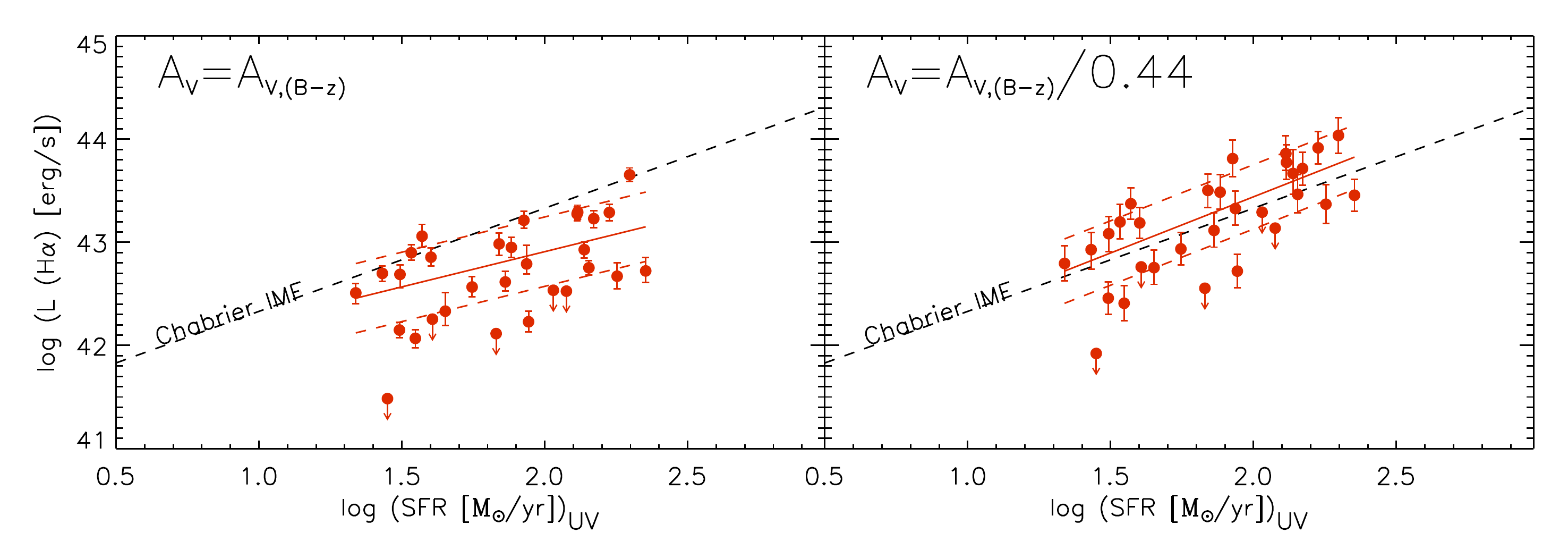}
\caption{\small{The H$\alpha$ luminosity L(H$\alpha$) $vs$ SFR$_{\rm UV}$  corrected for different amount of dust extinction. 
Symbols and lines have the same meaning as in Figure~\ref{fig:sfr_bc03}.}}\label{fig:sfr_uv_ha}
\end{figure*}

\subsubsection*{SFR from H$\alpha$ {\it vs} SFR from UV} 

In star-forming galaxies, the rest-frame UV luminosity over the spectral range of 1500-2800~\AA\ is dominated by the contribution of young (OB) stars, and does not suffer much contamination by the older stellar population. Hence, the L$_{\nu}$ luminosity over this wavelength range, corrected for dust extinction, scales linearly with the SFR and can be used as an efficient SFR indicator. Based on synthetic stellar population models, different calibrations have been obtained to convert UV luminosity to SFR, depending on the reference wavelength used, templates, and IMF \citep[see][and reference therein]{1998ARA&A..36..189K}. In this work, we used the relation relative to 1500~\AA, and adjusted for BC03 models \citep[following][equation 5]{2004ApJ...617..746D}, and for the Chabrier IMF (as in Section~\ref{sec:tredueuno}):

\begin{equation}
{\rm log}(SFR_{UV})= {\rm log}(L_{1500})-27.95-0.23, 
\label{eq:sfr_uv}
\end{equation}

where $SFR_{UV}$ is in unit of [$M_{\odot}\ yr^{-1}$], and $L_{1500}$, in [erg\,s$^{-1}$Hz$^{-1}$], is the intrinsic luminosity already corrected for dust extinction.
As for galaxies at $1.4\lesssim z\lesssim 2.5$ the $\lambda$=1500~\AA\ flux is redshifted in the $B$-band (which actually samples the 1250-1800~\AA\ rest-frame), we used the $B$-band magnitudes and the spectroscopic redshift information to derive the observed 1500~\AA\ luminosities. The intrinsic 1500~\AA\ luminosity, entering equation \ref{eq:sfr_uv}, was obtained by multiplying this observed luminosity by the factor $10^{0.4\,A_{1500}}$, where A$_{1500}$ is the extinction parameter at $\lambda$=1500~\AA. 
We computed A$_{1500}$ from the $(B-z)$ colors according the relation: 

\begin{equation}
A_{1500}=2.5\,(B-z+0.1), \label{eq:daddi_A1500} 
\end{equation}
derived from \citet[][]{2004ApJ...617..746D} Equation 4 (i.e., $E(B-V)_{UV}=0.25\,(B-z+0.1)$ and $A_{1500}=10\,E(B-V)_{UV}$, cf. \citealt{2007ApJ...670..156D}), which provides an approximate estimate for the extinction of star-forming $BzK$-selected galaxies at $z\sim2$.

In Figure~\ref{fig:comp-ebv} and Figure~\ref{fig:comp-sfr_uv} we compare the extinction parameter $E(B-V)_{UV}$, derived from the $(B-z)$ color, and the SFR from the UV rest-frame luminosities (both reported in Table~\ref{tbl-lum_sfr-6}) with those estimated by fitting the galaxy SED with MA05+CSFR, BC03+CSFR, and MA05+inv-$\tau$ models. From these figures it appears that SED fits and UV-derived values of $E(B-V)$ and SFR correlate quite tightly, with some small offset, $\lesssim 0.05$ in $E(B-V)$ and $\lesssim 0.2$ dex in SFR.
We note that in \citet{2004ApJ...617..746D} the relations used to derive both the $E(B-V)$ parameter, and the SFR$_{UV}$ were calibrated using BC03+CSFR models. This could be the cause of the slight offset in the SFR$_{UV}-$SFR(MA05+CSFR) and SFR$_{UV}-$SFR(MA05+inv-$\tau$) relations observed in the left and right panels of Figure~\ref{fig:comp-sfr_uv}. However, the SFRs derived with different stellar population models and SFHs agree within $\sim 0.2$ dex with the SFR$_{UV}$, and notice that the trend is opposite for CSFR and inverted-$\tau$ models, which give systematically lower SFR.

Figure~\ref{fig:sfr_uv_ha} is similar to Figure~\ref{fig:sfr_bc03}--\ref{fig:sfr_ma10}, but shows the $L^{0}(H\alpha)$ and $L^{00}(H\alpha)$ as a function of the SFR$_{UV}$, derived from Equation~\ref{eq:sfr_uv}, rather than from the SED fit. 
As expected, this figure confirms that for our zC-SINF sample the H{\small II} regions appear to be obscured by roughly twice as much compared to the stars. 
It is worth noticing that the $L^{0}(H\alpha)$ and $L^{00}(H\alpha)$ shown in this figure were obtained by using the extinction parameters derived from the $(B-z)$ colors instead of the ones derived from the SEDs. 
This was done to minimize the systematic effects related to different methods to measure the extinction. In fact, we verified that by correcting $L_{obs}(H\alpha)$ with A$_{V,SED}$, obtained using BC03+CSFR models, in Figure~\ref{fig:sfr_uv_ha} we would obtain a different slope (0.72 instead of 1.04) of the best-fit L(H$\alpha$)-SFR$_{UV}$ relation for the zC-SINF objects, and a slightly larger scatter.

\begin{figure*}[htbp]
\includegraphics[width=\textwidth]{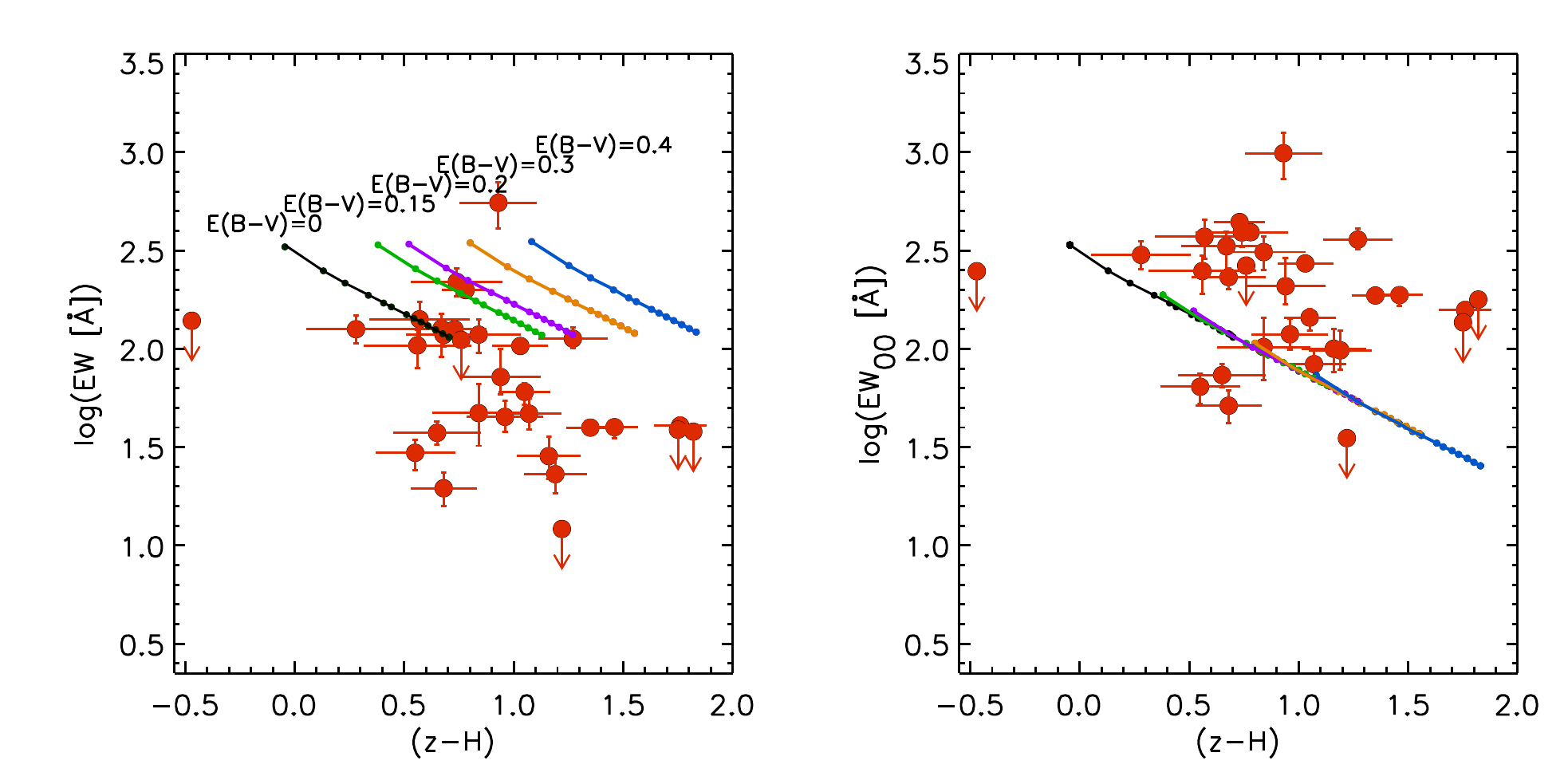}
\caption{\small{The two panels show $(z-H)$ colors {\it vs} H$\alpha$ equivalent widths (EW in logarithmic scale) for the zC-SINF sample (red filled circles), by assuming that nebular regions suffer  the same (left panel) or $\sim$double dust extinction (right panel) with respect to the stars which generate the continuum. The reported uncertainties correspond to the formal measurement uncertainties, and do not take into account other sources of uncertainties (such as those from the absolute flux calibration of the SINFONI and broad-band photometry, or from aperture matching). Equivalent widths of H$\alpha$ undetected sources are plotted at their 3 $\sigma$ limits (cf. Table~\ref{tbl-lum_sfr-6}). The curves show predicted EW and colors for BC03 stellar population models with constant SFR and Chabrier IMF, and different values of dust extinction ($E(B-V)$), as labeled in the left panel. Dots on the curves represent increasing ages in steps of 0.1 Gyr, in the range $0.1-1.5$ Gyr.}}\label{fig:ew} 
\end{figure*}

\subsubsection{H$\alpha$ Equivalent Width} 
For the 25 H$\alpha$-detected objects we computed the rest-frame H$\alpha$
equivalent width (EW(H$\alpha$)) as the ratio of the H$\alpha$ line integrated
flux to the flux density (expressed per unit wavelength)
of the stellar continuum. The latter quantity was derived from $H$, or $K$ broad-band magnitudes, for objects at $z<2$, and $z>2$, respectively, then corrected for the H$\alpha$ line fractional contribution (i.e., frac$_{\rm BB}$(H$\alpha$) in Table~\ref{tbl-Ha-5}). Note that for most of the objects 
frac$_{\rm BB}$(H$\alpha$)$< 10\%$, while only for two galaxies it reaches 20\%--40\%  (i.e., ZC410041, and ZC411737). 
The resulting EW(H$\alpha$)s are reported in Table~\ref{tbl-lum_sfr-6}, where for the H$\alpha$ undetections we list equivalent width upper limits.   
As for frac$_{\rm BB}$(H$\alpha$) (see \S~\ref{Ha}), the uncertainties
listed for EW(H$\alpha$) in the Table correspond to the formal measurements
uncertainties while the systematic uncertainties, dominated by those of the
SINFONI data, are typically $30\%$ and up to $50\%$ for
the faintest sources.
While the H$\alpha$ line integrated flux gives an estimate of the current SFR (as only stars with masses $>$ 10 $M_{\odot}$ and age $<20$~Myr contribute significantly to the ionizing flux), EW(H$\alpha$) probes the relative importance of the youngest most massive stars relative to the bulk of the stellar population, which is responsible for the continuum emission at the same H$\alpha$ wavelength. Therefore EW(H$\alpha$) is a stellar age indicator, being sensitive to the ratio of current over past (average) star formation rate. 

For galaxies with constant SFR over their lifetime, the equivalent width should be in inverse proportion to the galaxy stellar age, according to a power law, as the stellar continuum steadily increases with time, while the H$\alpha$ flux remains almost the same \citep[see also][]{2004ApJ...611..703V}. In fact, some authors suggested to derive constraints on galaxy star formation histories by studying EW(H$\alpha$) as a function of the age (e.g. FS09, \citealt{2006ApJ...647..128E}, see also \citealt{2011arXiv1103.0279K}). However, ages estimated from SED fitting strongly depend on the shape of the assumed SFH, and should be considered with some caution, also due to the parameter degeneracy affecting the best-fit choice \citep[cf.][]{2000A&A...363..476B}. In fact, as discussed in Section~\ref{sedfitting}, for our actively star-forming galaxies at $z\sim 2$,  when using constant SFR models, ages are underestimated due to the dominant contribution of the youngest stellar population to the stellar light, and cluster around the imposed age lower limit. On the other hand, ages estimated with inverted-$\tau$ models seem to be more realistic for these targets, but they are imposed by the choice of the formation redshift, $z_f$. 

Hence, we examined the equivalent width of our zC-SINF sample as a function of the $(z-H)$ color, i.e., a quantity tightly related to galaxy age, sampling the rest-frame 4000~\AA\ break for galaxies at $z=1.4-2.5$, and independent of the stellar population models employed (Figure~\ref{fig:ew}). 
However, as the $(z-H)$ color depends also on the dust extinction parameter, we had to take into account the $(z-H)$ color variation, as a function of both age  and of $E(B-V)$. 
The MA05 synthetic stellar population models with constant SFR, indicate that the $(z-H)$ color is nearly constant in the redshift interval $1.4\lesssim z\lesssim 2.5$ at a given age, and $E(B-V)$.
In Figure~\ref{fig:ew} we consider both the case of the same dust attenuation for H\,{\small II} regions and stellar continuum (EW(H$\alpha$), left panel), and the case of extra attenuation, according the Equation~\ref{eq:calz2000} from \citet{2000ApJ...533..682C} (EW$_{00}$(H$\alpha$), right panel, cf. Section~\ref{Ha_luminosity}).
For comparison, we show the region occupied by synthetic stellar population models with constant SFR, solar metallicity, and five different values of dust extinction ($E(B-V)$=0.0, 0.15, 0.2, 0.3, 0.4, chosen to include all values derived for the target galaxies). All the curves show the EW(H$\alpha$) and $(z-H)$ color evolution in the age range $0.1-1.5$ Gyr, and in steps of 0.1 Gyr (marked by small dots on the curves). 
The rest-frame $\rm EW(H\alpha)$ of a template at a given age and
$E(B-V)$ was computed as the ratio of the H$\alpha$ flux, derived from the model SFR via the Kennicutt (1998) conversion, to the $r$-band flux density of the template, assuming CSFR.  For consistency, the \citet{1998ARA&A..36..189K} conversion was corrected, as in Eq. (2), for the Chabrier IMF, and the BC03 for the Chabrier IMF were used for the $r$-band continuum and the $z - H$ colors.

If one assumes that stellar continuum and nebular regions are equally extincted by dust, the equivalent width does not depend on dust attenuation. In fact, in the left panel of Figure~\ref{fig:ew}, at a given age, and for increasing values of $E(B-V)$, the model curves have redder colors, but the same EW(H$\alpha$) values. However, under this assumption, we do not find a good agreement between data points and synthetic galaxy models. The measured EW(H$\alpha$) tend to lie below the model curves with $ E(B-V)>0$, in regions that would correspond to unphysical ages older than that of the universe at $z \sim 2$, and null (i.e., black curve) or negative color excesses. 
On the other hand, if one assumes extra dust attenuation for nebular regions with respect to the stars ($A_{V,HII}=A_{V}/0.44$; right panel of Figure~\ref{fig:ew}) all the model curves tighten on a locus that forms an extension of the unreddened curve, and at a given age, $EW_{00}$(H$\alpha$) exponentially decreases (as $10^{-0.33\ A_V(1/0.44 -1)}\sim 10^{-1.7 E(B-V)}$) for increasing values of $E(B-V)$. 
From Figure~\ref{fig:ew} it is evident that, although with a large scatter, for zC-SINF galaxies the case of extra dust attenuation better matches with the model predictions. 
So, these results confirm what we found in the previous sections by comparing H$\alpha$ luminosity with SFR from SED fitting. That is, for our sample H$\alpha$ seems to be more extincted than the UV stellar continuum, in agreement with the prescription of \citet{2000ApJ...533..682C}. The fact that a substantial fraction of the data points in the right panel of Figure~\ref{fig:ew} show larger equivalent width than the model curves suggests that some of our galaxies could have sub-solar metallicities, which would produce higher equivalent widths (higher H$\alpha$ luminosity) for a given SFR, age, and $E(B-V)$ \citep[cf.][]{2006ApJ...647..128E}.

\begin{figure*}[htbp]
\includegraphics[width=0.5\textwidth]{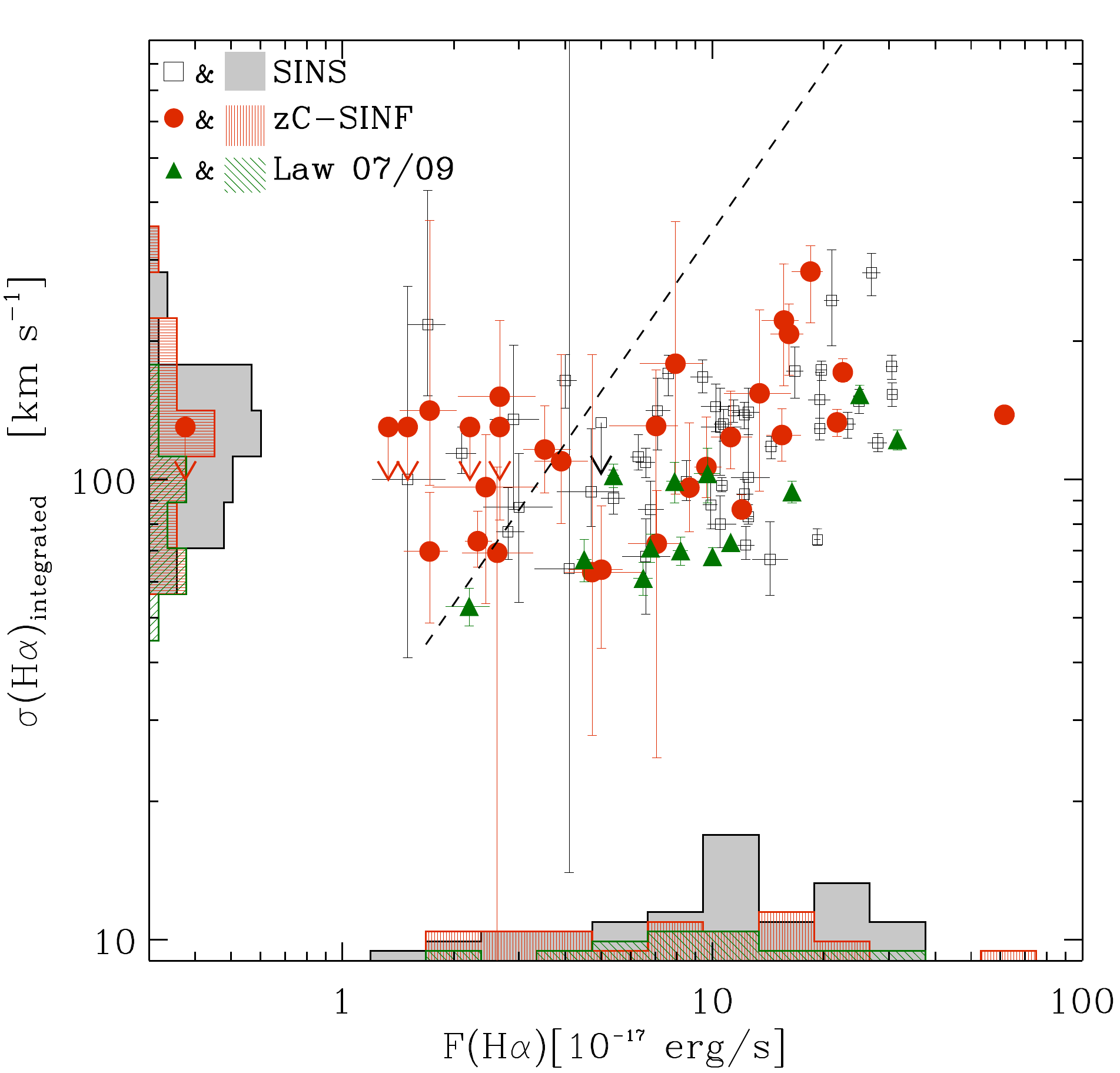}\includegraphics[width=0.5\textwidth]{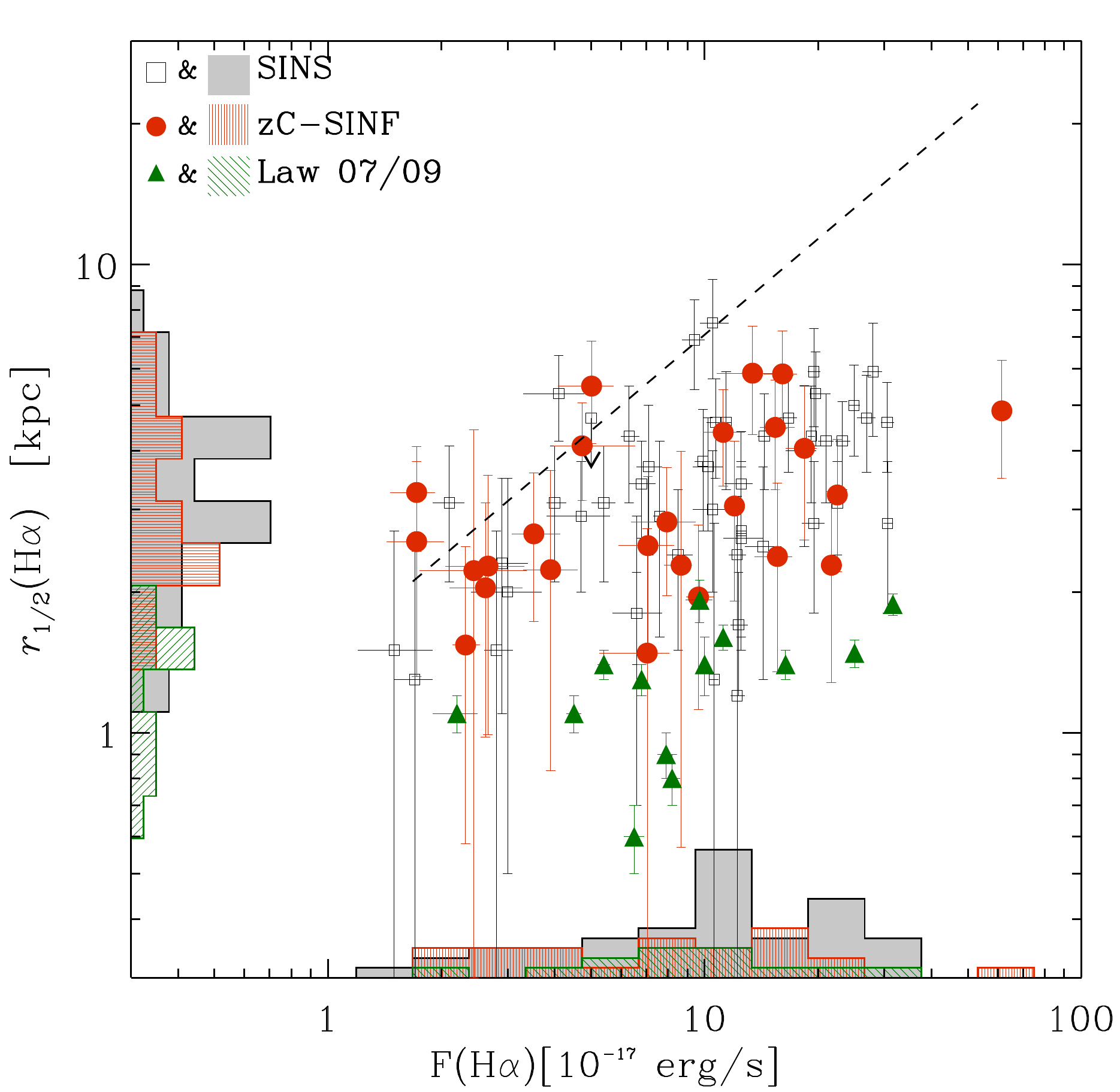}
\caption{\small{{\bf Left panel}: the integrated velocity dispersion $\sigma$(H$\alpha$)$_{integrated}$ vs. the H$\alpha$ flux F(H$\alpha$), where  $\sigma$(H$\alpha$)$_{integrated}$ is defined as the FWHM of the single Gaussian best fitting the integrated H$\alpha$ profile of individual galaxies, and is corrected for instrumental spectral resolution. Besides galaxies in the zC-SINF sample (red filled circles) also galaxies in the IFS samples of FS09 (i.e., the SINS sample, open squares), and of L09 (green triangles) are displayed. The three histograms (arbitrarily normalized) show the relative distributions in F(H$\alpha$) and $\sigma$(H$\alpha$)$_{integrated}$ for the three samples, as labeled on the panel, and exclude the upper limits.
{\bf Right panel}: the half-light radius, $r_{1/2}$, of the same galaxies vs. F(H$\alpha$), where $r_{1/2}$ is measured as the radius of the circular aperture centered on the centroid of the H$\alpha$ emission map, including 1/2 of the H$\alpha$ luminosity, and determined based on the curve-of-growth. Symbols and histograms are the same as in the left panel. In the two panels the dashed lines mark out the line widths (left panel) and the sizes (right panel) above which the galaxies would be undetected (i.e., $S/N < 3$ per spectral or spatial resolution element, respectively) in our data set, as a function of H$\alpha$ flux. 
}}\label{fig:FHa-sigma-re} 
\end{figure*}

\begin{figure*}[htbp]
\includegraphics[width=0.5\textwidth]{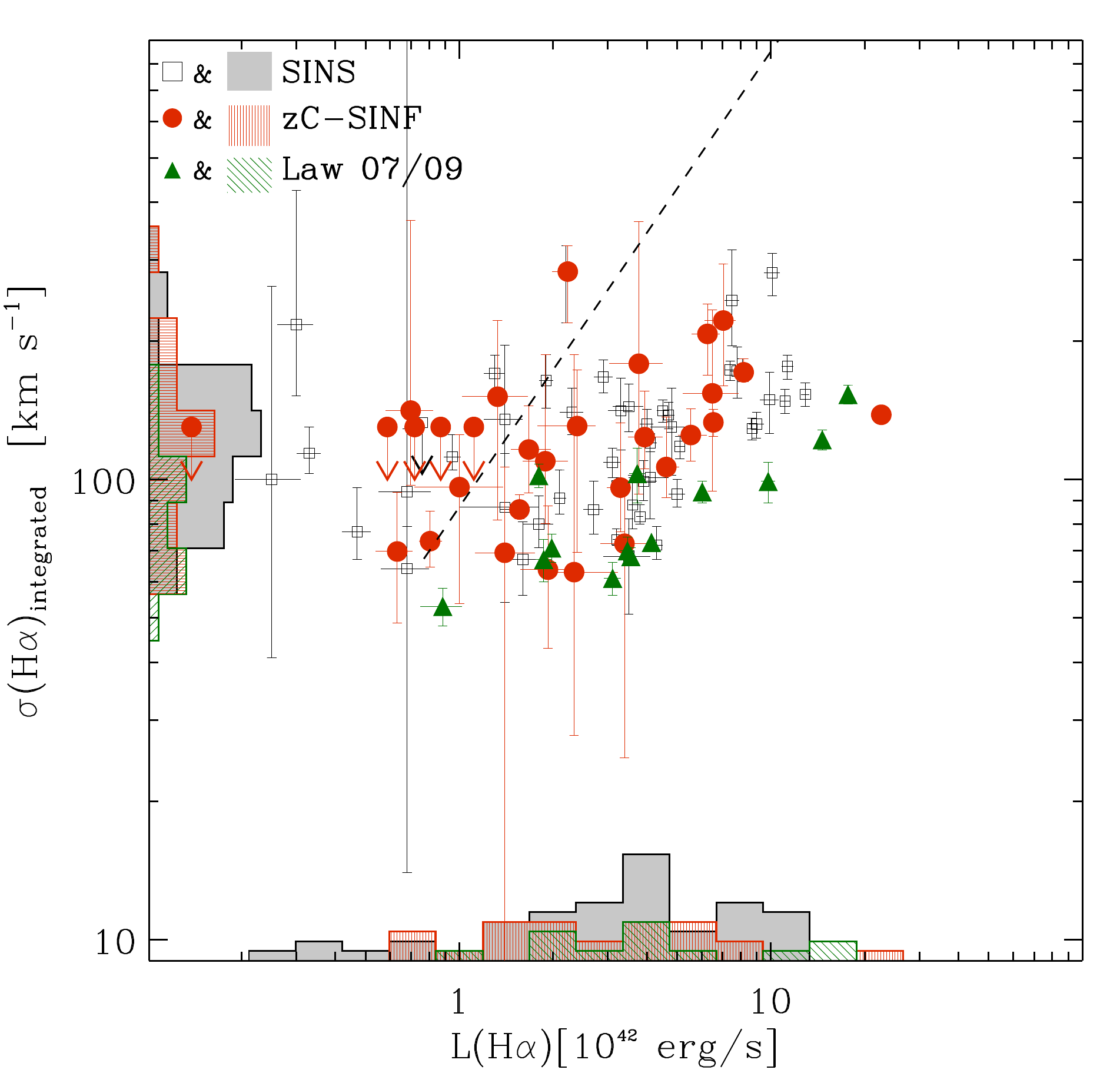}\includegraphics[width=0.5\textwidth]{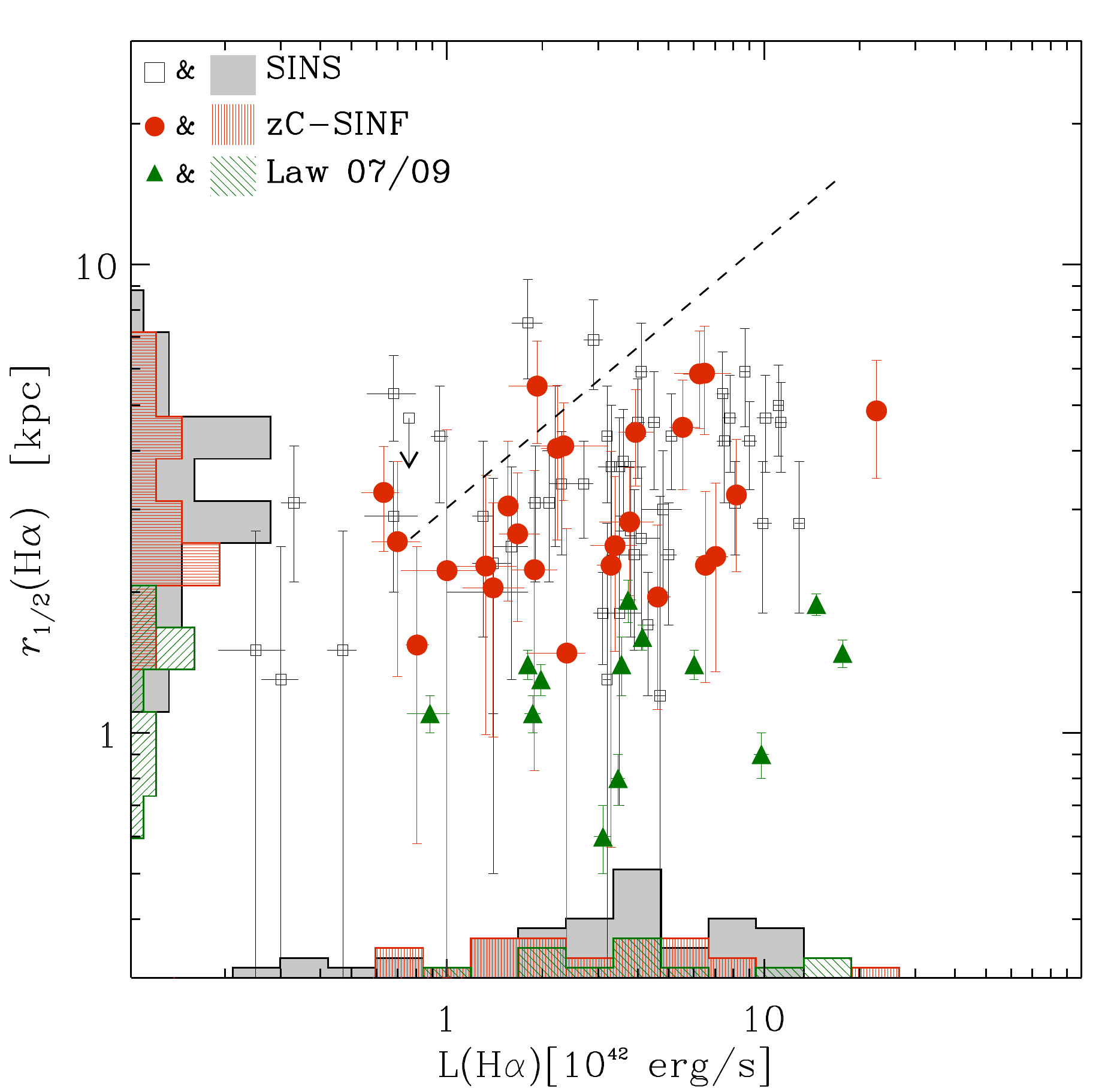}
\caption{\small{The same as in Figure~\ref{fig:FHa-sigma-re}, but now as a function of the H$\alpha$ luminosity.}}\label{fig:LHa-sigma-re} 
\end{figure*}

\begin{figure*}[htbp]
\includegraphics[width=0.5\textwidth]{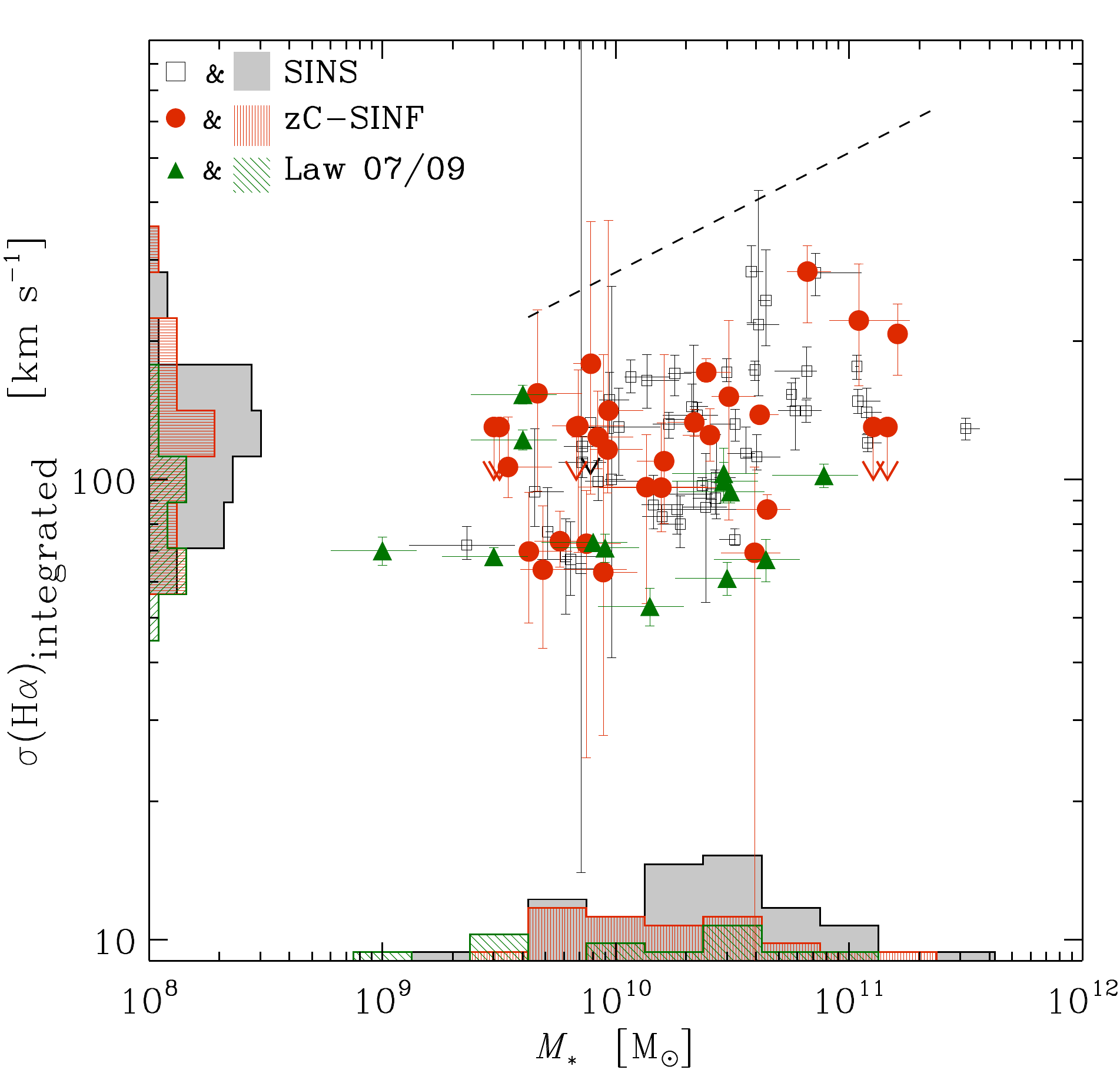}\includegraphics[width=0.5\textwidth]{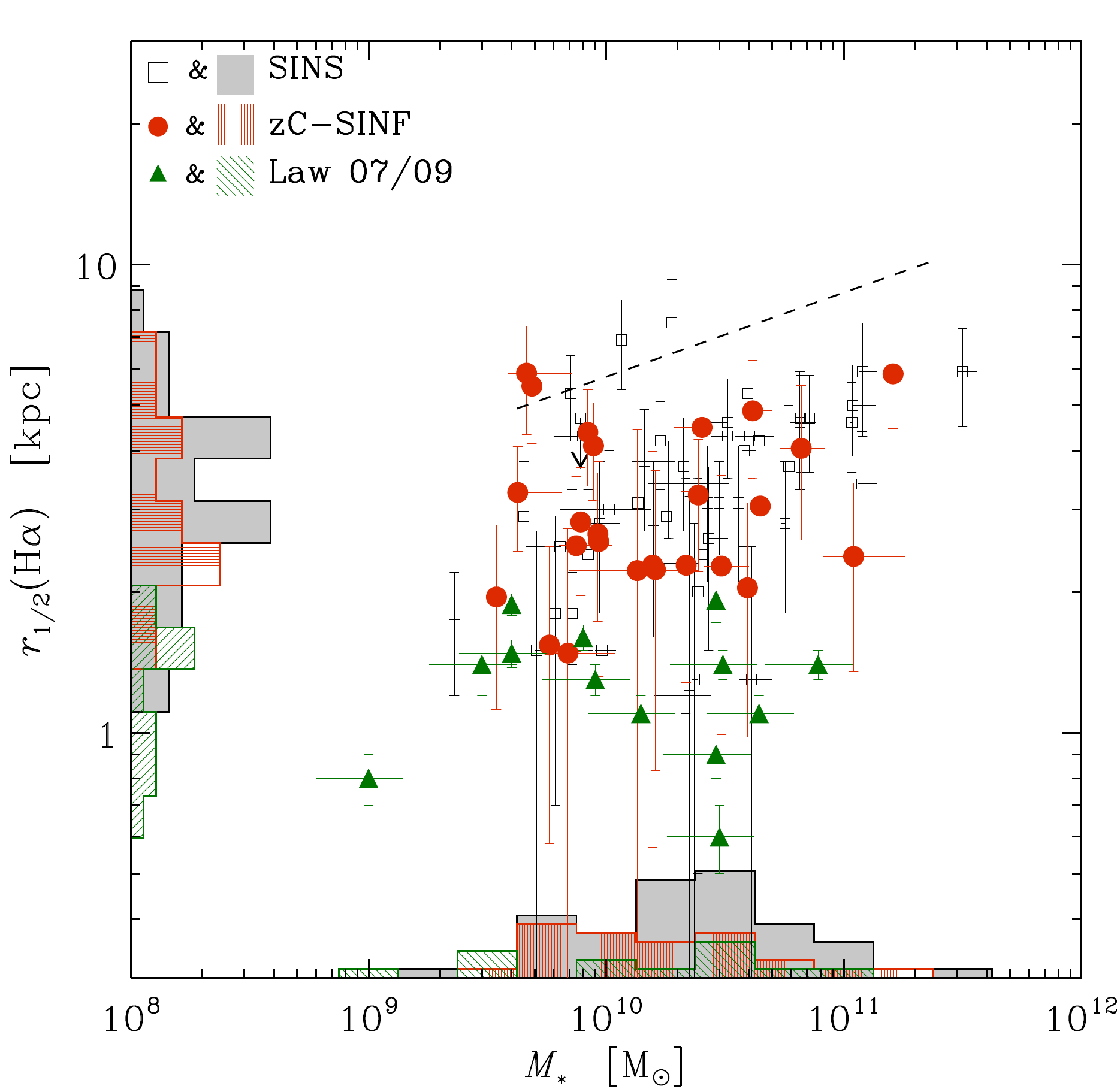}
\caption{\small{The same as in Figure~\ref{fig:FHa-sigma-re}, but now as a function of the stellar mass.}}\label{fig:M-sigma-re} 
\end{figure*}

\subsubsection{Comparison of the integrated H$\alpha$ properties of the zC-SINF sample with other IFS samples at $z\sim 2$}\label{sec:SINS-comparison} 

Figures~\ref{fig:FHa-sigma-re}--\ref{fig:M-sigma-re} show the distributions
of H$\alpha$ properties of the zC-SINF sample (red filled circles).  The
integrated velocity dispersions $\sigma$(H$\alpha$)$_{\rm integrated}$ and
the half-light radii $r_{1/2}({\rm H\alpha})$ are plotted as a function of the total observed fluxes $F({\rm H\alpha})$, of the luminosities $L({\rm H\alpha})$ (uncorrected for dust extinction), and of the stellar masses $M_{\star}$. Stellar masses are from BC03+CSFR models and Chabrier IMF, to allow a better comparison with other IFS samples using these assumptions (see below).
We also indicate in the Figures the typical sensitivity limits in terms
of size and velocity width.  These limits were determined following exactly
the same procedure as described in FS09.  In brief, we determined the average
S/N per spatial resolution element within the half-light radius for each galaxy
based on the H$\alpha$ maps.  We then calculated the necessary increase in
half-light radius for the S/N per resolution element to drop below $\rm S/N=3$,
keeping the total H$\alpha$ flux and integrated line width fixed.  These
derived sizes correspond to the actual surface brightness sensitivity of
each data set.  The average limit in terms of $r_{1/2}({\rm H\alpha})$ as
a function of $F({\rm H\alpha})$ was obtained by taking the running median
through the individual limits (and similarly as a function of
$L({\rm H\alpha})$ and $M_{\star}$).
The average limits in terms of $\sigma$(H$\alpha$)$_{\rm integrated}$
were computed in a similar manner, calculating the line width at which
the S/N per spectral resolution element drops below $\rm S/N = 3$ at
fixed $F({\rm H\alpha})$ and $r_{1/2}({\rm H\alpha})$.
Because the limits show significant scatter among the objects,
and some of the sources have an average S/N per resolution element
somewhat below 3, some of the individual measurements lie above the
dashed lines in Figures~\ref{fig:FHa-sigma-re} to \ref{fig:M-sigma-re},
which represent {\em typical} sensitivity limits.

The zC-SINF data are compared with the global H$\alpha$ properties of other
galaxy samples in the range $1.4 \lesssim z \lesssim 2.5$ also observed with integral field spectrographs, taken from the SINS survey with SINFONI (FS09, open squares)
and the sample observed with OSIRIS at the Keck~II telescope by
\citet[][hereafter L09; green filled triangles]{2009ApJ...697.2057L}.
The targets for all three studies were drawn from parent samples at $z \sim 2$
selected with photometric criteria designed to identify massive star-forming
galaxies, and are subject to similar biases introduced by the need for an
accurate optical spectroscopic redshift and a sufficiently bright H$\alpha$
emission.  Although we include the results from L09, one should keep in mind
that all targets of this sample were observed with the aid of AO and this may
introduce differences in the derived H$\alpha$ properties (see below).
The median properties for these three samples (obtained by considering only the H$\alpha$ detected objects) are shown in Table~\ref{tbl:comparisons}. 
The comparisons show that the zC-SINF and SINS samples span nearly the same
ranges in integrated H$\alpha$ flux, velocity width, and half-light radius.
They also overlap largely in terms of total H$\alpha$ luminosity and stellar
mass. Instead, although the L09 sample covers roughly the same range in H$\alpha$ flux and luminosity, and in stellar mass, the integrated velocity widths and half-light radii are systematically
and significantly smaller, by factors of approximately two for the ensemble of objects, as well as for a given H$\alpha$ flux/luminosity or a given stellar mass.
As discussed by FS09 and L09, these differences could arise from distinct
population properties, in particular, more compact sizes.  We emphasize
that no selection based on sizes was applied for our zC-SINF sample and
for the majority of the SINS sample.  The differences could also be
partly due to a combination of the higher resolution and poorer surface
brightness sensitivity for the AO-assisted OSIRIS data.  Indeed, this is quite likely the case, given that smaller sizes for given stellar mass would have implied higher velocities, which instead are lower (see Figure 18). This argues for the outer, high-velocity but low surface brightness regions having been lost.  Moreover, the very similar $M_{\star}$ distributions between the zC-SINF and L09 samples suggest that if the differences in sizes and velocity dispersions are attributed to intrinsic galaxy properties, stellar mass may not be the most important parameter driving these differences.

The new zC-SINF data follow a similar trend of increasing integrated H$\alpha$
velocity dispersion with increasing H$\alpha$ flux, H$\alpha$ luminosity, and
stellar mass as seen among the SINS and OSIRIS sample galaxies.  Likewise, the
H$\alpha$ half-light radii tend to be larger for brighter and more massive
sources, although the scatter is more significant.  As discussed by FS09,
the upper envelope of the distributions of $\sigma({\rm H\alpha})_{\rm integrated}$ and
$r_{1/2}({\rm H\alpha})$ versus $F({\rm H\alpha})$, $L({\rm H\alpha})$, and
$M_{\star}$ likely results from observational biases.  More specifically, for
SINFONI seeing-limited observations, the detection limits for 1\,hr integration
lies just above the locus of the measurements, implying that the zC-SINF sample
may be missing the most extended sources and/or those with broadest velocity
widths at a given H$\alpha$ flux/luminosity or stellar mass.  However, such a
bias cannot explain the lack of sources with small velocity dispersion at high
H$\alpha$ fluxes/luminosities and stellar masses seen in
Figs.~\ref{fig:FHa-sigma-re}, \ref{fig:LHa-sigma-re}, and
\ref{fig:M-sigma-re}, which could be the consequence of a real, physical
$\sigma({\rm H\alpha})_{\rm integrated} - M_{\star}$ relation. 
This issue, already pointed out by FS09 for the SINS sample, may not be
surprising given that both kinematic components that provide the total
dynamical support of galaxies (random motions and rotational/orbital
motions) contribute to the measured $\sigma({\rm H\alpha})_{\rm integrated}$,
which thus probes the full gravitational potential of the galaxies. Besides being due to inclination effects, the scatter in the $\sigma({\rm H\alpha})_{\rm integrated} - M_{\star}$ diagram may possibly reflect variations in the gas
and dark matter mass fraction among the sources.

\section{Kinematic classification}\label{kinem} 

For the zC-SINF pre-imaging data, the S/N ratio is generally too low
for a reliable quantitative kinematic classification of the objects
based on velocity fields and velocity dispersion maps.
The distinction between disk-like and merger-like systems through
kinemetry \citep{2008ApJ...682..231S} and detailed kinematic modeling 
will be performed based on the higher resolution data taken as part
of the follow-up AO observations of the program.  The results for the first few objects observed with
AO in early stages of our on-going SINFONI LP and previous programs
are presented elsewhere (Shapiro et al. 2008; Genzel et al. 2008, 2011;
Cresci et al. 2009).
Here, we attempt a first crude kinematic classification in
``rotation-dominated'' and ``dispersion-dominated'' galaxies. 

Ideally, the distinction between rotation- and dispersion-dominated objects
would rely on the ratio of the inclination-corrected circular or orbital
velocity and the local intrinsic velocity dispersion, $v_{\rm c}/\sigma_{0}$
\citep[with the transition at $v_{\rm c}/\sigma_{0} \sim 1$; e.g.,][]{2008ApJ...687...59G,2009ApJ...697..115C}.  Alternatively, the comparison between the contributions to
the dynamical mass from random and from rotational/orbital motions can also be
used \citep[e.g., ][]{2009A&A...504..789E}.  Both methods, however, require
modeling of high quality kinematic maps.
Other indicators have been used in the literature, notably based on averaging
over individual pixels or resolution elements across the galaxies to estimate
the intrinsic velocity dispersion \citep[e.g., L09,][]{2010Natur.467..684G}.
However, as demonstrated by \citealt{2011arXiv1108.0285D} these estimates remain largely biased by the surface brightness sensitivity of the data and by beam-smearing effects.

We therefore resorted to an alternative approach that can be followed for
the majority of the zC-SINF galaxies with the current no-AO data sets.
We used the working criterion introduced by FS09 and based on the ratio
of half of the maximum observed velocity difference across the source
(uncorrected for inclination) and the integrated velocity dispersion,
$v_{\rm obs} / 2\,\sigma({\rm H\alpha})_{\rm integrated}$ (each measured as
described in Section~\ref{PV} and reported in Table~\ref{tbl-Ha-5}).
Based on simulations of disk galaxies with a range of inclinations,
sizes, and masses, as well as of beam smearing appropriate for the SINS
sample, and thus also for the zC-SINF sample given the similar observing
conditions and global H$\alpha$ properties (Section~\ref{sec:SINS-comparison}), 
an intrinsic $v_{\rm c} / \sigma_{0} \sim 1$ corresponds roughly to
$v_{\rm obs} / 2\,\sigma({\rm H\alpha})_{\rm integrated} \sim 0.4$.
This criterion was devised in the framework of rotating disks; since
we do not distinguish between disk- and merger-like kinematics here,
a $v_{\rm obs} / 2\,\sigma({\rm H\alpha})_{\rm integrated} \ga 0.4$ could also
indicate dominant bulk orbital motions for merging/interacting systems.

\begin{figure}[htbp]
\centering
\includegraphics[width=\columnwidth]{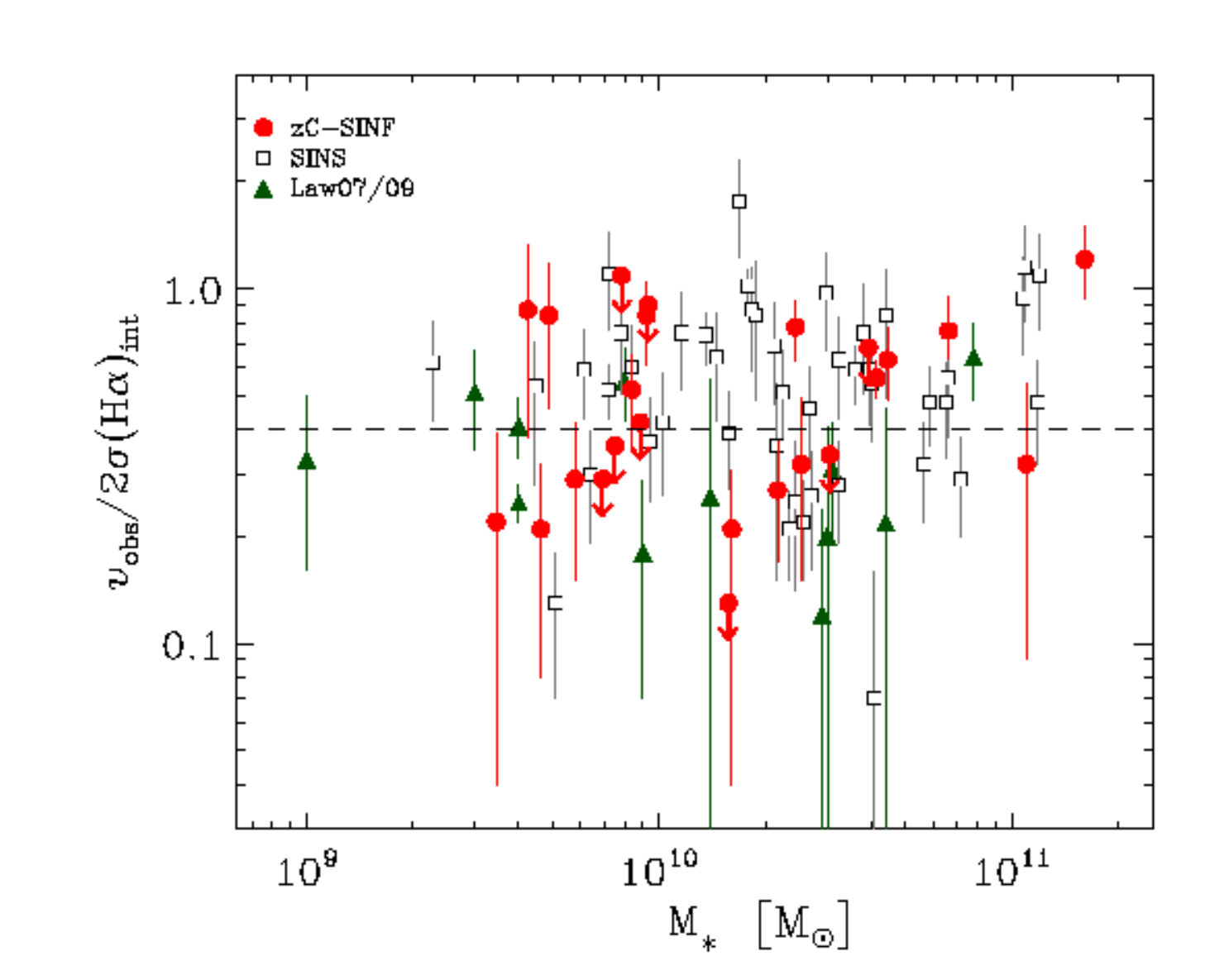}
\caption{\small{The ratio of the maximum observed velocity difference across the source, $v_{\rm obs}$, to twice the integrated velocity dispersion, $\sigma ({\rm H}\alpha)_{integrated}$, as a function of  galaxy stellar mass. The figure illustrates the working criterion of FS09, used here for a preliminary kinematic classification of the zC-SINF galaxies. The Results for the subset of 24 sources for which the S/N allowed us to derive  $v_{\rm obs}$ (or an upper limit thereof) are compared with those from the SINS H$\alpha$ sample, and from the sample of \citet[][]{2009ApJ...697.2057L}. Symbols are the same as in Figures \ref{fig:FHa-sigma-re}, \ref{fig:LHa-sigma-re}, and \ref{fig:M-sigma-re}. The dashed line (i.e., $v_{\rm obs} / 2\,\sigma({\rm H\alpha})_{\rm integrated} = 0.4$) represents the boundary between the dispersion-dominated and rotation-dominated regime. Galaxies above this line are tentatively identified as ``rotation-dominated'' according to the criterion developed by FS09.}}\label{fig:vobs}
\end{figure}

Figure~\ref{fig:vobs} shows the $v_{\rm obs}/2\,\sigma$(H$\alpha$)$_{\rm integrated}$ versus stellar mass for the subset of 24 zC-SINF objects for which we could
derive $v_{\rm obs}$ or an upper limit thereof.  The measurements for 46 objects from the SINS H$\alpha$ sample for which this ratio could be reliably determined are also shown, along with those taken from the sample of \citet{2009ApJ...697.2057L}.
According to the criterion above, 13 of the 24 objects have measured ratios or upper limits that lie in
the rotation-dominated regime (including two within 1 $\sigma$ of the dividing
line) and 11 objects fall in the dispersion-dominated regime (with three within $1\,\sigma$ of the dividing line).
Bearing in mind that the $v_{\rm obs} / 2\,\sigma({\rm H\alpha})_{\rm integrated}$
ratio is a crude indicator, and that for many of the sources the velocity
gradients can only be estimated over limited regions with sufficiently high S/N,
it is clear that the above results should be considered with the due caution.
Nevertheless, the $v_{\rm obs} / 2\,\sigma({\rm H\alpha})_{\rm integrated}$ estimates
suggest that roughly half of the zC-SINF galaxies appear to have kinematics
dominated by disk rotation or bulk orbital motions.
We point out that these candidates also include two objects from the
zCOSMOS ``Pilot Sample,'' already classified as rotating disks by means
of kinemetry and kinematic modeling (i.e., ZC407302, and ZC410542; Shapiro
et al. 2008; Cresci et al. 2009)\footnote{For one of the other two Pilot Sample objects, ZC407376, the S/N of the seeing-limited SINFONI data is insufficient for quantitative analysis through kinemetry and dynamical modeling, and the fourth one, ZC410116, is undetected in H$\alpha$ (Table~\ref{tbl-Ha-5}).}.

The relative proportions for the zC-SINF sample are roughly comparable
to those inferred for the SINS H$\alpha$ sample though with some small
differences. Using the same $v_{\rm obs} / 2\,\sigma({\rm H\alpha})_{\rm integrated}$ indicator, about one-third of the SINS H$\alpha$ sample has kinematics
dominated by dispersion, and two-thirds has kinematics dominated by
rotational/orbital motions (including the kinematically-identified
mergers).  This is also reflected in the median
$v_{\rm obs} / 2\,\sigma({\rm H\alpha})_{\rm integrated}$ of
0.44 (including upper limits) for zC-SINF, and 0.57 for SINS (or 0.54 when excluding the five objects for which the ratio was measured from AO data).
In contrast, the same indicator for the Law et al. (2007, 2009)
sample observed with AO implies a median of 0.28 and two-thirds of
dispersion-dominated objects.  As discussed by FS09, these differences
could be due to intrinsic differences between the samples studied and/or
to sensitivity limitations; in particular, there may be a weak trend of
increasing proportion of dispersion-dominated objects with decreasing
stellar mass but the scatter of the data is too large to assess the
trend robustly.
While clearly more details about the kinematic properties will be extracted
from sensitive AO-assisted observations and will allow an empirical calibration
of this indicator, seeing-limited observations remain important and are less
expensive in terms of observing time.  In the future, the multi-IFU instrument
KMOS at the VLT will provide much larger samples with kinematics but only in
seeing-limited mode.  The results from the zC-SINF no-AO data (and those from
SINS) thus constitute valuable references.
The analysis above suggest a significant fraction of objects dominated by
rotation/orbital motions, and it will be interesting to explore whether
the kinematic mix varies as a function of global galaxy
parameters once much larger samples allow to investigate this issue.

\section{Conclusions}\label{sec:summary}
This paper presents the first results from the `zCOSMOS - SINFONI project', based on an ESO VLT Large Program aimed at studying the kinematical properties of $z\sim2$ star-forming galaxies, as mapped by the H$\alpha$ emission line, with the near-infrared SINFONI AO-assisted integral field spectrograph. The zCOSMOS-SINFONI observing strategy includes (i) 'pre-imaging' natural seeing observations of 30 galaxies (i.e. the so-called zC-SINF sample), consisting in 1 hour of integration time per object, and (ii) deeper AO follow-up of $\sim 20$ sources, selected based on the pre-imaging results. 

The main distinct feature of this project with respect to the past consists in having assembled an unprecedented sample of galaxies at $z\sim2$, for which high quality kinematics resolved on spatial scales as small as $\sim$ 1 kpc will be soon available. The zC-SINF sample selection in the zCOSMOS-Deep survey, offered the advantage of reliable optical spectroscopic redshifts, and very wide area helping to cull targets with suitably bright and nearby stars enabling AO follow-up, as well as of the extensive multi-wavelength information from the COSMOS survey. Moreover, our sample is uniform in terms of color criteria, quality of optical spectroscopic follow-up, and coverage in M$_{\star}$ and SFR.  Last, but not least, the 20 galaxies observed with AO, will roughly double the current number of "benchmark" galaxies at $z\sim 2$ with spatially resolved kinematics on $\sim 1$~kpc scale.

In this work the selection criteria used to assemble the full zC-SINF sample, the general galaxy properties derived from SED fitting analysis, and the results from the first part of the SINFONI program, based on natural-seeing observations, were described. The results were then compared with the literature available for the broader population of star-forming galaxies at $z\sim2$. 
The main conclusions are summarized in the following.
\begin{enumerate}

\item The zC-SINF sample was culled from the zCOSMOS survey, so to span a wide mass range ($3\times 10^9<(M_{\star}/M_{\odot})<2\times10^{11}$), to give the best possible representation of the star-forming galaxy population at $z\sim 2$. We showed that the sample is inevitably biased by the zCOSMOS pre-selection at $ B_{AB}\lesssim25$, needed to obtain an adequate S/N in the VIMOS spectra, and by the lower limit that we imposed to the SFR ($\sim 10 M_{\odot}\ yr^{-1}$) of the selected objects, to match the typical detection limit for H$\alpha$ line emission of $z\sim 2$ galaxies in 1 hour of pre-imaging observations. However, the sample includes galaxies in a fairly wide range of SFRs (i.e., $\sim 40-300~M_{\odot}$/yr).   

\item Using the \citet{2010ApJ...708..202M} $K$-band selected multi-wavelength catalog, for each object we derived stellar mass, SFR, dust extinction $A_V$,  and age, by fitting the photometric SED. We checked the results against different libraries of synthetic stellar population models, and different star formation histories (i.e., constant SFR, and ``inverted-$\tau$'' models), and found that zC-SINF masses, and SFRs, always agree within $\sim 0.2-0.3$ dex. However, SFRs decrease and stellar masses increase when switching from constant SFR to inverted-$\tau$ models, in such a way that the latter give a specific SFR $\sim 3$ times lower than the former models. For various astrophysical reasons, and for giving SFRs in better agreement with those derived from the H$\alpha$ luminosity, for the $z\sim 2$ galaxies in our sample we favor inverted-$\tau$ models. In any way, secularly increasing SFRs are to be preferred compared to constant or declining  SFRs.
We used the \citet{2009ApJ...690.1236I} $I$-band selected catalog to check also the SED fitting results against different photometric catalogs, and also found that they are stable within $\sim 0.2$ dex.

\item  Out of the initial 30, we detected 25 galaxies with SINFONI, and extracted H$\alpha$ flux, integrated velocity dispersion, and half-light radius (i.e., F(H$\alpha$), $\sigma$(H$\alpha$)$_{\rm integrated}$, and $r_{1/2}$(H$\alpha$)). The H$\alpha$ integrated properties were then compared with those derived from IFS for other samples at $z\sim 2$. 
We found that the zC-SINF and the SINS sample (FS09; also based on SINFONI data) span the same range in F(H$\alpha$), $\sigma$(H$\alpha$)$_{\rm integrated}$, and $r_{1/2}$(H$\alpha$). 
The fact that the results from this more uniform zC-SINF sample are similar to those found with the more heterogeneous SINS H$\alpha$ sample provides an important confirmation in terms of the general properties of massive star-forming galaxies in the relevant M$_{\star}$ and SFR range.

\item Based on the H$\alpha$ integrated fluxes, H$\alpha$ luminosities (L(H$\alpha$)) were derived, and used to determine SFRs through the \citet{1998ARA&A..36..189K} relation. SFRs were also derived from the UV luminosity (L$_{1500}$), redshifted in the $B$-band for galaxies at $z\sim 2$, as well as from the SED fitting, and then compared with those from H$\alpha$. In this comparison, we considered two possibilities to account for dust extinction: (i) the H{\small II} regions are as extincted as the stellar continuum; (ii) they are $\sim 2$ times more extincted, as parametrized by \citet{2000ApJ...533..682C}. For zC-SINF galaxies, we found that the additional dust extinction for nebular regions produces a better agreement among the H$\alpha$-derived SFR and other indicators. 

These results from the comparison of SFR(H$\alpha$) with other SFR estimates are further supported by our analysis of the H${\alpha}$ equivalent widths. Surface brightness limitations may affect our conclusions, but we estimate that these limitations are unlikely to account for the entire systematic offset towards lower  SFR(H$\alpha$)s and EW(H$\alpha$)s that emerge when assuming A$_{V,H{\small II}}$ = A$_{V,SED}$.
The evidence from our zC-SINF sample for extra attenuation towards the H{\small II} regions with respect to the bulk of the stellar population dominating the broad-band continuum emission agrees with earlier findings following similar arguments for the SINS H$\alpha$ sample (FS09), and with the first direct measurements of the Balmer decrement in small samples of other massive, actively star-forming galaxies at $z \sim 2$.

\item We presented a preliminary kinematic classification using a working criterion developed by FS09, and based on the ratio of two observed quantities, i.e., half of the maximum observed velocity difference across the source (uncorrected for inclination) ($v_{obs}$), and the H$\alpha$ line width (i.e., the integrated velocity dispersion, $\sigma$(H$\alpha$)$_{\rm integrated}$). Following this criterion, we found that, out of the 24 galaxies for which S/N ratio allowed us to retrieve a velocity gradient value, 14 objects appear to be rotation-dominated disk candidates, comparable to the fraction inferred in other IFS samples at high-$z$. 
The higher S/N and higher spatial resolution of the follow-up AO-assisted SINFONI data, to be presented in upcoming papers, will allow a more robust classification and a detailed assessment of the reliability of our pre-classification.
\end{enumerate}

\begin{acknowledgements}
We acknowledge the great efforts of the entire zCOSMOS team in analyzing the data and creating the spectroscopic database, from which our targets were drawn.
N.M.F.S. acknowledges support by the Minerva program of the MPG.
N.B. is supported by the Marie Curie grant PIOF-GA-2009-236012 from the European Commission. G.C. acknowledges support by the ASI-INAF grant I/009/10/0.
This work has been partly supported by the ASI grant "COFIS-Analisi Dati" and by the INAF grant "PRIN-2008".
\end{acknowledgements}

\appendix

\section*{APPENDIX A}
\subsection*{Spectral Energy Distributions for the zC-SINF sample}\label{appendix-A}
\begin{figure*}[h]
\centering
\includegraphics[width=0.9\textwidth]{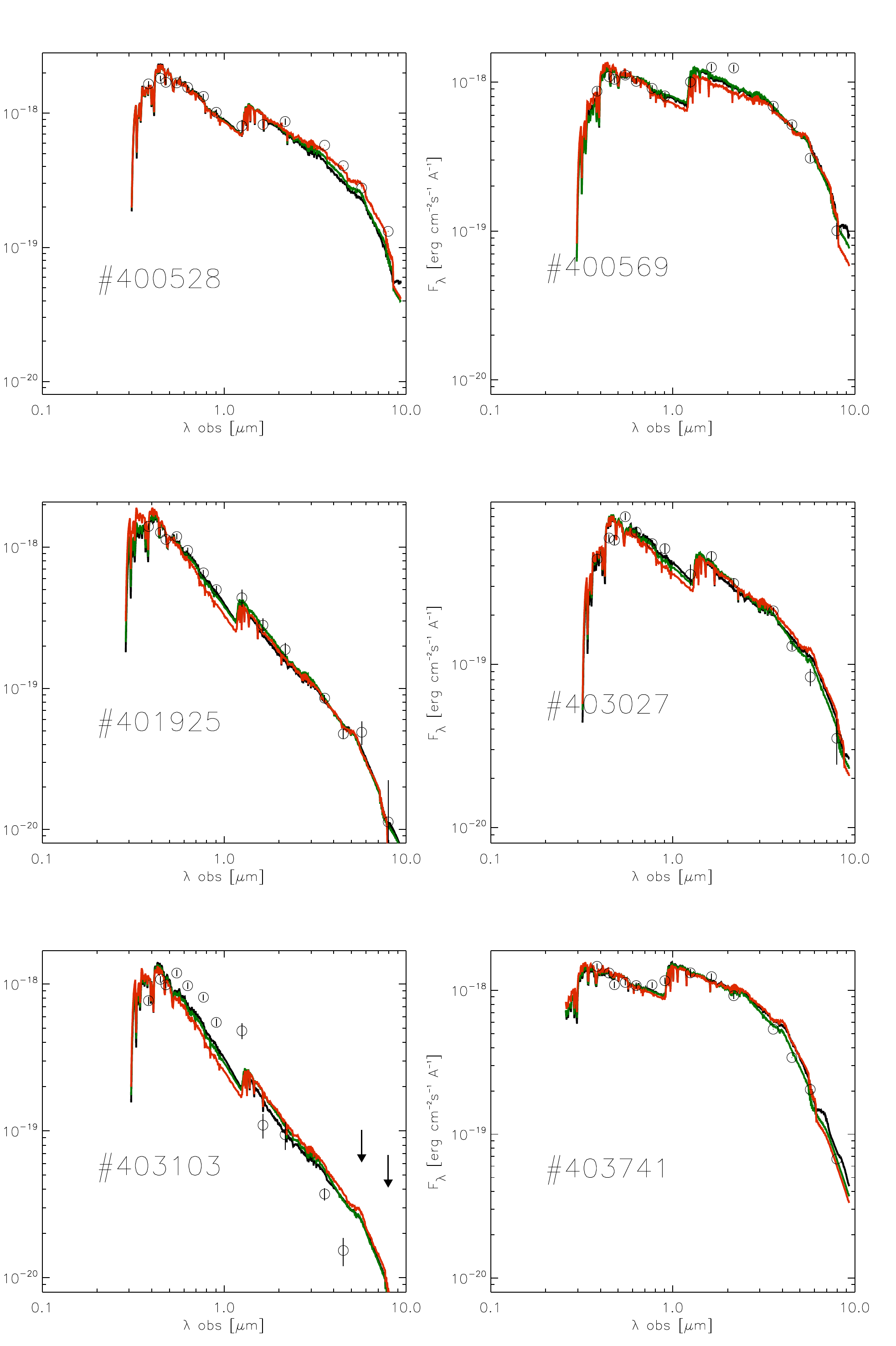}
\caption{\small{Best-fit SEDs for the whole zC-SINF sample in units of F$_{\lambda}$[erg/cm$^2$/s/\AA] vs observed wavelengths ($\lambda$ obs)[$\mu$m]. 
Open circles are the observed fluxes from the photometric data. The black, green, and red curves are the best-fit SEDs built with BC03+CSFR, MA05+CSFR, and MA05+inv-$\tau$ SFH models, respectively. The black arrows are 2$\sigma$ flux upper limits (see also Table~\ref{tbl-phot-3}).}}\label{fig:sed}
\end{figure*}

Figure~\ref{fig:sed} shows the best-fit spectral energy distributions (SED) for the 30 galaxies of the zC-SINF sample. The open circles represent the photometric data points from $u^*$ to IRAC-bands. In each panel the best-fit curves obtained from BC03 constant SFR models are shown in black, together with those from MA05 constant SFR (green), and MA05 inverted-$\tau$ (red) models. The best-fit curves were obtained by fixing the redshift to the spectroscopic value, as measured from the H$\alpha$ emission line (i.e., $z$(H$\alpha$)) for the 25 H$\alpha$-detected objects. For the remaining 5 H$\alpha$ undetected objects the optical zCOSMOS redshift was used. Best fit parameters are shown in Table~\ref{tbl-sedfit-4}.

\section*{APPENDIX B}
\subsection*{H$\alpha$ maps, position-velocity diagrams and integrated spectra}\label{appendix-B}

Figure~\ref{fig:sinfoni} shows the $I$-band emission from HST+ACS (F814W), the velocity-integrated H$\alpha$ linemap, the position-velocity diagram, and the integrated spectrum for each of the 25 H$\alpha$-detected galaxies of the zC-SINF sample. The H$\alpha$ maps, position-velocity diagrams, and spectra were derived from 1 hour integration time for all objects, except ZC405081 (1.7 hour), and ZC407302 and ZC407376 (2 hours), and natural-seeing SINFONI observations (no-AO).

The position-velocity $(p-v)$ diagrams were extracted from the data cubes, without additional smoothing from median-filtering, using a synthetic slit
6 pixels wide (0$\farcs$75) oriented along the major axis of the galaxies, and indicated by the rectangles on the H$\alpha$ maps. Whenever possible, the galaxy major axis was identified based on the velocity field, as the direction of the steepest observed velocity gradient across the source (H$\alpha$ kinematic major axis). For those sources with no clearly apparent velocity gradient, we took the morphological major axis. It is worth noticing that for the galaxies with a clear velocity gradient the kinematic major axis in general was found to be consistent with the morphological one (see Section~\ref{PV}). 

The integrated spectra were extracted from the unsmoothed data cubes in circular apertures shown (as black filled circles) on the H$\alpha$ maps, using the aperture radii ($r_{ap}$) listed in Table~\ref{tbl-Ha-5}. $r_{ap}$ were derived based on the curve-of-growth method, so to enclose the total H$\alpha$ flux (i.e., more than 90\%), as also explained in Section~\ref{Ha}.

\bibliography{zC-SINF_aph}

\clearpage

\begin{deluxetable}{cccccccc}
\tabletypesize{\footnotesize}
\centering
\tablecaption{zCOSMOS-SINFONI `pre-imaging' observing runs.\label{tbl-observ-1}}
\tablewidth{0pt}
\tablehead{
\colhead{Source}&\colhead{RA} &\colhead{DEC}& \colhead{Date}& \colhead{Period}&\colhead{Band}&\colhead{PSF FWHM [arcs]} &\colhead{T$_{int}$ [s]}}
\startdata
ZC407376\tablenotemark{a} & 10:00:45.1  &  02:07:05.2 &  16-23 Apr 2007   & P79    &  K    & 0.60 & 7200\\
ZC407302\tablenotemark{a} & 09:59:56.0  &  02:06:51.3 &  16-23 Apr 2007   & P79    &  K    & 0.56 & 7200\\
ZC410116\tablenotemark{a} & 09:59:36.5  &  02:16:14.5 &  18-20 Aug 2007   & P79    &  K    & 0.71 & 3600\\
ZC410542\tablenotemark{a} & 09:59:54.7  &  02:17:48.3 &  18-20 Aug  2007  & P79    &  K    & 0.82 & 3600\\
ZC409985 		  & 09:59:14.2  &  02:15:46.9 &  21-22 Feb 2008   & P81    &  K    & 0.69 & 3600\\
ZC410123 		  & 10:02:06.5  &  02:16:15.6 & 22-23 Feb 2008    & P81    &  K	   & 0.29 & 3600\\ 
ZC405226 		  & 10:02:19.5  &  02:00:18.2 & 22-23 Feb 2008    & P81    &  K	   & 0.62 & 3600\\
ZC404073		  & 10:01:42.8  &  01:56:12.7 & 22-23 Feb 2008    & P81    &  K	   & 0.57 & 3600\\ 
ZC403027 		  & 10:02:21.3  &  01:52:34.3 & 23-24 Feb 2008    & P81    &  K    & 0.70 & 3600\\	       
ZC400528 		  & 09:59:47.6  &  01:44:19.0 & 26-27 Feb 2009    & P83    &  K    & 0.49 & 3600\\ 
ZC400569 		  & 10:01:08.7  &  01:44:28.3 & 26-27 Feb 2009    & P83    &  K    & 0.60 & 3600\\
 ZC403103\tablenotemark{b} & 09:59:19.0  &  01:52:50.1 & 1-2 Mar 2009     & P83    &  K	   & 0.64 & 3600\\	 
ZC405545 		  & 10:00:32.8  &  02:01:17.7 & 14-15 Mar 2009    & P83    &  K	   & 0.84 & 3600\\ 
ZC415876 		  & 10:00:09.4  &  02:36:58.3 & 14-15 Mar 2009    & P83    &  K	   & 0.49 & 3600\\ 
ZC405081 		  &10:00:22.7  &  01:59:46.2  & 16-17 Mar 2009    & P83    &  K	   & 0.54 & 6000\\
ZC401925 		  &10:01:01.7  &  01:48:38.2  & 16-17 Mar 2009    & P83    &  K	   & 0.52 & 3600\\ 
ZC404221 		  &10:01:41.3  &  01:56:42.8  & 16-17 Mar 2009    & P83    &  K	   & 0.55 & 3600\\
ZC403741                  &10:00:18.4  &  01:55:08.1  & 20-21 Mar 2009    & P83    &  H    & 0.60 & 3600\\      
ZC404987                  &10:00:40.4  &   01:59:27.7 & 24-25 Mar 2009    & P83    &  K	   & 0.58 & 3600\\
ZC412369                  &10:01:46.9 &   02:23:24.6 & 24-25 Mar 2009     & P83    &  K	   & 0.56 & 3600\\
ZC406690                  &09:58:59.1  &  02:05:04.2& 29-30 Dec 2009      & P84    &  K	   & 0.68 & 3600\\  
ZC411737                  &10:00:32.4  &  02:21:20.9 & 01-02 Jan 2010     & P84    &  K	   & 0.48 & 3600\\
ZC413597 		  &09:59:36.4  &  02:27:59.1 & 01-02 Jan 2010     & P84    &  K	   & 0.54 & 3600\\
ZC405501                  &09:59:53.7   & 02:01:08.8  & 25-26 Jan 2010    & P84    &  K	   & 0.43 & 3600\\
ZC407928 		  & 09:59:53.4 &   02:08:41.5 & 25-26 Jan 2010    & P84    &  K	   & 0.47 & 3600\\
ZC415087 		  & 10:00:32.4  &  02:33:40.6 & 08-09 Feb 2010    & P84    &  K    & 0.76 & 3600\\	 
ZC413507 		  & 10:00:24.2  &  02:27:41.3 & 10-11 Feb 2010    & P84    &  K	   & 0.43 & 3600\\	      
ZC409129 		  & 10:01:28.0 &   02:12:46.6 & 07-08 Mar 2010    & P84    &  K	   & 0.80 & 3600\\
ZC410041 		  & 10:00:44.3  &  02:15:58.4 & 07-08 Mar 2010    & P84    &  K	   & 0.78 & 3600\\
ZC405430 		  & 10:01:32.9  &  02:00:56.5& 01-02 Apr 2010     & P84    &  K    & 0.65 & 3600\\	    
\enddata
\tablenotetext{a}{Targets observed as a 'pilot sample' of the zCOSMOS and SINS teams collaboration, that are also included in the SINS sample presented in FS09. They are named here with the {\it new} zCOSMOS IDs. They had provisional IDs in 2007, respectively, ZC-772759, ZC-782941, ZC-946803, and ZC-1101592, and so they were named in FS09.}
\tablenotetext{b}{This galaxy is the only one in the zC-SINF sample being pre-selected with the $UGR$/BX criterion in zCOSMOS-Deep, instead of the $BzK$ criterion (see also L07, Sect.~\ref{sec:sel_bias}, and Fig.~\ref{fig:BzK}).}
\end{deluxetable}

\begin{deluxetable}{cccccccc}
\tabletypesize{\small}
\centering
\tablecaption{Optical photometry of the zC-SINF sample\label{tbl-phot-2}}
\tablewidth{0pt}
\tablehead{
\colhead{ID}&\colhead{$u^*$} &\colhead{$B_J$}& \colhead{$g^+$}& \colhead{$V_J$}&\colhead{$r^+$}&\colhead{$i^+$}& \colhead{$z^+$}
}
\startdata
ZC400528  & 24.14 {\Tiny $\pm$ 0.02} &  23.71  {\Tiny $\pm$ 0.01}  &  23.64  {\Tiny $\pm$ 0.02 }&  23.34  {\Tiny $\pm$ 0.01 }& 23.13   {\Tiny $\pm$ 0.01 } &22.86    {\Tiny  $\pm$ 0.01 }&  22.80 {\Tiny  $\pm$ 0.03 }\\
ZC400569  & 24.84 {\Tiny $\pm$ 0.05} &  24.27  {\Tiny $\pm$ 0.02}  &  24.17  {\Tiny $\pm$ 0.03 }&  23.79  {\Tiny $\pm$ 0.02 }& 23.60   {\Tiny $\pm$ 0.01 }  &23.28   {\Tiny  $\pm$ 0.02 }&  23.05 {\Tiny  $\pm$ 0.04 }\\
ZC401925  & 24.32 {\Tiny $\pm$ 0.02} &  24.10  {\Tiny $\pm$ 0.01}   & 24.08  {\Tiny $\pm$ 0.02 }&  23.72  {\Tiny $\pm$ 0.02 }& 23.67   {\Tiny $\pm$ 0.01 } &23.63    {\Tiny  $\pm$ 0.02 }&  23.57 {\Tiny  $\pm$ 0.05 }\\
ZC403027  & 25.58 {\Tiny $\pm$ 0.07} &  24.93  {\Tiny $\pm$ 0.03}  &  24.82  {\Tiny $\pm$ 0.06 }&  24.15  {\Tiny $\pm$ 0.03 }& 24.09   {\Tiny $\pm$ 0.02 } &23.83    {\Tiny  $\pm$ 0.03 }&  23.55 {\Tiny  $\pm$ 0.05 }\\
ZC403103  & 24.97 {\Tiny $\pm$ 0.04} &  24.28  {\Tiny $\pm$ 0.01}  &  24.23  {\Tiny $\pm$ 0.03 }&  23.72  {\Tiny $\pm$ 0.02 }& 23.64   {\Tiny $\pm$ 0.01 } &23.40    {\Tiny  $\pm$ 0.02 }&  23.47 {\Tiny  $\pm$ 0.05 }\\
ZC403741  & 24.27 {\Tiny $\pm$ 0.02} &  24.06  {\Tiny $\pm$ 0.01}  &  24.12  {\Tiny $\pm$ 0.02 }&  23.77  {\Tiny $\pm$ 0.02 }& 23.53   {\Tiny $\pm$ 0.01 } &23.08    {\Tiny  $\pm$ 0.01 }&  22.65 {\Tiny  $\pm$ 0.02 }\\
ZC404073  & 25.95 {\Tiny $\pm$ 0.10} &  24.99  {\Tiny $\pm$ 0.03}  &  24.83  {\Tiny $\pm$ 0.05 }&  24.12  {\Tiny $\pm$ 0.02 }& 23.95   {\Tiny $\pm$ 0.02 } &23.71    {\Tiny  $\pm$ 0.02 }&  23.54 {\Tiny  $\pm$ 0.06 }\\
ZC404221  & 24.09 {\Tiny $\pm$ 0.02} &  23.86  {\Tiny $\pm$ 0.01}  &  23.75  {\Tiny $\pm$ 0.02 }&  23.49  {\Tiny $\pm$ 0.02 }& 23.49   {\Tiny $\pm$ 0.01 } &23.35    {\Tiny  $\pm$ 0.02 }&  23.51 {\Tiny  $\pm$ 0.06 }\\
ZC404987  & 24.85 {\Tiny $\pm$ 0.03} &  24.57  {\Tiny $\pm$ 0.02}  &  24.37  {\Tiny $\pm$ 0.02 }&  24.06  {\Tiny $\pm$ 0.02 }& 24.08   {\Tiny $\pm$ 0.02 } &23.97    {\Tiny  $\pm$ 0.03 }&  23.86 {\Tiny  $\pm$ 0.07 }\\
ZC405081  & 24.78 {\Tiny $\pm$ 0.02} &  24.44  {\Tiny $\pm$ 0.02}  &  24.46  {\Tiny $\pm$ 0.03 }&  24.06  {\Tiny $\pm$ 0.02 }& 24.06   {\Tiny $\pm$ 0.02 } &23.91    {\Tiny  $\pm$ 0.03 }&  23.96 {\Tiny  $\pm$ 0.08 }\\
ZC405226  & 24.77 {\Tiny $\pm$ 0.03} &  24.21  {\Tiny $\pm$ 0.01}  &  24.04  {\Tiny $\pm$ 0.03 }&  23.58  {\Tiny $\pm$ 0.02 }& 23.55   {\Tiny $\pm$ 0.01 } &23.33    {\Tiny  $\pm$ 0.02 }&  23.29 {\Tiny  $\pm$ 0.04 }\\
ZC405430  & 25.99 {\Tiny $\pm$ 0.10} &  25.33  {\Tiny $\pm$ 0.04}  &  25.18  {\Tiny $\pm$ 0.07 }&  24.76  {\Tiny $\pm$ 0.04 }& 24.69   {\Tiny $\pm$ 0.04 } &24.51    {\Tiny  $\pm$ 0.05 }&  24.37 {\Tiny  $\pm$ 0.12 }\\
ZC405501  & 24.22 {\Tiny $\pm$ 0.02} &  23.97  {\Tiny $\pm$ 0.01}  &  24.00  {\Tiny $\pm$ 0.03 }&  23.61  {\Tiny $\pm$ 0.02 }& 23.62   {\Tiny $\pm$ 0.02 } &23.41    {\Tiny  $\pm$ 0.02 }&  23.40 {\Tiny  $\pm$ 0.06 }\\
ZC405545  & 25.65 {\Tiny $\pm$ 0.06} &  25.02  {\Tiny $\pm$ 0.03}  &  24.95  {\Tiny $\pm$ 0.05 }&  24.47  {\Tiny $\pm$ 0.03 }& 24.50   {\Tiny $\pm$ 0.03 }   &24.19  {\Tiny  $\pm$ 0.04 }&  24.21 {\Tiny  $\pm$ 0.11 }\\
ZC406690  & 22.98 {\Tiny $\pm$ 0.01} &  22.58  {\Tiny $\pm$ 0.01}  &  22.62  {\Tiny $\pm$ 0.01 }&  22.15  {\Tiny $\pm$ 0.01 }& 22.05   {\Tiny $\pm$ 0.00 } &21.88    {\Tiny  $\pm$ 0.01 }&  21.89 {\Tiny  $\pm$ 0.02 }\\
ZC407302  & 24.03 {\Tiny $\pm$ 0.01} &  23.62  {\Tiny $\pm$ 0.01}  &  23.51  {\Tiny $\pm$ 0.01 }&  23.05  {\Tiny $\pm$ 0.01 }& 23.06   {\Tiny $\pm$ 0.01 } &22.85    {\Tiny  $\pm$ 0.01 }&  22.71 {\Tiny  $\pm$ 0.02 }\\
ZC407376  & 25.28 {\Tiny $\pm$ 0.04} &  24.75  {\Tiny $\pm$ 0.03}  &  24.73  {\Tiny $\pm$ 0.05 }&  24.10  {\Tiny $\pm$ 0.03 } & 24.11  {\Tiny $\pm$ 0.03 }   &23.78  {\Tiny  $\pm$ 0.03 }&  23.59 {\Tiny  $\pm$ 0.07 }\\   
ZC407928  & 25.47 {\Tiny $\pm$ 0.05} &  24.81  {\Tiny $\pm$ 0.02}  &  24.75  {\Tiny $\pm$ 0.06 }&  24.17  {\Tiny $\pm$ 0.03 }& 24.09   {\Tiny $\pm$ 0.03 } &23.80    {\Tiny  $\pm$ 0.03 }&  23.82 {\Tiny  $\pm$ 0.08 }\\
ZC409129  & 25.66 {\Tiny $\pm$ 0.06} &  24.85  {\Tiny $\pm$ 0.02}  &  24.77  {\Tiny $\pm$ 0.04 }&  24.32  {\Tiny $\pm$ 0.02 }& 24.17   {\Tiny $\pm$ 0.02 } &23.87    {\Tiny  $\pm$ 0.03 }&  23.67 {\Tiny  $\pm$ 0.06 }\\	  
ZC409985  & 24.41 {\Tiny $\pm$ 0.02} &  24.16  {\Tiny $\pm$ 0.02}  &  24.29  {\Tiny $\pm$ 0.03 }&  23.66  {\Tiny $\pm$ 0.02 }& 23.55   {\Tiny $\pm$ 0.01 } &23.46    {\Tiny  $\pm$ 0.02 }&  23.49 {\Tiny  $\pm$ 0.05 }\\
ZC410041  & 24.71 {\Tiny $\pm$ 0.02} &  24.24  {\Tiny $\pm$ 0.02}  &  24.24  {\Tiny $\pm$ 0.03 }&  23.83  {\Tiny $\pm$ 0.02 }& 23.82   {\Tiny $\pm$ 0.02 } &23.67    {\Tiny  $\pm$ 0.02 }&  23.70 {\Tiny  $\pm$ 0.07 }\\
ZC410116  & 25.41 {\Tiny $\pm$ 0.05} &  25.31  {\Tiny $\pm$ 0.04}  &  25.24  {\Tiny $\pm$ 0.06 }&  24.69  {\Tiny $\pm$ 0.03 }& 24.59   {\Tiny $\pm$ 0.03 } &24.15    {\Tiny  $\pm$ 0.03 }&  23.81 {\Tiny  $\pm$ 0.06 }\\  
ZC410123  & 24.62 {\Tiny $\pm$ 0.03} &  24.21  {\Tiny $\pm$ 0.01}  &  24.23  {\Tiny $\pm$ 0.03 }&  23.75  {\Tiny $\pm$ 0.02 }& 23.84   {\Tiny $\pm$ 0.02 } &23.63    {\Tiny  $\pm$ 0.02 }&  23.61 {\Tiny  $\pm$ 0.06 }\\
ZC410542  & 23.97 {\Tiny $\pm$ 0.01} &  23.78  {\Tiny $\pm$ 0.01}  &  23.82  {\Tiny $\pm$ 0.02 }&  23.46  {\Tiny $\pm$ 0.02 }& 23.26   {\Tiny $\pm$ 0.01 } &22.78    {\Tiny  $\pm$ 0.01 }&  22.32 {\Tiny  $\pm$ 0.02 }\\
ZC411737  & 24.70 {\Tiny $\pm$ 0.03} &  24.25  {\Tiny $\pm$ 0.02}  &  24.25  {\Tiny $\pm$ 0.03 }&  23.84  {\Tiny $\pm$ 0.02 }& 23.77   {\Tiny $\pm$ 0.02 } &23.63    {\Tiny  $\pm$ 0.02 }&  23.74 {\Tiny  $\pm$ 0.07 }\\
ZC412369  & 24.37 {\Tiny $\pm$ 0.03} &  24.01  {\Tiny $\pm$ 0.01}  &  23.89  {\Tiny $\pm$ 0.03 }&  23.38  {\Tiny $\pm$ 0.02 }& 23.40   {\Tiny $\pm$ 0.01 } &23.06    {\Tiny  $\pm$ 0.02 }&  22.88 {\Tiny  $\pm$ 0.04 }\\
ZC413507  & 25.64 {\Tiny $\pm$ 0.08} &  24.73  {\Tiny $\pm$ 0.03}  &  24.78  {\Tiny $\pm$ 0.05 }&  24.06  {\Tiny $\pm$ 0.03 }& 24.02   {\Tiny $\pm$ 0.02 } &23.82    {\Tiny  $\pm$ 0.03 }&  23.71 {\Tiny  $\pm$ 0.07 }\\
ZC413597  & 25.28 {\Tiny $\pm$ 0.06} &  24.82  {\Tiny $\pm$ 0.02}  &  24.70  {\Tiny $\pm$ 0.04 }&  24.18  {\Tiny $\pm$ 0.02 }& 23.95   {\Tiny $\pm$ 0.02 } &23.90    {\Tiny  $\pm$ 0.03 }&  23.80 {\Tiny  $\pm$ 0.07 }\\
ZC415087  & 24.80 {\Tiny $\pm$ 0.03} &  24.56  {\Tiny $\pm$ 0.02}  &  24.50  {\Tiny $\pm$ 0.04 }&  24.04  {\Tiny $\pm$ 0.03 }& 24.21   {\Tiny $\pm$ 0.03 } &23.86    {\Tiny  $\pm$ 0.03 }&  23.78 {\Tiny  $\pm$ 0.08 }\\
ZC415876  & 25.23 {\Tiny $\pm$ 0.04} &  24.64  {\Tiny $\pm$ 0.02}  &  24.57  {\Tiny $\pm$ 0.04 }&  23.99  {\Tiny $\pm$ 0.02 }& 23.95   {\Tiny $\pm$ 0.02 } &23.72    {\Tiny  $\pm$ 0.02 }&  23.69 {\Tiny  $\pm$ 0.06 }\\
\enddata																									      
\end{deluxetable}

\begin{deluxetable}{ccccccccc}
\tabletypesize{\small}
\centering
\tablecaption{near-IR/mid-IR photometry of the zC-SINF sample\label{tbl-phot-3}}
\tablewidth{0pt}
\tablehead{
\colhead{ID}&\colhead{$J$}&\colhead{$H$}& \colhead{$K$}&\colhead{$m_{3.6}$}&\colhead{$m_{4.5}$}&\colhead{$m_{5.8}$}&\colhead{$m_{8.0}$}&\colhead{$S_{24}$[$\mu$Jy]}
}
\startdata
ZC400528  &  22.35 {\Tiny  $\pm$ 0.11 }&  21.75  {\Tiny $\pm$ 0.04  }&  21.08 {\Tiny $\pm$ 0.03  }&  20.43{\Tiny$\pm$ 0.10}   &  20.30  {\Tiny $\pm $0.10  } &  20.21  {\Tiny$\pm$ 0.10  } &  20.29 {\Tiny $\pm $0.15 } &   228  {\Tiny$\pm$  18}  \\	
ZC400569  &  22.11 {\Tiny  $\pm$ 0.09 }&  21.29  {\Tiny $\pm$ 0.03  }&  20.69 {\Tiny $\pm$ 0.02  }&  20.25{\Tiny$\pm$ 0.10}   &  20.05  {\Tiny $\pm $0.10  } &  20.11  {\Tiny$\pm$ 0.11  } &  20.62 {\Tiny $\pm $0.16 } &   141   {\Tiny$\pm$ 29}  \\	
ZC401925  &  23.01 {\Tiny  $\pm$ 0.14 }&  22.92  {\Tiny $\pm$ 0.12  }&  22.74 {\Tiny $\pm$ 0.12  }&  22.52{\Tiny$\pm$ 0.11}   &  22.64  {\Tiny $\pm $0.12  } &  22.11  {\Tiny$\pm$ 0.23  } &  22.99 {\Tiny $\pm $1.06 } &   $<80$\tablenotemark{c}  \\
ZC403027  &  23.24 {\Tiny  $\pm$ 0.15 }&  22.39  {\Tiny $\pm$ 0.05  }&  22.19 {\Tiny $\pm$ 0.07  }&  21.53{\Tiny$\pm$ 0.10}   &  21.56  {\Tiny $\pm $0.10  } &  21.53  {\Tiny$\pm$ 0.16  } &  21.76 {\Tiny $\pm $0.35 } &    84   {\Tiny$\pm$ 16}  \\ 
ZC403103  &  22.91 {\Tiny  $\pm$ 0.13 }&  23.94  {\Tiny $\pm$ 0.20  }&  23.49 {\Tiny $\pm$ 0.23  }&  23.42{\Tiny$\pm$ 0.14}   &  23.87  {\Tiny $\pm $0.25  } &  $>$21.3\tablenotemark{b}			   & $>$21.0\tablenotemark{b}			&   $<80$\tablenotemark{c}  	       \\ 
ZC403741  &  21.81 {\Tiny  $\pm$ 0.05 }&  21.30  {\Tiny $\pm$ 0.02  }&  21.02 {\Tiny $\pm$ 0.02  }&  20.51{\Tiny$\pm$ 0.10}   &  20.49  {\Tiny $\pm $0.11  } &  20.54  {\Tiny$\pm$ 0.11  } &  21.02{\Tiny $\pm$ 0.15 }&   $<80$\tablenotemark{c}       \\ 
ZC404073  &  22.63 {\Tiny  $\pm$ 0.12 }&  22.35  {\Tiny $\pm$ 0.04  }&  21.96 {\Tiny $\pm$ 0.05  }&  21.34{\Tiny$\pm$ 0.10}   &  21.20  {\Tiny $\pm $0.10  } &  20.87  {\Tiny$\pm$ 0.12  } &  21.39{\Tiny $\pm$ 0.27 }&    82	{\Tiny$\pm$ 18}  \\  
ZC404221  &  22.93 {\Tiny  $\pm$ 0.17 }&  22.83  {\Tiny $\pm$ 0.08  }&  22.44 {\Tiny $\pm$ 0.09  }&  21.86{\Tiny$\pm$ 0.10}   &  21.80  {\Tiny $\pm$ 0.11  } &  21.78  {\Tiny$\pm$ 0.17  } &  21.03{\Tiny $\pm$ 0.19 }&   $<80$\tablenotemark{c}  \\ 
ZC404987  &  23.31 {\Tiny  $\pm$ 0.15 }&  23.19  {\Tiny $\pm$ 0.11  }&  22.77 {\Tiny $\pm$ 0.13  }&  22.27{\Tiny$\pm$ 0.11}   &  22.39  {\Tiny $\pm$ 0.12  } &  23.04  {\Tiny$\pm$ 0.52  } &  22.81{\Tiny $\pm$ 1.01 }&   $<80$\tablenotemark{c}  \\ 
ZC405081  &  23.27 {\Tiny  $\pm$ 0.15 }&  23.40  {\Tiny $\pm$ 0.15  }&  22.91 {\Tiny $\pm$ 0.16  }&  -1.00\tablenotemark{a}   &  -1.00\tablenotemark{a}  		    &  -1.00\tablenotemark{a}			   &  -1.00\tablenotemark{a}		      &   $<80$\tablenotemark{c}	    \\ 
ZC405226  &  23.57 {\Tiny  $\pm$ 0.44 }&  22.61  {\Tiny $\pm$ 0.07  }&  22.33 {\Tiny $\pm$ 0.09  }&  21.97{\Tiny$\pm$ 0.10}   &  22.07  {\Tiny $\pm$ 0.11  } &  22.32  {\Tiny $\pm$ 0.27 }  &  21.73  {\Tiny $\pm$ 0.36 }&   $<80$\tablenotemark{c}  \\
ZC405430  &  25.04 {\Tiny  $\pm$ 1.01 }&  23.61  {\Tiny $\pm$ 0.16  }&  23.10 {\Tiny $\pm$ 0.17  }&  23.41{\Tiny$\pm$ 0.14}   &  23.79  {\Tiny $\pm$ 0.25  } &  $>$21.3\tablenotemark{b}    		    & $>$21.0\tablenotemark{b} 		         &   $<80$\tablenotemark{c}  \\    
ZC405501  &  22.81 {\Tiny  $\pm$ 0.19 }&  23.12  {\Tiny $\pm$ 0.14  }&  22.25 {\Tiny $\pm$ 0.11  }&  22.05{\Tiny$\pm$ 0.10}   &  21.99  {\Tiny $\pm$ 0.11  } &  22.70  {\Tiny $\pm$ 0.39 }  & $>$21.0\tablenotemark{b} 			 &   $<80$\tablenotemark{c}	      \\ 
ZC405545  &  23.74 {\Tiny  $\pm$ 0.26 }&  22.99  {\Tiny $\pm$ 0.12  }&  22.32 {\Tiny $\pm$ 0.11  }&  22.64{\Tiny$\pm$ 0.11}   &  22.50  {\Tiny $\pm$ 0.12  } &  23.65  {\Tiny $\pm$ 0.89 }  & $>$21.0\tablenotemark{b} 			 &   $<80$\tablenotemark{c}	\\ 
ZC406690  &  21.41 {\Tiny  $\pm$ 0.03 }&  21.11  {\Tiny $\pm$ 0.02  }&  20.81 {\Tiny $\pm$ 0.03  }&  20.93{\Tiny$\pm$ 0.10}   &  20.89  {\Tiny $\pm$ 0.10  } &  20.72  {\Tiny $\pm$ 0.12 }  &  20.84  {\Tiny $\pm$ 0.18 }&   209  {\Tiny$\pm$  16}  \\   
ZC407302  &  22.17 {\Tiny  $\pm$ 0.05 }&  21.98  {\Tiny $\pm$ 0.03  }&  21.48 {\Tiny $\pm$ 0.04  }&  -1.00\tablenotemark{a}   &  -1.00\tablenotemark{a}  		     & -1.00\tablenotemark{a}			     &  	-1.00\tablenotemark{a}	  &   295 $\pm$  16  \\    
ZC407376  &  22.67 {\Tiny  $\pm$ 0.10 }&  22.32  {\Tiny $\pm$ 0.05  }&  21.79 {\Tiny $\pm$ 0.06  }&  21.86{\Tiny$\pm$ 0.10}   &  21.69  {\Tiny $\pm$ 0.10  } &  20.97  {\Tiny $\pm $ 0.12 }  &  21.53 {\Tiny  $\pm$ 0.27} &   103  {\Tiny$\pm$  16}  \\  
ZC407928  &  23.62 {\Tiny  $\pm$ 0.23 }&  23.25  {\Tiny $\pm$ 0.12  }&  22.63 {\Tiny $\pm$ 0.13  }&  22.55{\Tiny$\pm$ 0.11}   &  22.48  {\Tiny $\pm$ 0.12  } &  22.51  {\Tiny $\pm $ 0.31 }  & $>$21.0\tablenotemark{b}  	          &   $<80$\tablenotemark{c}  \\ 
ZC409129  &  22.97 {\Tiny  $\pm$ 0.11 }&  21.92  {\Tiny $\pm$ 0.03  }&  21.55 {\Tiny $\pm$ 0.04  }&  20.83{\Tiny$\pm$ 0.10}   &  20.62  {\Tiny $\pm$ 0.10  } &  20.36  {\Tiny $\pm $ 0.11 }  &  20.38 {\Tiny  $\pm$ 0.14} &    92   {\Tiny$\pm$ 15}    \\
ZC409985  &  23.04 {\Tiny  $\pm$ 0.11 }&  22.42  {\Tiny $\pm$ 0.05  }&  22.30 {\Tiny $\pm$ 0.08  }&  22.39{\Tiny$\pm$ 0.11}   &  22.32  {\Tiny $\pm$ 0.12  } &  21.84  {\Tiny $\pm $ 0.20 }  & $>$21.0\tablenotemark{b}  	          &   $<80$\tablenotemark{c}  \\ 
ZC410041  &  23.40 {\Tiny  $\pm$ 0.17 }&  22.77  {\Tiny $\pm$ 0.08  }&  23.16 {\Tiny $\pm$ 0.18  }&  22.95{\Tiny$\pm$ 0.12}   &  23.38  {\Tiny $\pm$ 0.19  } &  23.03  {\Tiny $\pm $ 0.52 }  &  22.08 {\Tiny  $\pm$ 0.50 }&   $<80$\tablenotemark{c}  \\
ZC410116  &  22.86 {\Tiny  $\pm$ 0.09 }&  21.99  {\Tiny $\pm$ 0.04  }&  21.46 {\Tiny $\pm$ 0.04  }&  20.76{\Tiny$\pm$ 0.10}   &  20.48  {\Tiny $\pm$ 0.10  } &  20.24  {\Tiny $\pm $ 0.11 }  &  19.94  {\Tiny $\pm$ 0.12} &   146  {\Tiny$\pm$  12}  \\  
ZC410123  &  22.73 {\Tiny  $\pm$ 0.08 }&  23.06  {\Tiny $\pm$ 0.09  }&  22.80 {\Tiny $\pm$ 0.11  }&  22.81{\Tiny$\pm$ 0.11}   &  22.59  {\Tiny $\pm$ 0.13  } & $>$21.3\tablenotemark{b}  			     &  22.87  {\Tiny $\pm$ 1.10} &   $<80$\tablenotemark{c}		\\ 
ZC410542  &  21.42 {\Tiny  $\pm$ 0.03 }&  20.86  {\Tiny $\pm$ 0.02  }&  20.60 {\Tiny $\pm$ 0.02  }&  20.02{\Tiny$\pm$ 0.10}   &  19.89  {\Tiny $\pm$ 0.10  } &  20.00  {\Tiny $\pm $ 0.10 }  &  20.14  {\Tiny $\pm$ 0.13} &   165  {\Tiny$\pm$  12}\\  
ZC411737  &  23.73 {\Tiny  $\pm$ 0.22 }&  23.00  {\Tiny $\pm$ 0.12  }&  22.81 {\Tiny $\pm$ 0.15  }&  23.13{\Tiny$\pm$ 0.12}   &  23.31  {\Tiny $\pm$ 0.17  } & $>$21.3\tablenotemark{b}   			     & $>$21.0\tablenotemark{b}                   &   $<80$\tablenotemark{c}  \\
ZC412369  &  22.45 {\Tiny  $\pm$ 0.08 }&  21.85  {\Tiny $\pm$ 0.04  }&  21.39 {\Tiny $\pm$ 0.05  }&  21.65{\Tiny$\pm$ 0.10}   &  21.58  {\Tiny $\pm$ 0.10  } &  21.30  {\Tiny $\pm $ 0.15 }  &  22.77  {\Tiny $\pm$ 0.94} &   $<80$\tablenotemark{c}  \\ 
ZC413507  &  23.19 {\Tiny  $\pm$ 0.14 }&  22.77  {\Tiny $\pm$ 0.08  }&  22.52 {\Tiny $\pm$ 0.11  }&  22.43{\Tiny$\pm$ 0.11}   &  22.37  {\Tiny $\pm$ 0.12  } &  21.23  {\Tiny $\pm $ 0.14 }  & $>$21.0\tablenotemark{b} &   $<80$\tablenotemark{c}  \\ 
ZC413597  &  22.99 {\Tiny  $\pm$ 0.13 }&  22.96  {\Tiny $\pm$ 0.09  }&  22.58 {\Tiny $\pm$ 0.11  }&  22.55{\Tiny$\pm$ 0.11}   &  22.56  {\Tiny $\pm$ 0.12  } &  22.88  {\Tiny $\pm $ 0.45 }  &  21.98  {\Tiny $\pm$ 0.43} &   $<80$\tablenotemark{c}  \\ 
ZC415087  &  23.22 {\Tiny  $\pm$ 0.14 }&  22.94  {\Tiny $\pm$ 0.11  }&  22.87 {\Tiny $\pm$ 0.16  }& -1.00\tablenotemark{a}    &  -1.00\tablenotemark{a}  		     & -1.00\tablenotemark{a}			       &  -1.00\tablenotemark{a} 	          &	$<80$\tablenotemark{c}	 \\
ZC415876  &  23.23 {\Tiny  $\pm$ 0.15 }&  22.73  {\Tiny $\pm$ 0.08  }&  22.38 {\Tiny $\pm$ 0.11  }& 22.38{\Tiny $\pm$ 0.10}  &  22.34  {\Tiny $\pm$ 0.12   }&  22.35   {\Tiny $\pm $0.24 }  &  21.21  {\Tiny $\pm$ 0.26} & $<80$\tablenotemark{c} \\ 
\enddata
\tablenotetext{a}{Sources with IRAC fluxes equal to -1.00 are blended in all the IRAC bands. This was deduced based both on the visual inspection of IRAC image stamps, and on the empirical criterion, tested on extensive simulations by the GOODS team, which consists in classifying as blended all the galaxies for which the angular separation between the IRAC and the $K$-band position exceeds $0\farcs6$ \citep[cf.][]{2007ApJ...670..156D}.}
\tablenotetext{b}{For IRAC and undetected sources we list the 5$\sigma$ magnitude lower limits \citep{2007ApJS..172...86S}, corresponding to the flux upper limits shown in Fig.~\ref{fig:sed} (black arrows)}.
\tablenotetext{c}{5$\sigma$ detection limit for the MIPS 24~$\mu$m catalog \citep{2009ApJ...703..222L}.}
\end{deluxetable}

\begin{turnpage} 
\begin{deluxetable}{c|rrr|rrr|rrr|rrr}
\tabletypesize{\scriptsize}
\tablecaption{SED fitting results\label{tbl-sedfit-4}}
\tablewidth{0pt}
\tablehead{
ID & \multicolumn{3}{|c|}{M* [$10^{10} M_{\odot}$]} & \multicolumn{3}{|c|}{SFR [$\frac{M_{\odot}}{yr}$]} &\multicolumn{3}{|c|}{Age [$10^8$yr]}& \multicolumn{3}{|c}{$A_{\rm V}$ [mag]}\\
   &\footnotesize{MA05} & \footnotesize{MA05}&\footnotesize{BC03} & \footnotesize{MA05}& \footnotesize{MA05} & \footnotesize{BC03} & \footnotesize{MA05}& \footnotesize{MA05} & \footnotesize{BC03}& \footnotesize{MA05}& \footnotesize{MA05}& \footnotesize{BC03}\\
   & \footnotesize{(CSFR)} &\footnotesize{(inv-$\tau$)} &\footnotesize{(CSFR)} & \footnotesize{(CSFR)}& \footnotesize{(inv-$\tau$)} & \footnotesize{(CSFR)} & \footnotesize{(CSFR)}& \footnotesize{(inv-$\tau$)} &\footnotesize{(CSFR)}&\footnotesize{(CSFR)}& \footnotesize{(inv-$\tau$)}&\footnotesize{(CSFR)}\\ 
} 
\startdata
ZC400528& 7.81 &  9.12  & 11.00     & 206.0 &163.0 &148.0   & 5.1 & 15.0 &11.4  &       1.0  &  1.0 & 0.9   \\
ZC400569& 8.65 &  11.50  & 16.10     & 436.0 &213.0 &241.0   & 2.6 & 17.0 &10.1  &       1.6  &  1.4 & 1.4   \\
ZC401925& 0.49 &  0.85  & 0.58    & 52.7  & 20.0 & 47.3   & 1.1 & 17.0 & 1.6  &       0.7  &  0.4 & 0.7   \\
ZC403027& 1.81 &  3.63  & 3.05     & 242.0 &108.0 &164.0   & 1.0 & 15.0 & 2.5  &       1.4  &  1.2 & 1.3   \\
ZC403103& 0.32 &  0.65  & 0.30    & 43.1  & 18.8 & 41.8   & 1.0 & 15.0 & 1.0  &       0.6  &  0.4 & 0.6   \\
ZC403741& 2.92 &  3.63  & 4.45     & 164.0 &105.0 &113.0   & 2.3 & 30.0 & 5.7  &       1.7  &  1.6 & 1.6   \\
ZC404073& 3.03 & 5.50   & 3.93     & 326.0 &126.0 &290.0   & 1.1 & 11.0 & 1.8  &       1.5  &  1.2 & 1.5   \\
ZC404221& 0.85 & 1.62   & 1.57     & 114.0 & 42.4 & 61.2   & 1.0 & 17.0 & 3.6  &       0.9  &  0.6 & 0.7   \\
ZC404987& 0.54 & 1.12   & 0.69    & 71.6  & 32.4 & 63.0   & 1.0 & 17.0 & 1.4  &       1.0  &  0.8 & 1.0   \\
ZC405081& 0.50 & 1.32   & 0.49    & 66.4  & 38.3 & 61.0   & 1.0 & 17.0 & 1.0  &       0.9  &  0.8 & 0.9   \\
ZC405226& 0.92 & 1.86   & 0.93    & 123.0 & 54.2 &117.0   & 1.0 & 15.0 & 1.0  &       1.0  &  0.8 & 1.0   \\
ZC405430& 0.32 & 0.54   & 0.32    & 42.6  & 15.3 & 39.8   & 1.0 & 15.0 & 1.0  &       0.9  &  0.6 & 0.9   \\
ZC405501& 0.71 &  1.35  & 0.84    & 94.6  & 35.3 & 84.9   & 1.0 & 17.0 & 1.3  &       0.9  &  0.6 & 0.9   \\
ZC405545& 0.57 &  0.98  & 0.68    & 61.4  & 23.0 & 55.6   & 1.1 & 17.0 & 1.6  &       1.1  &  0.8 & 1.1   \\
ZC406690& 2.68 &  4.57  & 4.14     & 288.0 &134.0 &200.0   & 1.1 & 17.0 & 2.9  &       0.8  &  0.6 & 0.7   \\
ZC407302& 2.20 &  4.57  & 2.44     & 294.0 &132.0 &340.0   & 1.0 & 17.0 & 1.0  &       1.2  &  1.0 & 1.3   \\
ZC407376& 1.36 &  2.51  & 2.53     & 163.0 & 74.2 & 88.7   & 1.0 & 17.0 & 4.0  &       1.4  &  1.2 & 1.2   \\
ZC407928& 0.65 &  1.29  & 0.78    & 86.2  & 37.7 & 79.3   & 1.0 & 15.0 & 1.3  &       1.0  &  0.8 & 1.0   \\
ZC409129& 8.44 &  9.55  & 14.60     & 223.0 &188.0 &159.0   & 5.1 & 11.0 & 14.3 &       1.4  &  1.4 & 1.3   \\
ZC409985& 0.79 &  1.48  & 1.61     & 95.0  & 43.6 & 50.8   & 1.0 & 15.0 & 4.5  &       0.8  &  0.6 & 0.6   \\
ZC410041& 0.38 &  0.78 & 0.46    & 51.3  & 23.0 & 46.9   & 1.0 & 15.0 & 1.3  &       0.6  &  0.4 & 0.6   \\
ZC410116& 6.68 &  8.32  & 12.70     & 219.0 &110.0 &124.0   & 4.0 & 17.0 & 16.1 &       1.8  &  1.6 & 1.6   \\
ZC410123& 0.42 &  0.78 & 0.42    & 50.2  & 18.3 & 59.1   & 1.0 & 17.0 & 1.0  &       0.7  &  0.4 & 0.8   \\
ZC410542& 3.16 &  5.01  & 6.62     & 304.0 &128.0 &136.0   & 1.3 & 30.0 & 7.2  &       1.9  &  1.6 & 1.6   \\
ZC411737& 0.38 &  0.76 & 0.34    & 50.2  & 22.4 & 48.1   & 1.0 & 15.0 & 1.0  &       0.6  &  0.4 & 0.6   \\
ZC412369& 1.41 &  2.24  & 2.17     & 136.0 & 51.8 & 94.1   & 1.3 & 19.0 & 3.2  &       1.1  &  0.8 & 1.0   \\
ZC413507& 0.87 &  1.48  & 0.88    & 116.0 & 43.0 &111.0   & 1.0 & 15.0 & 1.0  &       1.1  &  0.8 & 1.1   \\
ZC413597& 0.68 &  1.38  & 0.75    & 90.6  & 40.0 & 84.4   & 1.0 & 15.0 & 1.1  &       1.0  &  0.8 & 1.0   \\
ZC415087& 0.74 &  1.41  & 1.35     & 88.1  & 41.2 & 47.4   & 1.0 & 15.0 & 4.0  &       1.0  &  0.8 & 0.8   \\
ZC415876& 0.77 &  1.55  & 0.92    & 102.0 & 45.3 & 93.7   & 1.0 & 15.0 & 1.3  &       1.0  &  0.8 & 1.0   \\
\enddata
\tablecomments
{We do not list the uncertainties on the SED best-fit parameters, since the formal errors derived by the $\chi^2$ procedure are often too small to be reliable, 
as already pointed out in Section ~\ref{sec:const_vs_invt}, and clearly contrast with the much larger systematic differences highlighted 
in the same Section (see also Fig.~\ref{fig:comp-mass}-\ref{fig:comp-sfr}).} 
\end{deluxetable}
\end{turnpage}

\begin{turnpage}

\begin{deluxetable}{ccccccccc}
\tabletypesize{\footnotesize}
\tablecaption{H$\alpha$ integrated properties\label{tbl-Ha-5}}
\tablewidth{0pt}
\tablehead{
\colhead{ID}&
\colhead{z$_{opt}$} &
\colhead{z(H$\alpha$)} & 
\colhead{$r_{\rm ap}$}&
\colhead{F(H$\alpha$)} & 
\colhead{$\sigma$(H$\alpha$)$_{\rm integrated}$}&
\colhead{r$_{1/2}$(H$\alpha$)}&
\colhead{$v_{obs}$}&
\colhead{frac$_{\rm BB}$(H$\alpha$)}\\
\colhead{ }&
\colhead{ }&
\colhead{ } & 
\colhead{[arcsec]}&
\colhead{[$10^{-17}$ erg cm$^{-2}$ s$^{-1}$]} & 
\colhead{[km s$^{-1}$]}&
\colhead{[kpc]}&
\colhead{[km s$^{-1}$]}&
\colhead{}
}
\startdata
ZC400528  &  2.377  &  2.3876  & 1.00  &   $15.6^{+1.5}_{-2.0}$  &  $221^{+72}_{-62}$  &  $2.4 \pm 1.0$  &   $143 \pm 90$  	& $ 0.065^{+0.007}_{-0.009}$\\  
ZC400569  &   2.2433  &  2.2405  & 1.25  &   $16.1^{+1.5}_{-1.8}$  &  $207^{+34}_{-39}$  &  $5.8 \pm 1.4$  &   $502 \pm 84$  	& $ 0.043^{+0.004}_{-0.005}$\\  
ZC401925  &  2.1394 &  2.1411  & 0.75  & $2.32^{+0.22}_{-0.18}$  &  $73^{+12}_{-9}$  &  $1.5 \pm 1.0$  &   $43 \pm 19$  	& $ 0.038^{+0.005}_{-0.005}$\\  
ZC403027  &  2.480  &  2.4848  & 1.00  &   $2.66^{+0.67}_{-0.61}$  &  $151 \pm 70$  &  $2.3 \pm 1.3$  &   $< 94$  		& $ 0.032^{+0.008}_{-0.008}$\\  
ZC403103  &  2.3613  &  \ldots  & 1.00  &   $< 3.99$                &  \ldots           &  \ldots         &   \ldots        	& $<0.050		   $\\  
ZC403741  &  1.6526 &  1.4455  & 1.00   &  $12.0 \pm 0.1$        &  $86^{+7}_{-4}$  &  $3.1 \pm 1.1$  &   $108 \pm 25$  	& $ 0.033^{+0.002}_{-0.002}$\\  
ZC404073  &  2.5514  &  2.5571  & 0.75  &   $2.62^{+0.66}_{-0.52}$  &  $69^{+37}_{-62}$  &  $2.0 \pm 1.1$  &   $< 70$  		& $ 0.027^{+0.007}_{-0.006}$\\  
ZC404221  &  2.2192  &  2.2201  & 1.00  &   $8.66^{+0.60}_{-0.85}$  &  $96^{+37}_{-19}$  &  $2.3 \pm 1.7$  &   $< 24$  		& $ 0.114^{+0.012}_{-0.015}$\\  
ZC404987  &  2.1174 &  2.1239  & 0.75  &   $7.04^{+1.02}_{-1.80}$  &  $131^{+43}_{-61}$  &  $1.5 \pm 1.3$  &  $< 69$  		& $ 0.118^{+0.022}_{-0.033}$\\  
ZC405081  &  2.2239  &  2.2344  & 1.25  &   $5.01^{+0.71}_{-0.93}$  &  $64^{+24}_{-21}$  &  $5.5 \pm 1.4$  &   $107 \pm 26$  	& $ 0.102^{+0.021}_{-0.024}$\\  
ZC405226 &  2.2875  &  2.2872  & 1.00  & $1.72^{+0.31}_{-0.29}$  &  $141^{+224}_{-44}$  &  $2.6 \pm 1.2$  &   $189 \pm 27$  	& $ 0.021^{+0.004}_{-0.004}$\\  
ZC405430  &  2.4466  &  \ldots  & 1.00  &   $< 4.50$                &  \ldots           &  \ldots         &   \ldots  		& $<0.041		   $\\  
ZC405501  &  2.1530  &  2.1543  & 1.25  &   $11.2^{+1.6}_{-1.3}$  &  $124^{+32}_{-18}$  &  $4.4 \pm 1.0$  &   $129 \pm 26$  	& $ 0.118^{+0.021}_{-0.018}$\\  
ZC405545  &  2.195  &  \ldots  & 1.00  &   $< 1.13$                &  \ldots           &  \ldots         &   \ldots  		& $<0.004		   $\\  
ZC406690  &  2.1934  &  2.1949  & 1.50  &   $61.5^{+1.4}_{-1.1}$  &  $138 \pm 4$  &  $4.9 \pm 1.4$  &   $154 \pm 19$  		& $ 0.177^{+0.009}_{-0.009}$\\  
ZC407302  &  2.1803  &  2.1814  & 1.00  &   $22.5^{+0.9}_{-0.8}$  &  $171^{+12}_{-8}$  &  $3.2 \pm 1.0$  &   $266 \pm 51$  	& $ 0.119^{+0.007}_{-0.007}$\\  
ZC407376  &  2.1792  &  2.1733  & 1.25  &   $15.4^{+2.0}_{-1.4}$  &  $125^{+18}_{-15}$  &  $4.5 \pm 1.2$  &   $79 \pm 42$  	& $ 0.108^{+0.015}_{-0.011}$\\  
ZC407928  &  2.4295  &  2.4309  & 1.00  &   $7.94^{+1.52}_{-1.57}$  &  $178^{+185}_{-86}$  &  $2.8 \pm 0.9$  &   $233 \pm 110$  & $ 0.141^{+0.032}_{-0.032}$\\  
ZC409129\tablenotemark{a}  &  2.496  &  2.5002  & 1.00  &   $< 6.63$                &  \ldots        &  \ldots    &   \ldots    & $<0.015		   $\\  
ZC409985  &  2.4505 &  2.4577  & 1.00  &   $3.90^{+0.71}_{-0.60}$  &  $110^{+77}_{-29}$  &  $2.2 \pm 1.4$  &   $45 \pm 19$  	& $ 0.052^{+0.010}_{-0.009}$\\  
ZC410041  &  2.4500 &  2.4539  & 1.25  &   $13.4^{+2.9}_{-2.7}$  &  $154^{+80}_{-60}$  &  $5.9 \pm 1.5$  &   $65 \pm 21$  	& $ 0.393^{+0.107}_{-0.101}$\\  
ZC410116  &  2.0903  &  \ldots  & 1.00  &   $< 7.98$                &  \ldots           &  \ldots         &   \ldots  		& $<0.013		   $\\  
ZC410123  &  2.1993  &  2.1987  & 1.00  &  $1.72^{+0.21}_{-0.26}$  &  $70^{+24}_{-21}$  &  $3.3 \pm 0.8$  &   $121 \pm 54$  	& $ 0.031^{+0.005}_{-0.006}$\\  
ZC410542  &  1.4035 &  1.4049  & 1.25  &   $18.4^{+1.4}_{-2.0}$  &  $283^{+39}_{-64}$  &  $4.1 \pm 1.5$  &   $431 \pm 44$  	& $ 0.033^{+0.003}_{-0.004}$\\  
ZC411737  &  2.4473 &  2.4443  & 0.75  &   $9.63^{+0.95}_{-0.69}$  &  $106^{+30}_{-15}$  &  $2.0 \pm 0.8$  &   $46 \pm 35$  	& $ 0.203^{+0.034}_{-0.032}$\\  
ZC412369  &  2.0285  &  2.0283  & 1.25  &   $21.7^{+1.0}_{-1.0}$  &  $133 \pm 9$  &  $2.3 \pm 1.0$  &   $73 \pm 26$  		& $ 0.096^{+0.006}_{-0.006}$\\  
ZC413507  &  2.4794  &  2.4794  & 0.75  &   $4.73^{+1.81}_{-0.75}$  &  $63^{+124}_{-35}$  &  $4.1 \pm 1.0$  &   $< 46$  	& $ 0.078^{+0.031}_{-0.015}$\\  
ZC413597  &  2.4451  &  2.4498  & 1.00  &   $7.07^{+1.22}_{-1.16}$  &  $73^{+22}_{-48}$  &  $2.5 \pm 1.0$  &   $< 44$  		& $ 0.121^{+0.024}_{-0.023}$\\  
ZC415087  &  2.2947  &  2.2986  & 0.75  &  $2.44^{+0.93}_{-0.69}$  &  $96^{+29}_{-42}$  &  $< 2.2$  &   \ldots  		& $ 0.050^{+0.020}_{-0.016}$\\  
ZC415876  &  2.4308 &  2.4362  & 1.00  &   $3.52^{+0.62}_{-0.44}$  &  $116^{+28}_{-23}$  &  $2.7 \pm 0.9$  &   $194 \pm 26$  	& $ 0.050^{+0.010}_{-0.008}$\\  
\enddata
\tablecomments{
The integrated H$\alpha$ properties are computed from the spatially-integrated
spectrum in a circular aperture of radius $r_{\rm ap}$.
For undetected sources, the $3\,\sigma$ upper limit on the H$\alpha$ flux is
listed; it is based on the noise spectrum for a circular aperture of radius
$1^{\prime\prime}$, assuming an H$\alpha$ redshift equal to the optical redshift
and an intrinsic line width of $\rm 130~km\,s^{-1}$.
The integrated velocity width is corrected for instrumental velocity broadening,
and the half-light radius is corrected for the angular resolution of the respective
data sets.
The half-light radius upper limit for ZC415087 corresponds to the observed
measurement (uncorrected for resolution), which is smaller than the PSF FWHM
(see text).
}
\tablenotetext{a}
{
For ZC409129, no narrow line component as expected from an
origin predominantly from star formation is detected, and,
as for the other undetected sources, the upper limit refers
to a fiducial line width with sigma = 130 km/s, an aperture
of 1.0" in radius, and a line centroid corresponding to the
optical redshift.  However, a very broad component is detected,
consistent with the Type 2 AGN features clearly
seen in the rest-UV spectrum from zCOSMOS.  The estimated
flux and width of this broad component are
$\rm \approx 2.4 \times 10^{-16}~erg\,^{-1}\,cm^{-2}$ and
$\rm \approx 1000~km\,s^{-1}$.  For the purpose of the
analysis presented in this paper, which focuses on the
star formation properties, we use the upper limit on a
narrow component.
}
\end{deluxetable}															     
\end{turnpage}

\begin{turnpage}

\begin{deluxetable}{ccccccccccc}
\tabletypesize{\footnotesize}
\tablecaption{zC-SINF H$\alpha$ luminosity and SFR from H$\alpha$ and UV continuum \label{tbl-lum_sfr-6}}
\tablewidth{0pt}
\tablehead{
\colhead{\scriptsize{ID}}&\colhead{\scriptsize{L$_{obs}$}}&\colhead{\scriptsize{L$^0$(H$\alpha$)}}&\colhead{\scriptsize{$L^{00}$(H$\alpha$)}}&\colhead{\scriptsize{SFR$^0$(H$\alpha$)}}&\colhead{\scriptsize{SFR$^{00}$(H$\alpha$)}}&\colhead{\scriptsize{EW(H$\alpha$})}&\colhead{\scriptsize{EW$^{00}$(H$\alpha$)}}&\colhead{\scriptsize{SFR$_{UV}$}}&\colhead{\scriptsize{$E(B-V)_{UV}$}}\\
\colhead{ }&\colhead{\scriptsize{[$10^{42}$ erg s$^{-1}$]}}&\colhead{\scriptsize{[$10^{42}$ erg s$^{-1}$]}}&\colhead{\scriptsize{[$10^{42}$ erg s$^{-1}$]}}&\colhead{\scriptsize{[$M_{\odot}$ yr$^{-1}$]}}&\colhead{\scriptsize{[$M_{\odot}$ yr$^{-1}$]}}&\colhead{\scriptsize{[\AA]}}&\colhead{\scriptsize{[\AA]}}&\colhead{\scriptsize{[$M_{\odot}$ yr$^{-1}$]}}&\colhead{ }
} 
\startdata
    ZC400528    &     7.03  \scriptsize{$^{+0.67}_{-0.92}$  } &     13.9\scriptsize{$^{+2.51}_{-2.81}$    } &  33.3\scriptsize{$^{+12.0}_{-12.3}$}&    65.18\scriptsize{$^{+11.76}_{-13.13}$} & 	  155.67\scriptsize{$^{+56.06}_{-57.73}$}     &    60\scriptsize{$^{+6}_{-8}$}      & 144   \scriptsize{$^{+14}_{-19}$     }	    &148.41$\pm$    77.37 &    0.28$\pm{0.03}$  \\
    ZC400569    &     6.25  \scriptsize{$^{+0.58}_{-0.69}$  } &     18.1\scriptsize{$^{+3.25}_{-3.42}$    } &  70.2\scriptsize{$^{+25.2}_{-25.6}$}&    84.77\scriptsize{$^{+15.19}_{-15.99}$} & 	  328.35\scriptsize{$^{+118.02}_{-119.57}$}   &    40\scriptsize{$^{+3}_{-4}$}      & 157   \scriptsize{$^{+15}_{-17}$     }	    &168.23$\pm$    75.44 &    0.37$\pm{0.03}$  \\
    ZC401925    &     0.805 \scriptsize{$^{+0.07}_{-0.06}$  } &    1.37 \scriptsize{$^{+0.24}_{-0.23}$    } &  2.70\scriptsize{$^{+0.96}_{-0.95}$}&    6.41\scriptsize{$^{+1.14}_{-1.10}$}    & 	  12.62\scriptsize{$^{+4.53}_{-4.487}$}       &    37\scriptsize{$^{+5}_{-4}$}      &  73   \scriptsize{$^{+10}_{-9}$     }	    &31.02$\pm$     12.43 &    0.18$\pm{0.03}$  \\
    ZC403027    &     1.33  \scriptsize{$^{+0.33}_{-0.30}$  } &    3.56 \scriptsize{$^{+1.05}_{-0.99}$    } &  12.5\scriptsize{$^{+5.46}_{-5.31}$}&    16.66\scriptsize{$^{+4.93}_{-4.63}$}   & 	  58.59\scriptsize{$^{+25.52}_{-24.84}$}      &    28\scriptsize{$^{+7}_{-6}$}      & 100   \scriptsize{$^{+25}_{-23}$     }	    &179.18$\pm$    98.17 &    0.41$\pm{0.03}$  \\
    ZC403103    &     $<$1.7  			      &   $<$2.8  				    &  $<$5.0 				  &    $<$13.0		                      &  	  $<$23.2 				      &   $<$139 			    & $<$248					   &114.90$\pm$     34.11 &    0.26$\pm{0.03}$  \\
    ZC403741    &     1.56  \scriptsize{$^{+0.06}_{-0.05}$  } &    5.26 \scriptsize{$^{+0.83}_{-0.81}$    } &  24.7\scriptsize{$^{+8.62}_{-8.59}$}&    24.60\scriptsize{$^{+3.90}_{-3.83}$}   & 	 115.62\scriptsize{$^{+40.31}_{-40.17}$}      &    39\scriptsize{$^{+1}_{-1}$}      & 186   \scriptsize{$^{+8}_{-7}$     }	    &142.71 $\pm$   21.70 &    0.42$\pm{0.03}$  \\
    ZC404073    &     1.40  \scriptsize{$^{+0.35}_{-0.27}$  } &    4.37 \scriptsize{$^{+1.30}_{-1.10}$    } &  18.6\scriptsize{$^{+8.14}_{-7.60}$}&    20.45\scriptsize{$^{+6.07}_{-5.14}$}   & 	 87.26\scriptsize{$^{+38.07}_{-35.57}$}       &    23\scriptsize{$^{+5}_{-4}$}      &  98   \scriptsize{$^{+25}_{-20}$     }	    &225.53$\pm$    129.12&   0.43 $\pm{0.03}$ \\   
    ZC404221    &     3.30  \scriptsize{$^{+0.22}_{-0.32}$  } &    5.60 \scriptsize{$^{+0.93}_{-1.02}$    } &  11.0\scriptsize{$^{+3.89}_{-3.97}$}&    26.20\scriptsize{$^{+4.38}_{-4.76}$}   & 	 51.56\scriptsize{$^{+18.20}_{-18.56}$}       &   117\scriptsize{$^{+12}_{-15}$}    & 231   \scriptsize{$^{+25}_{-29}$     }	    &27.02$\pm$     11.90&    0.13 $\pm{0.03}$ \\   
    ZC404987    &     2.40  \scriptsize{$^{+0.34}_{-0.60}$  } &    5.11 \scriptsize{$^{+1.08}_{-1.53}$    } &  13.4\scriptsize{$^{+5.08}_{-5.80}$}&    23.88\scriptsize{$^{+5.04}_{-7.13}$}   & 	 62.83\scriptsize{$^{+23.74}_{-27.13}$}       &   126\scriptsize{$^{+24}_{-35}$}    & 332    \scriptsize{$^{+63}_{-94}$  }	    &31.15$\pm$     12.38&    0.23 $\pm{0.03}$ \\    
    ZC405081    &     1.92  \scriptsize{$^{+0.27}_{-0.35}$  } &    3.81 \scriptsize{$^{+0.79}_{-0.92}$    } &  9.11\scriptsize{$^{+3.43}_{-3.60}$}&    17.84\scriptsize{$^{+3.73}_{-4.30}$}   & 	 42.61\scriptsize{$^{+16.06}_{-16.83}$}       &   104\scriptsize{$^{+20}_{-24} $}   & 248    \scriptsize{$^{+50}_{-58}$   }	    &21.82$\pm$     9.46&     0.17 $\pm{0.04}$ \\  
ZC405226    &     0.69  \scriptsize{$^{+0.13}_{-0.12}$  } &    1.49 \scriptsize{$^{+0.36}_{-0.34}$    } &  3.92\scriptsize{$^{+1.55}_{-1.53}$}&    7.0\scriptsize{$^{+1.6}_{-1.6}$}   & 	 18.4\scriptsize{$^{+7.3}_{-7.2}$}       &    19\scriptsize{$^{+3}_{-3}$}      & 51\scriptsize{$^{+10}_{-9} $   }	    &87.76 $\pm$    40.95&    0.29 $\pm{0.03}$ \\    
ZC405430    &    $<2.2$   &   $<4.3$  & $<10.2$ &    $<20.0$			              &       		$<47.7$				      &   $<111$  &  $<265$	    & 68.78$\pm$    21.9&     0.30 $\pm{0.05}$ \\
ZC405501    &    3.93  \scriptsize{$^{+0.55}_{-0.46}$  } &   7.79 \scriptsize{$^{+1.62}_{-1.50}$    } & 18.61\scriptsize{$^{+7.00}_{-6.85}$}&  36.4\scriptsize{$^{+7.6}_{-7.0}$}			              &	      	87.1\scriptsize{$^{+32.7}_{32.0}$} &   125\scriptsize{$^{+21}_{-19}$}    & 300\scriptsize{$^{+52}_{-46}$  }	    & 67.96$\pm$16.21&    0.19 $\pm{0.03}$  \\  
    ZC405545    &     $<$0.4  			      &    $<$0.9 				    & $<$2.7				  &    $<$4.5			              &	    		$<$13.0	  			      &    $<$12			            & $<$35  				            &28.16$\pm$     12.06&    0.26 $\pm{0.04}$ \\
ZC406690    &   22.60\scriptsize{$^{+0.51}_{-0.40}$  } &    38.47\scriptsize{$^{+5.91}_{-5.90}$} & 75.7\scriptsize{$^{+26.2}_{-26.2}$}&   179.9\scriptsize{$^{+27.6}_{-27.5}$}			 	      &	              354\scriptsize{$^{+122.6}_{-122.5}$}				      &   199\scriptsize{$^{+6}_{-6}$}	    &  391 \scriptsize{$^{+13}_{-12}$	   }	    & 337.00$\pm$   83.40 &   0.22  $\pm{0.03}$ \\ 
    ZC407302    &     8.17  \scriptsize{$^{+0.34}_{-0.27}$  } &     21.9\scriptsize{$^{+3.46}_{-3.42}$    } &  77.2\scriptsize{$^{+26.9}_{-26.8}$}&    102.70\scriptsize{$^{+16.20}_{-16.00}$}&	361.14\scriptsize{$^{+125.78}_{- 125.46}$}    	      &   125\scriptsize{$^{+6}_{-6}$    }  &  441 \scriptsize{$^{+23}_{-21} $  }	    &130.41$\pm$    54.72 &   0.28  $\pm{0.03}$\\ 
    ZC407376    &     5.53  \scriptsize{$^{+0.70}_{-0.50}$  } &     13.8\scriptsize{$^{+2.74}_{-2.45}$    } &  44.0\scriptsize{$^{+16.3}_{-15.8}$}&    64.41\scriptsize{$^{+12.80}_{-11.44}$} & 205.63\scriptsize{$^{+76.19}_{-74.00}$}	      	      &   112\scriptsize{$^{+15}_{-11}$  }  &  359 \scriptsize{$^{+49}_{-38}$     }	    &84.53$\pm$     35.76 &   0.35  $\pm{0.03}$\\ 
ZC407928    &   3.77   \scriptsize{$^{+0.72}_{-0.74}$  } &    8.05\scriptsize{$^{+1.98}_{-2.02}$} & 21.18\scriptsize{$^{+8.48}_{-8.55}$}&   37.67\scriptsize{$^{+9.27}_{-9.46}$} &	  99.09\scriptsize{$^{+39.70}_{-40.00}$} 	&   141\scriptsize{$^{+31}_{-32}$  }  & 371\scriptsize{$^{+83}_{-85}$}          & 117.70$\pm$    37.24 &   0.30  $\pm{0.04}$ \\  

    ZC409129    &     $<$3.3   			      &    $<$9.0     			    &  $<$31.5 				  &    $<$41.9	 			      & $<$147.5	      				      &    $<$39			    &  $<$136					    &119.06$\pm$    66.64 &                                                                                                                                                    0.36  $\pm{0.03}$\\  

ZC409985    &     1.89\scriptsize{$^{+0.34}_{-0.29}$  } &    2.98\scriptsize{$^{+0.71}_{-0.65}$    } &  5.32\scriptsize{$^{+2.10}_{-2.04}$}&    13.93\scriptsize{$^{+3.31}_{-3.03}$}   & 	24.89\scriptsize{$^{+9.83}_{-9.54}$}       &   46\scriptsize{$^{+9}_{-8}$ }&  83\scriptsize{$^{+16}_{-14}$} 	 &55.70$\pm$30.86 &    0.22$\pm{0.03}$ \\
    ZC410041    &     6.48  \scriptsize{$^{+1.39}_{-1.28}$  } &    10.2 \scriptsize{$^{+2.71}_{-2.57}$    } &  18.3\scriptsize{$^{+7.56}_{-7.40}$}&    47.86\scriptsize{$^{+12.69}_{-12.02}$} &85.50\scriptsize{$^{+35.36 }_{-34.62}$}	              &   553 \scriptsize{$^{+150 }_{-142}$}&  988   \scriptsize{$^{+ 268}_{-255} $  }	 &37.22$\pm$	 20.80 &  0.18  $\pm{0.03}$\\
    ZC410116    &     $<$2.6  			      &    $<$8.8  				    &  $<$41.4 				  &    $<$41.2  			      &    $<$193.8	      				      &    $<$38			    &  $<$178					    &107.18$\pm$    43.23 &                                                                                                                                                            0.44  $\pm{0.03}$\\ 
    ZC410123    &     0.632 \scriptsize{$^{+0.07}_{-0.09}$  } &     1.16\scriptsize{$^{+0.22}_{-0.24}$    } &  2.52\scriptsize{$^{+0.92}_{-0.95}$}&    5.43\scriptsize{$^{+1.05}_{-1.17}$}    &       11.77\scriptsize{$^{ +4.33}_{-4.46}$}	      &    29  \scriptsize{$^{+4}_{-5}$    }&   64  \scriptsize{$^{+10}_{-11}$} 	    &35.26$\pm$     14.85 &  0.20  $\pm{0.03}$\\ 
    ZC410542    &     2.23  \scriptsize{$^{+0.17}_{-0.24}$  } &    7.52 \scriptsize{$^{+1.29}_{-1.41}$    } &  35.3\scriptsize{$^{+12.6}_{-12.8}$}&    35.15\scriptsize{$^{+6.02}_{-6.62}$}   & 	  165.24\scriptsize{$^{+58.70}_{-60.06}$}     &    39  \scriptsize{$^{+3}_{-4}$    }&  187  \scriptsize{$^{+ 14}_{-20}$    }	    &137.47$\pm$    29.10 &  0.43  $\pm{0.03}$\\  
    ZC411737    &     4.61  \scriptsize{$^{+0.45}_{-0.33}$  } &    7.28 \scriptsize{$^{+1.32}_{-1.23}$    } &  13.0\scriptsize{$^{+4.69}_{-4.61}$}&    34.05\scriptsize{$^{+6.18}_{-5.73}$}   & 	   60.84\scriptsize{$^{+21.95}_{-21.56}$ }    &   219  \scriptsize{$^{+37}_{-34}$  }&  391  \scriptsize{$^{+66}_{-60} $   }	    &34.11$\pm$     18.83 &  0.17  $\pm{0.03}$\\  
ZC412369    &     6.53\scriptsize{$^{+0.31}_{-0.29}$  } &    13.97\scriptsize{$^{+2.23}_{-2.21}$    } &  36.75\scriptsize{$^{+12.83}_{-12.81}$}&    65.34\scriptsize{$^{+10.42}_{-10.34}$}   &        171.90\scriptsize{$^{+60.02}_{-59.93}$}        &    103\scriptsize{$^{+6}_{-6}$} &  271\scriptsize{$^{+17}_{-16}$}	    &129.76$\pm$47.29 &	     0.34$\pm{0.03}$  \\

    ZC413507    &     2.33  \scriptsize{$^{+0.89}_{-0.37}$  } &     5.39\scriptsize{$^{+2.24}_{-1.19}$    } &  15.6\scriptsize{$^{+8.31}_{-6.28}$}&    25.20\scriptsize{$^{+10.48}_{-5.57}$}  &   73.02\scriptsize{$^{+38.86}_{-29.39}$}      	      &    71  \scriptsize{$^{+28}_{-13}$  }&  208  \scriptsize{$^{+82}_{-39}$    }	    &86.41$\pm$     47.68	 &   0.31  $\pm{0.03}$\\      
    ZC413597    &     3.39  \scriptsize{$^{+0.58}_{-0.55}$  } &     7.24\scriptsize{$^{+1.68}_{-1.63}$    } &  19.0\scriptsize{$^{+7.45}_{-7.38}$}&    33.88\scriptsize{$^{+7.84}_{-7.62}$}   &   89.13\scriptsize{$^{+34.82}_{-34.50}$}     	      &   118  \scriptsize{$^{+23}_{-22}$  }&  310  \scriptsize{$^{+61}_{-59} $  }	    &76.41$\pm$     41.01 &	     0.31  $\pm{0.03}$\\      
    ZC415087    &     1.00  \scriptsize{$^{+0.38}_{-0.28}$  } &     1.84\scriptsize{$^{+0.76}_{-0.59}$    } &  3.98\scriptsize{$^{+2.12}_{-1.85}$}&    8.60\scriptsize{$^{+3.56}_{-2.78}$}    &        18.64\scriptsize{$^{+9.90}_{-8.67}$}    	      &    47  \scriptsize{$^{+19}_{-14}$  }&  102  \scriptsize{$^{+41}_{-32}$     }	    &44.87$\pm$     21.20 &	     0.25  $\pm{0.04}$\\      
    ZC415876    &      1.67 \scriptsize{$^{+0.29}_{-0.20}$  } &   3.57  \scriptsize{$^{+0.83}_{-0.70}$    }&  9.39 \scriptsize{$^{+3.68}_{-3.50}$}&    16.70\scriptsize{$^{+3.90}_{-3.30}$ }  &	   43.93\scriptsize{$^{+17.23}_{-16.35}$}  	      &	 45  \scriptsize{$^{+9}_{-7}$	 }  &  118  \scriptsize{$^{+24}_{-19}$   }	    &72.76$\pm$ 39.09 &      0.30  $\pm{0.03}$\\   	
\enddata		     
\tablecomments
{
For sources undetected in H$\alpha$, 3 $\sigma$ upper limits are given for the H$\alpha$ luminosities, the corresponding SFRs, and for the H$\alpha$ equivalent widths.
}
\end{deluxetable}	     						    
\end{turnpage}

\clearpage

\begin{table}
\begin{center}
\caption{\small{Median properties for the zC-SINF, SINS, and L09 samples compared in Figures~\ref{fig:FHa-sigma-re}--\ref{fig:M-sigma-re}.}\label{tbl:comparisons}}
\begin{tabular}{l|c|c|c|c|c}
\tableline

Sample	&	$M_{\rm \star,med}$      &  $F({\rm H\alpha})_{\rm med}$        &     $L({\rm H\alpha})_{\rm med}$ &   $\sigma({\rm H\alpha})_{\rm integrated,med}$ &  $r_{1/2}({\rm H\alpha})_{\rm med}$ \\
        &[$10^{10} M_{\odot}$]	 & [$10^{-17} \rm erg\ cm^{-2}\ s^{-1}$]  &     [$10^{42} \rm erg\ s^{-1}$]   &    [km $\rm s^{-1}$]                           &         [kpc]                       \\
\tableline
zC-SINF	& 1.3    &   7.9   &   2.4  & 123   &  2.7  \\

SINS	&  2.6   &   11.0  &   3.5  & 130   &  3.1   \\

L09	& 1.2	  &   9.0   &   3.4  & 70    &  1.4    \\
\tableline
\end{tabular}
\end{center}
\end{table}

\clearpage

\addtocounter{figure}{-1}
\begin{figure*}[t]
\centering
\includegraphics[width=0.9\textwidth]{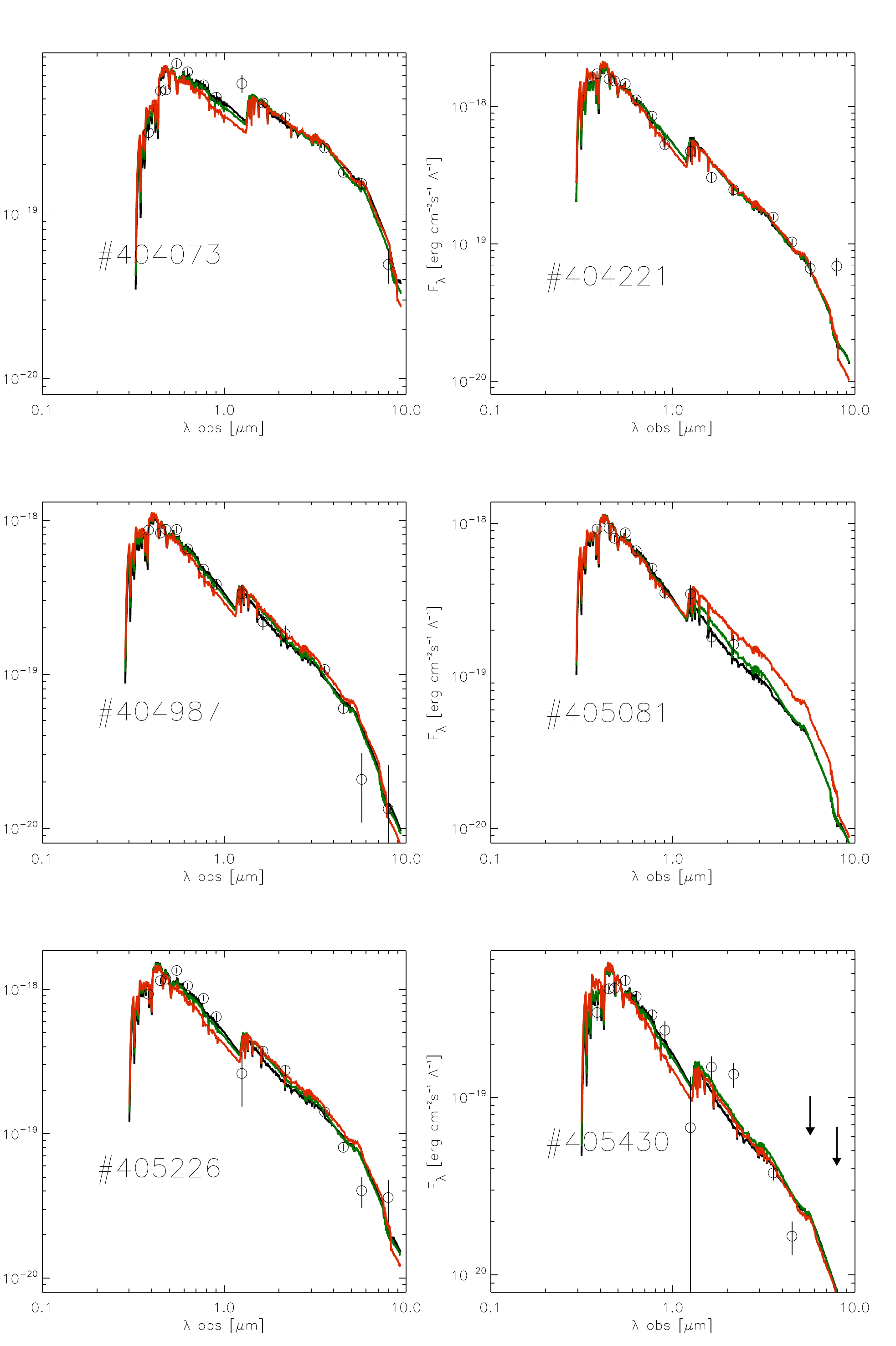}
\caption{\small{Continue. }}
\end{figure*}

\addtocounter{figure}{-1}
\begin{figure*}[t]
\centering
\includegraphics[width=0.9\textwidth]{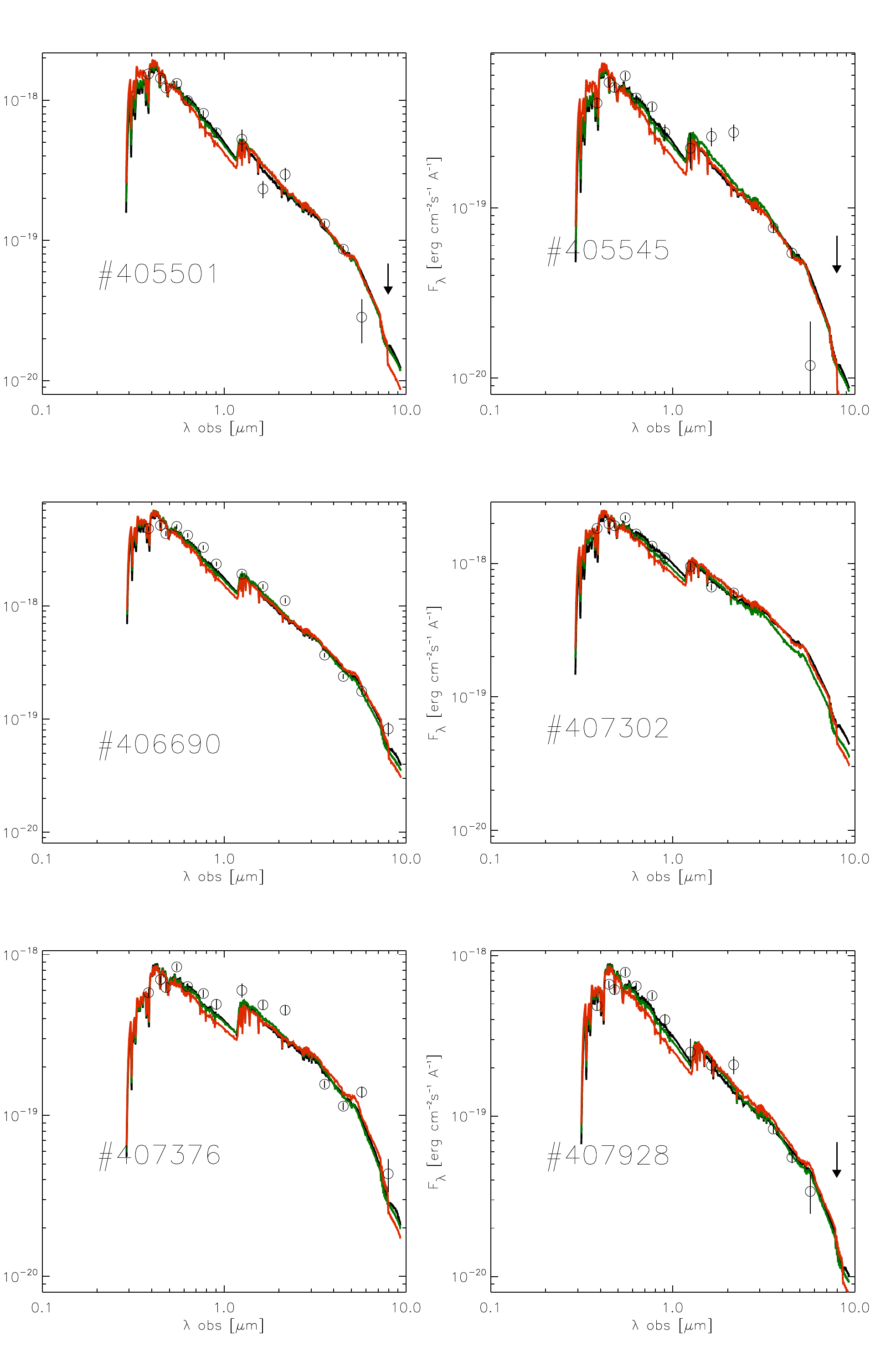}
\caption{\small{Continue.}}
\end{figure*}

\addtocounter{figure}{-1}
\begin{figure*}[t]
\centering
\includegraphics[width=0.9\textwidth]{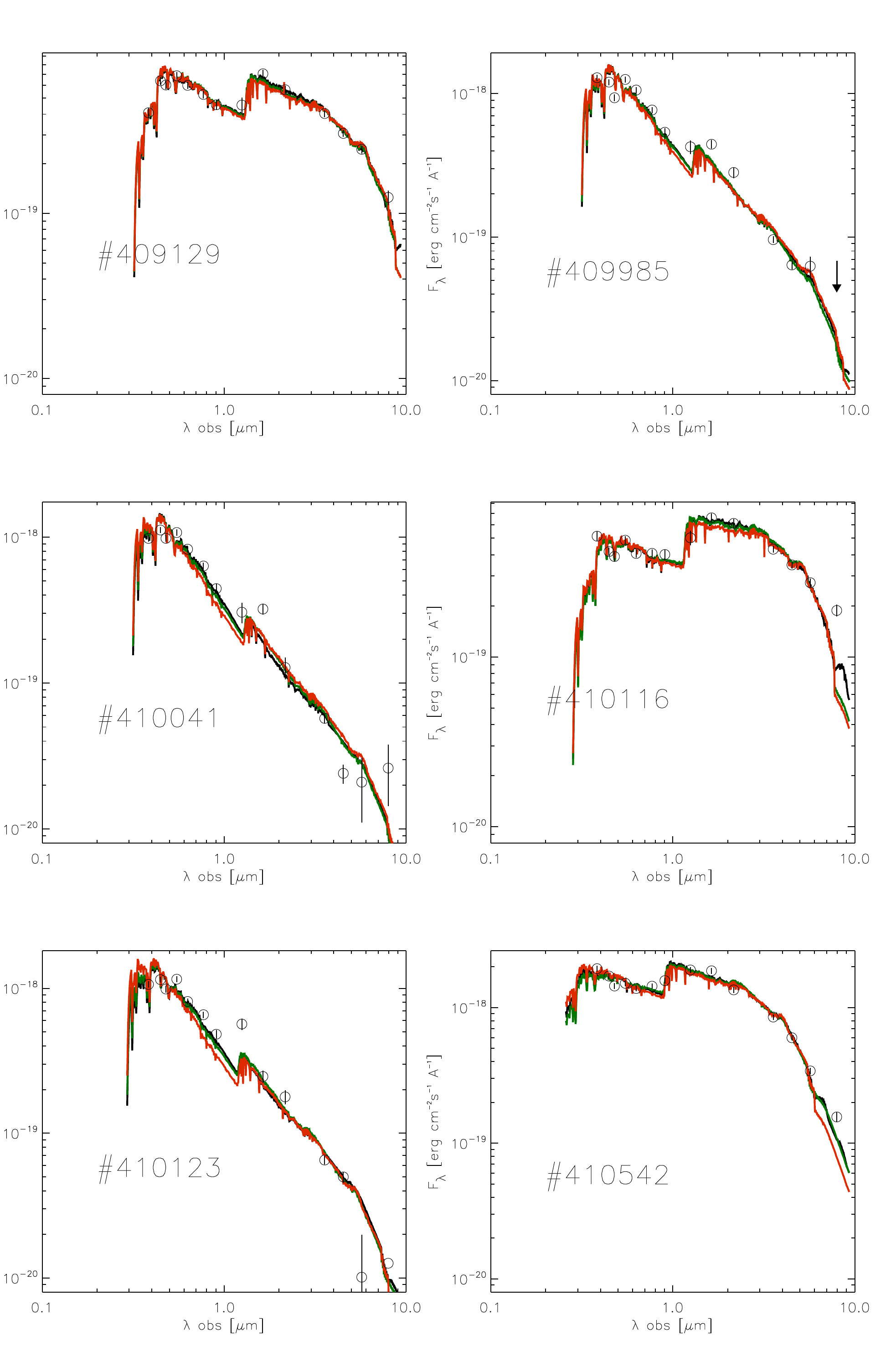}
\caption{\small{Continue.}}
\end{figure*}

\addtocounter{figure}{-1}
\begin{figure*}[t]
\centering
\includegraphics[width=0.9\textwidth]{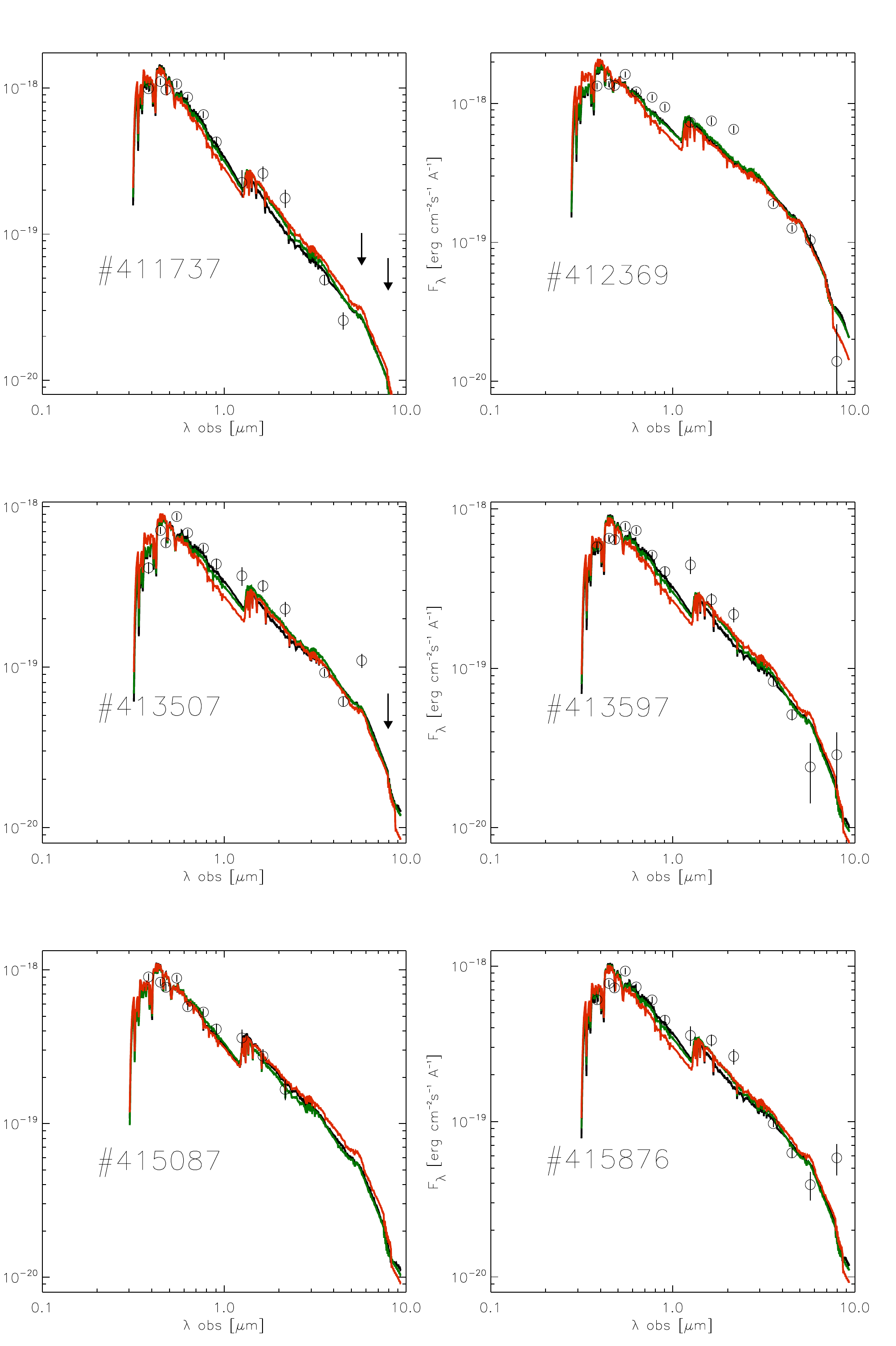}
\caption{\small{Continue.}}
\end{figure*}

\begin{figure*}[h]
\centering
\includegraphics[width=\textwidth]{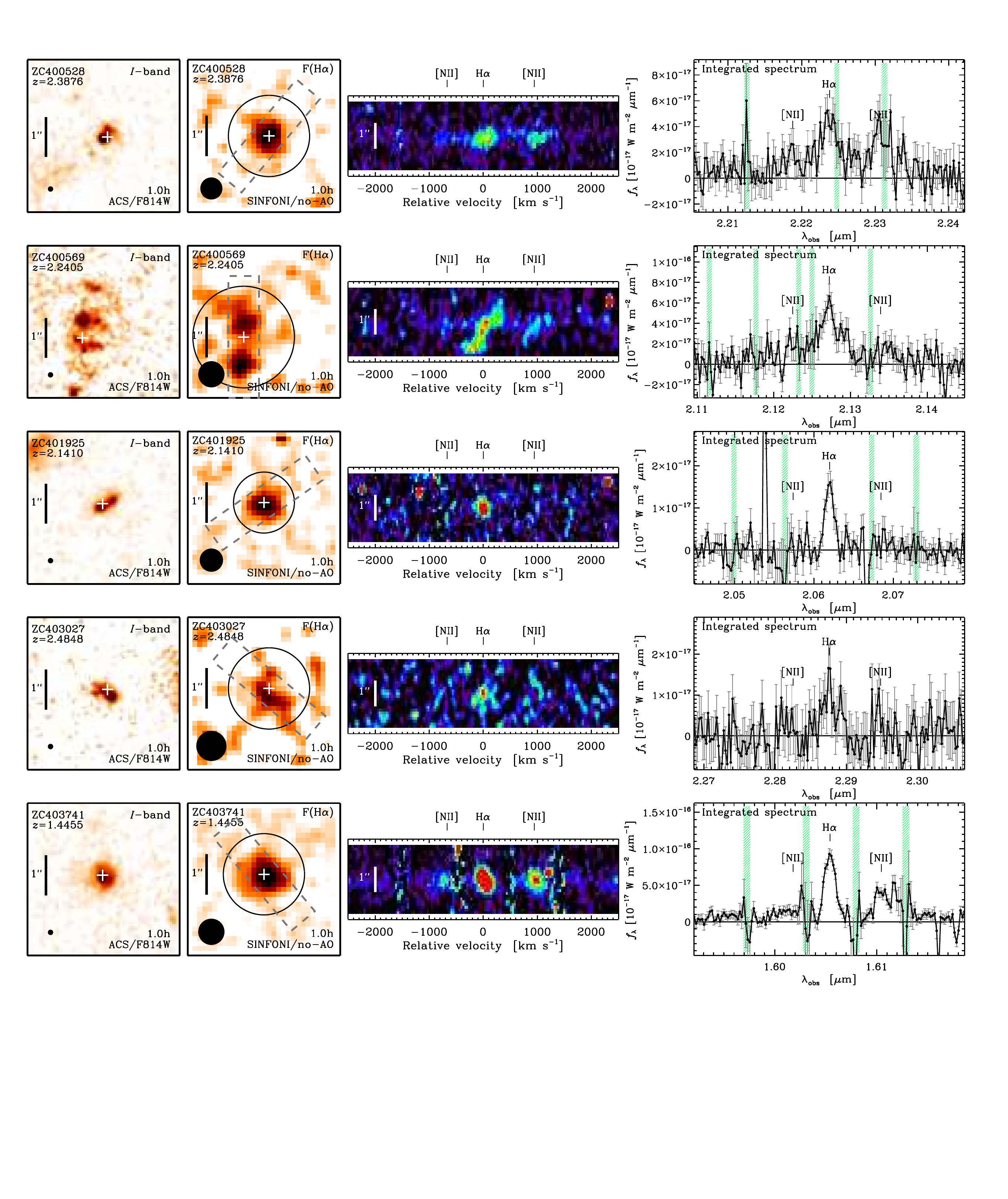}
\caption{\small{From left to right we show the $I$-band stamps from the COSMOS HST+ACS/F814W images, and the H$\alpha$ line maps, position-velocity diagrams, and integrated spectra obtained from SINFONI natural seeing (no-AO) observations, for the H$\alpha$-detected zC-SINF sources. The circular aperture for the integrated spectrum and the synthetic slit used to extract the $p-v$ diagrams along the major axis are indicated on the H$\alpha$ maps.  Vertical green hatched bars in the spectra show the locations of bright night sky lines, with width corresponding to the FWHM of the effective spectral resolution of the data.}}\label{fig:sinfoni}
\end{figure*}
\addtocounter{figure}{-1}

\begin{figure*}[t]
\centering
\includegraphics[width=\textwidth]{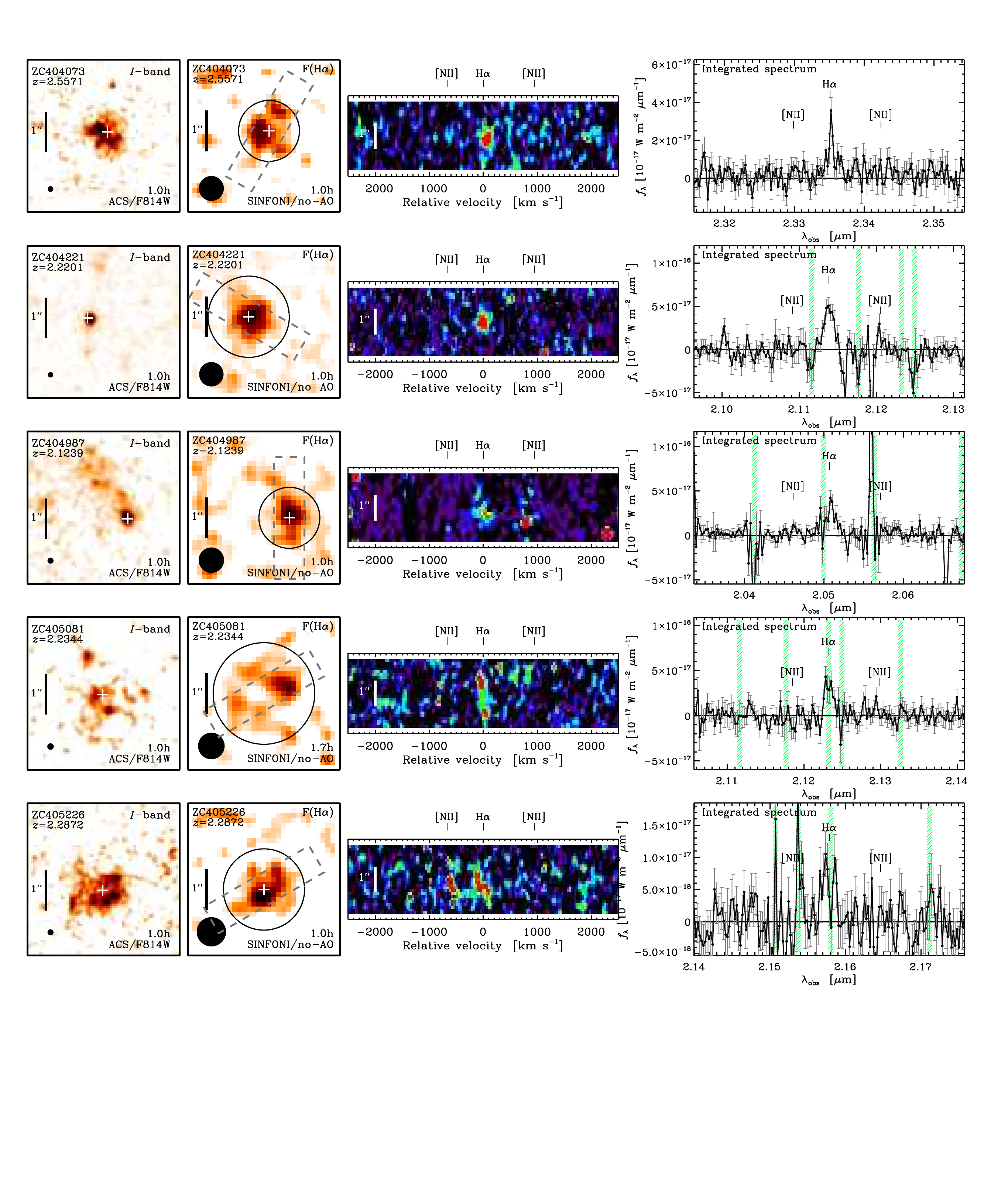}
\caption{\small{Continue.}}
\end{figure*}

\addtocounter{figure}{-1}
\centering
\begin{figure*}[t]
\includegraphics[width=\textwidth]{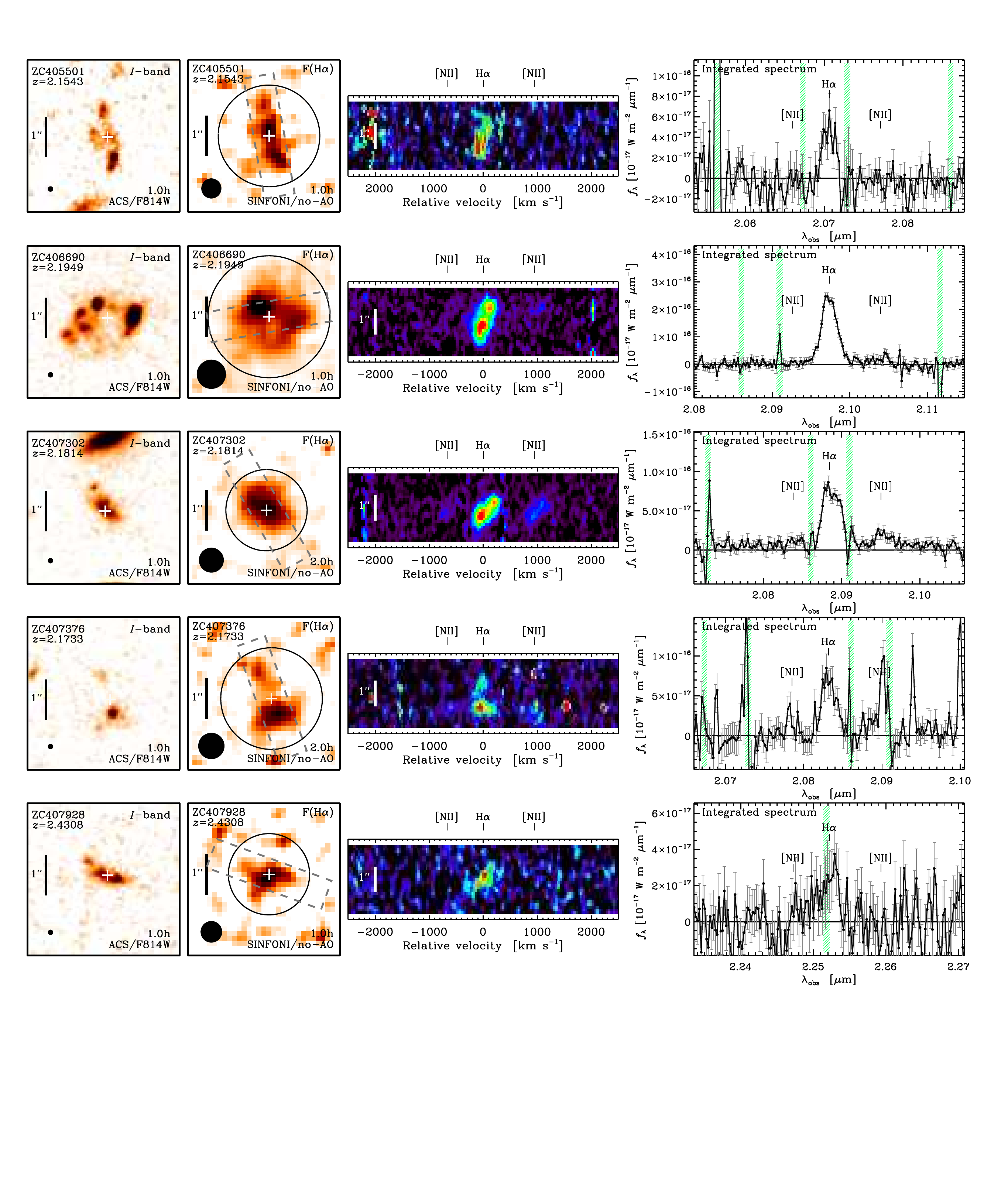}
\caption{\small{Continue.}}
\end{figure*}
\addtocounter{figure}{-1}

\begin{figure*}[t]
\centering
\includegraphics[width=\textwidth]{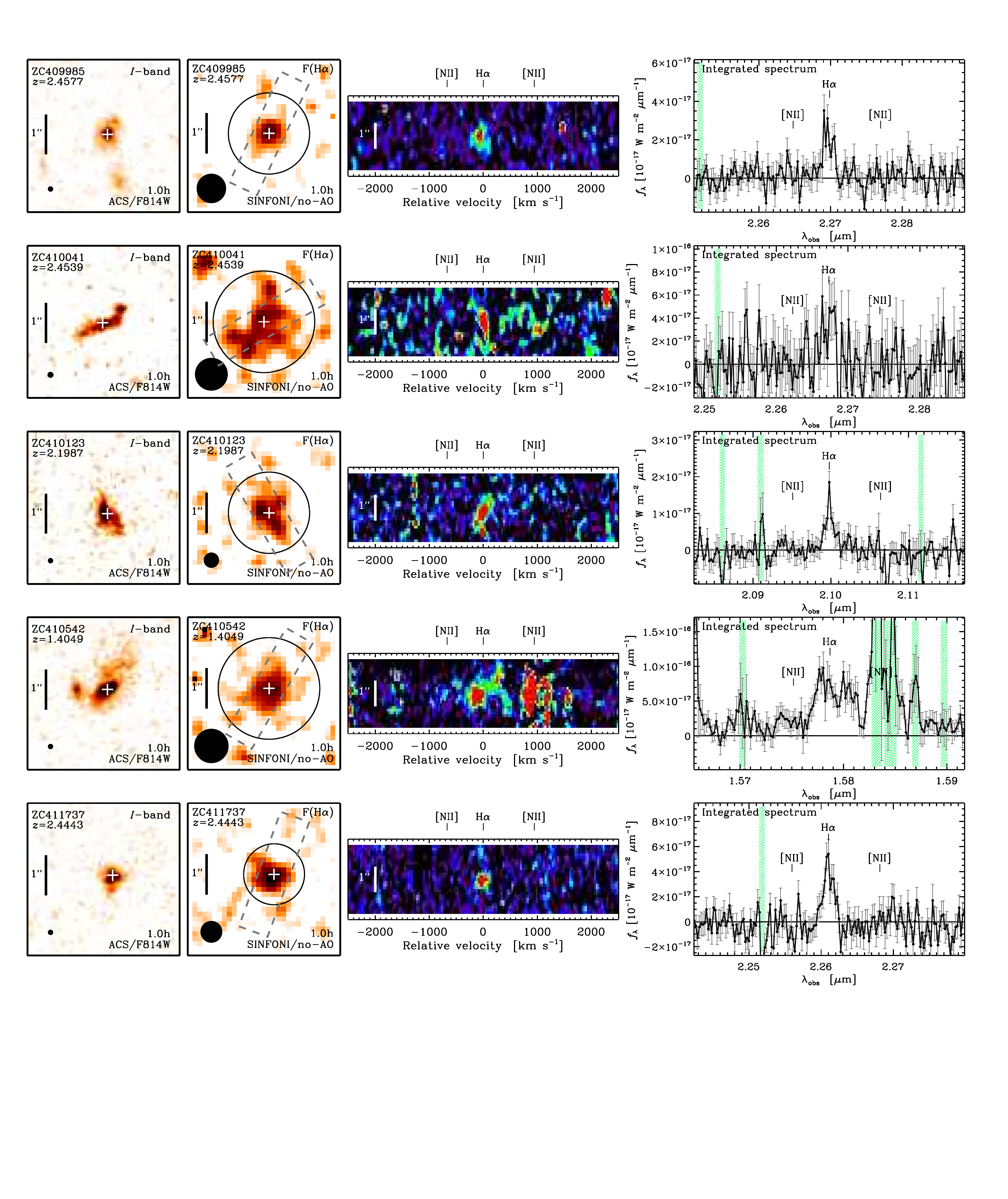}
\caption{\small{Continue.}}
\end{figure*}
\addtocounter{figure}{-1}

\begin{figure*}[t]
\centering
\includegraphics[width=\textwidth]{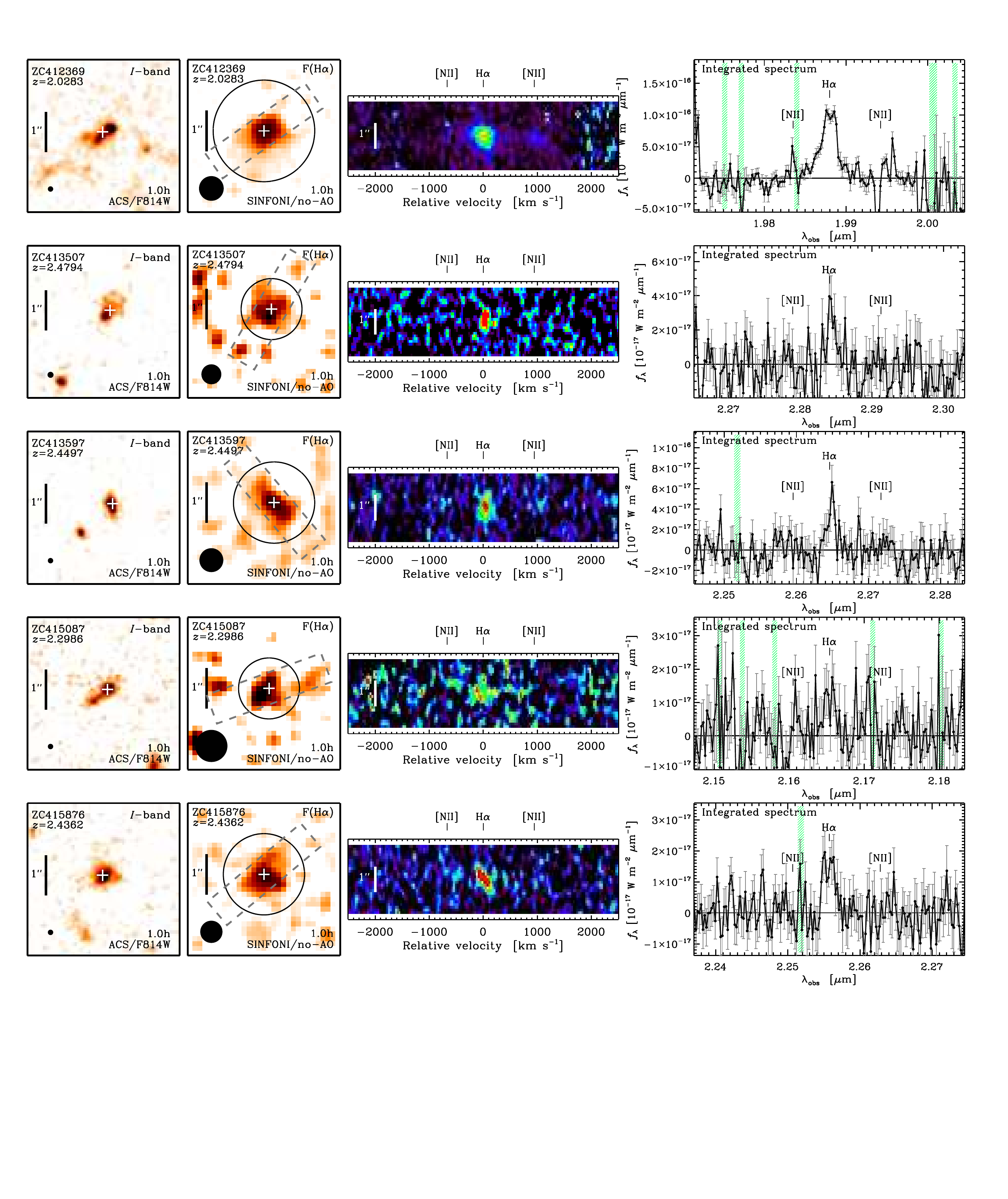}
\caption{\small{Continue.}}
\end{figure*}

\end{document}